\documentclass[prd,aps,showpacs,epsf,floats,onecolumn]{revtex4}%
\usepackage{amssymb}
\usepackage{amsfonts}
\usepackage{amsmath}
\usepackage{graphicx}%
\providecommand{\U}[1]{\protect\rule{.1in}{.1in}}
\begin{document}
\title{\textbf{From the Classical Frenet-Serret Apparatus to the Curvature and
Torsion of Quantum-Mechanical Evolutions. Part I. Stationary Hamiltonians}}
\author{\textbf{Paul M. Alsing}$^{1}$ and \textbf{Carlo Cafaro}$^{2,3}$}
\affiliation{$^{1}$Air Force Research Laboratory, Information Directorate, Rome, NY 13441, USA}
\affiliation{$^{2}$University at Albany-SUNY, Albany, NY 12222, USA}
\affiliation{$^{3}$SUNY Polytechnic Institute, Utica, NY 13502, USA}

\begin{abstract}
It is known that the Frenet-Serret apparatus of a space curve in
three-dimensional Euclidean space determines the local geometry of curves. In
particular, the Frenet-Serret apparatus specifies important geometric
invariants, including the curvature and the torsion of a curve. It is also
acknowledged in quantum information science that low complexity and high
efficiency are essential features to achieve when cleverly manipulating
quantum states that encode quantum information about a physical system.

In this paper, we propose a geometric perspective on how to quantify the
bending and the twisting of quantum curves traced by dynamically evolving
state vectors. Specifically, we propose a quantum version of the Frenet-Serret
apparatus for a quantum trajectory in projective Hilbert space traced by a
parallel-transported pure quantum state evolving unitarily under a stationary
Hamiltonian specifying the Schr\"{o}dinger equation. Our proposed constant
curvature coefficient is given by the magnitude squared of the covariant
derivative of the tangent vector $\left\vert T\right\rangle $ to the state
vector $\left\vert \Psi\right\rangle $ and represents a useful measure of the
bending of the quantum curve. Our proposed constant torsion coefficient,
instead, is defined in terms of the magnitude squared of the projection of the
covariant derivative of the tangent vector $\left\vert T\right\rangle $,
orthogonal to both $\left\vert T\right\rangle $ and $\left\vert \Psi
\right\rangle $. The torsion coefficient provides a convenient measure of the
twisting of the quantum curve. Remarkably, we show that our proposed curvature
and torsion coefficients coincide with those existing in the literature,
although introduced in a completely different manner. Interestingly, not only
we establish that zero curvature corresponds to unit geodesic efficiency
during the quantum transportation in projective Hilbert space, but we also
find that the concepts of curvature and torsion help enlighten the statistical
structure of quantum theory. Indeed, while the former concept can be
essentially defined in terms of the concept of kurtosis, the positivity of the
latter can be regarded as a restatement of the well-known Pearson inequality
that involves both the concepts of kurtosis and skewness in mathematical
statistics. Finally, not only do we present illustrative examples with nonzero
curvature for single-qubit time-independent Hamiltonian evolutions for which
it is impossible to generate torsion, but we also discuss physical
applications extended to two-qubit stationary Hamiltonians that generate
curves with both nonzero curvature and nonvanishing torsion traced by quantum
states with different degrees of entanglement, ranging from separable states
to maximally entangled Bell states. In an appendix, we examine the different
curvature and torsion characteristics of the three qubit $|GHZ\rangle$ and
$|W\rangle$ states under evolution by a quantum Heisenberg Hamiltonian.

\end{abstract}

\maketitle

\section{Introduction}

Geometric reasoning is a powerful tool in theoretical physics to improve our
description and, to a certain extent, to sharpen our comprehension of physical
phenomena in both classical and quantum settings by providing deep physical
insights \cite{pettini07}. For instance, it is acknowledged that the classical
Frenet-Serret apparatus of a space curve in three-dimensional Euclidean space
determines the local geometry of curves and specifies important geometric
invariants, including the curvature and the torsion of a curve \cite{parker77}%
. In classical mechanics, the concepts of curvature and torsion introduced
within the Frenet-Serret apparatus can be helpful in studying geometric
properties of classical Newtonian trajectories of a particle. For instance,
curvature and torsion can be used to specify the geometry of the cylindrical
helix motion of an electron in a homogeneous external magnetic field
\cite{consa18}. It is also known that the clever manipulation of quantum
states that encode quantum information about a physical system is an extremely
valuable skill in quantum information science \cite{nielsen00}. In quantum
mechanics, motivated by the problem of parameter estimation, the concept of
curvature of a quantum Schr\"{o}dinger trajectory was originally introduced in
Ref. \cite{brody96} as a generalization of the notion of curvature of a
classical exponential family of distributions of relevance to statistical
mechanics. In the context of geometry of quantum statistical inference of Ref.
\cite{brody96}, the curvature of a curve can be expressed in terms of the
suitably defined squared acceleration vector of the curve and is a measure of
the parametric sensitivity \cite{brody13} that specifies the particular
parametric estimation problem being under consideration. In Ref.
\cite{laba17}, instead, Laba and Tkachuk proposed a definition of both
curvature and torsion of quantum evolutions for pure quantum state undergoing
a time-independent Hamiltonian evolution. In their work, focusing on
single-qubit quantum states, curvature measured the deviation of the
dynamically evolving state vector from the geodesic line on the Bloch sphere.
Instead, their proposed torsion concept quantified the deviation of the
dynamically evolving state vector from a two-dimensional subspace specified by
the instantaneous plane of evolution. Interestingly, building on the formalism
developed in Ref. \cite{laba17}, the concepts of curvature and torsion have
been recently used to study the geometric properties of different types of
graph states of spin systems evolving under Ising-like interactions in Ref.
\cite{laba22}.

In this paper, we are interested in quantifying the concepts of curvature and
torsion of a quantum trajectory for several reasons.\ First, we are interested
in understanding how to \emph{bend} and \emph{twist }quantum-mechanical
evolutions of quantum states that encode relevant quantum information about
the physical systems being observed. This is not only important from a
conceptual standpoint, it can also be especially relevant in experimental
quantum laboratory settings \cite{zutic04,hanson07}. Second, we are interested
in comprehending the possible link between the curvature (and/or the torsion)
and the complexity of a path traced out by a quantum state driven from a
source state to a target state \cite{carloprd22,carlopre22}. Finally, we are
interested in finding out if we can suitably manipulate via bending and
twisting a trajectory traced out by a quantum state in an efficient manner so
that one optimizes the travel time, maximizes the speed of evolution and,
possibly, minimizes possible dissipative effects of thermodynamical origin
that can emerge in the physical system being analyzed
\cite{carloijqi19,carlopra20,carlocqg23}.

In this paper, we present a geometric approach to characterize the
\emph{bending} and the \emph{twisting} of quantum curves traced out by
evolving state vectors. More precisely, we offer a quantum version of the
classical Frenet-Serret apparatus for a quantum trajectory on the Bloch sphere
traced out by a parallel-transported pure quantum state developing unitarily
subject to a time-independent Hamiltonian that specifies the Schr\"{o}dinger
equation. Indeed, we remark that our formalism is not limited to single-qubit
two-dimensional complex Hilbert spaces and to quantum curves on the standard
Bloch sphere $%
\mathbb{C}
P^{1}=\mathcal{S}^{3}/\mathcal{S}^{1}=\mathcal{S}^{2}$ with $\mathcal{S}^{k}$
denoting the $k$-sphere. Instead, it applies in principle to any
$N$-dimensional complex Hilbert space and to quantum curves on generalized
\textquotedblleft Bloch spheres\textquotedblright\ $%
\mathbb{C}
P^{N-1}=\mathcal{S}^{2N-1}/\mathcal{S}^{1}$ \cite{karol17}. We propose a
curvature coefficient defined as the magnitude squared of the covariant
derivative of the tangent vector to the state vector and represents a suitable
measure of the bending of the quantum curve. We also suggest a concept of
torsion coefficient, instead, specified by means of the magnitude squared of
the projection of the covariant derivative of the tangent vector, orthogonal
to the state vector as well as to the tangent vector to the state vector. Our
proposed torsion coefficient is a good measure of the twisting of the quantum
curve. Remarkably, as a by-product, we demonstrate that our proposed curvature
and torsion coefficients are identical to those proposed by Laba and Tkachuk
in Ref. \cite{laba17}, although we justify our proposals inspired by the
classical Frenet-Serret apparatus. Finally, not only we consider illustrative
examples for pedagogical purposes, we also discuss the generalization of our
theoretical construct to time-dependent quantum-mechanical scenarios where
both curvature and torsion coefficients play a key role.

The layout of the rest of this paper is as follows. In\ Sec. II, we present
some background material that focuses on the curvature and torsion
coefficients of a quantum evolution as originally proposed by Laba and Tkachuk
in Ref. \cite{laba17}. In Sec. III, in preparation of our newly proposed
theoretical construct in Sec. IV, we recall the essential ingredients of a
classical Frenet-Serret apparatus with special emphasis on the notions of
bending and twisting as captured by the curvature and the torsion of a curve
in three-dimensional Euclidean space. In Sec. IV, we present our quantum
version of the classical Frenet-Serret apparatus suitable for quantifying the
bending and the twisting of a quantum curve traced out by a
parallel-transported unit state vector that evolves under the action of a
time-independent Hamiltonian. Remarkably, although obtained from an
alternative perspective that mimics a classical apparatus, we find that our
proposed curvature and torsion coefficients coincide with the ones proposed by
Laba and Tkachuk in Ref. \cite{laba17}. In Sec. V, we present several points
of discussion. First, we discuss the statistical interpretation of the
curvature and torsion coefficients.\ Second, we illustrate the usefulness of
recasting the expressions of these two coefficients in terms of the Bloch
vector for two-level systems. Third, we elaborate on several challenges that
can emerge in higher-dimensional Hilbert spaces with quantum evolutions
governed by nonstationary Hamiltonians. Finally, we present in Sec. V a
comparison between our proposed quantum apparatus and the classical
Frenet-Serret one. In Sec. VI, we demonstrate simple illustrative examples of
the behavior of curvature and torsion coefficients \ for quantum evolutions
specified by single-qubit and two-qubit time-independent Hamiltonians. In
Section VII, we present our conclusive remarks. Finally, an illustrative
example on how to frame a quantum curve can be found in Appendix A, a link
between the concept of geodesic curvature \cite{krey91,kazaras12,erdos00} and
our proposed curvature coefficient appears in Appendix B and, lastly, in
Appendix C we report on the behavior of curvature and torsion coefficients for
quantum curves traced by three-qubit quantum states evolving under a quantum
Heisenberg model Hamiltonian.

\section{The quantum Laba-Tkachuk framework}

In this section, we report on some relevant background material that puts the
emphasis on how to introduce suitable measures of curvature and torsion of a
quantum evolution. Specifically, we revisit the work that was originally
proposed by Laba and Tkachuk in Ref. \cite{laba17}.

\subsection{Curvature}

In Ref. \cite{laba17}, Laba and Tkachuk propose a concept of curvature
coefficient for a Schr\"{o}dinger quantum-mechanical evolution specified by a
time-independent Hamiltonian \textrm{H}. The curvature coefficient proposed in
Ref. \cite{laba17} emerges by quantifying the departure of the unit evolution
vector $\left\vert \psi\left(  t\right)  \right\rangle =e^{(-i/\hslash
)\mathrm{H}t}\left\vert \psi\left(  0\right)  \right\rangle $ with $0\leq
t\leq t_{f}$ from the geodesic path $\left\vert \psi\left(  \xi\right)
\right\rangle $ connecting the initial state $\left\vert \psi\left(  0\right)
\right\rangle \overset{\text{def}}{=}\left\vert \psi_{i}\right\rangle $ and
the final state $\left\vert \psi\left(  t_{f}\right)  \right\rangle
\overset{\text{def}}{=}\left\vert \psi_{f}\right\rangle $. The real parameter
$\xi\in\left[  0\text{, }1\right]  $ and $\left\vert \psi\left(  \xi\right)
\right\rangle $ is given by \cite{laba17},%
\begin{equation}
\left\vert \psi\left(  \xi\right)  \right\rangle \overset{\text{def}}{=}%
\frac{1}{\sqrt{1-2\xi\left(  1-\xi\right)  \left(  1-\left\vert \left\langle
\psi_{f}\left\vert \psi_{i}\right.  \right\rangle \right\vert \right)  }%
}\left[  \left(  1-\xi\right)  \left\vert \psi_{i}\right\rangle +\xi
\frac{\left\langle \psi_{f}\left\vert \psi_{i}\right.  \right\rangle
}{\left\vert \left\langle \psi_{f}\left\vert \psi_{i}\right.  \right\rangle
\right\vert }\right]  \left\vert \psi_{f}\right\rangle \text{.} \label{geo}%
\end{equation}
More specifically, the focus about the temporal evolution that occurs during
the temporal interval $\left[  0\text{, }t_{f}\right]  $ is on two stages. In
the first stage, the state evolves from $\left\vert \psi\left(  0\right)
\right\rangle $ to $\left\vert \psi\left(  \Delta t\right)  \right\rangle
=e^{(-i/\hslash)\mathrm{H}\Delta t}\left\vert \psi\left(  0\right)
\right\rangle $ with $0\leq\Delta t\leq t_{f}$. In the second stage, the state
evolves from $\left\vert \psi\left(  \Delta t\right)  \right\rangle $ to
$\left\vert \psi\left(  \Delta t+\Delta t^{\prime}\right)  \right\rangle
=e^{(-i/\hslash)\mathrm{H}\left(  \Delta t+\Delta t^{\prime}\right)
}\left\vert \psi\left(  0\right)  \right\rangle =e^{(-i/\hslash)\mathrm{H}%
\Delta t^{\prime}}\left\vert \psi\left(  \Delta t\right)  \right\rangle $ with
$\Delta t^{\prime}\geq0$ and $0\leq\Delta t+\Delta t^{\prime}\leq t_{f}$.
Then, setting $\Delta t^{\prime}=\Delta t$ for simplicity, Laba and Tkachuk
argue that a departure of the actual quantum-mechanical evolution from the
geodesic evolution from $\left\vert \psi_{i}\right\rangle $ to $\left\vert
\psi_{f}\right\rangle $ can be detected by considering the minimal squared
distance $d_{\min}^{2}$ between $\left\vert \psi\left(  \Delta t\right)
\right\rangle $ and $\left\vert \psi\left(  \xi\right)  \right\rangle $
in\ Eq. (\ref{geo}),%
\begin{equation}
d_{\min}^{2}\overset{\text{def}}{=}\min_{0\leq\xi\leq1}d^{2}\left(
\xi\right)  =\min_{0\leq\xi\leq1}\left\{  \gamma^{2}\left[  1-\left\vert
\left\langle \psi\left(  \Delta t\right)  \left\vert \psi\left(  \xi\right)
\right.  \right\rangle \right\vert ^{2}\right]  \right\}  \text{.}
\label{minidi}%
\end{equation}
Note that $\gamma$ is an arbitrary real constant that, for convenience, can be
set equal to $\gamma=2$. Moreover, $d^{2}\left(  \xi\right)  $ is the squared
Fubini-Study distance between $\left\vert \psi\left(  \Delta t\right)
\right\rangle $ and $\left\vert \psi\left(  \xi\right)  \right\rangle $.
Performing a Taylor series expansion in $\Delta t$ and keeping terms up to the
fourth order, it is found in Ref. \cite{laba17} that%
\begin{equation}
d_{\min}^{2}=\kappa_{\mathrm{LT}}\frac{\gamma^{2}}{4\hslash^{2}}\left(  \Delta
t\right)  ^{4}+O\left(  \Delta t^{4}\right)  \text{.} \label{curva}%
\end{equation}
The subscript \textquotedblleft\textrm{LT}\textquotedblright\textrm{ }means
Laba and Tkachuk. The quantity $\kappa_{\mathrm{LT}}$ in Eq. (\ref{curva}) is
the so-called curvature coefficient introduced by Laba and Tkachuk in Ref.
\cite{laba17} and is given by,%
\begin{equation}
\kappa_{\mathrm{LT}}\overset{\text{def}}{=}\left\langle (\Delta\mathrm{H}%
)^{4}\right\rangle -\left\langle (\Delta\mathrm{H})^{2}\right\rangle
^{2}\text{,} \label{labacurva}%
\end{equation}
where $\Delta\mathrm{H}\overset{\text{def}}{=}\mathrm{H}-\left\langle
\mathrm{H}\right\rangle $ with $\left\langle \mathrm{H}\right\rangle $ being
the expectation value of the constant Hamiltonian $\mathrm{H}$. For
completeness, we emphasize that it is possible to define a unitless curvature
coefficient $\bar{\kappa}_{\mathrm{LT}}\overset{\text{def}}{=}\kappa
_{\mathrm{LT}}/\left\langle (\Delta\mathrm{H})^{2}\right\rangle ^{2}$, that is%
\begin{equation}
\bar{\kappa}_{\mathrm{LT}}\overset{\text{def}}{=}\frac{\left\langle
(\Delta\mathrm{H})^{4}\right\rangle -\left\langle (\Delta\mathrm{H}%
)^{2}\right\rangle ^{2}}{\left\langle (\Delta\mathrm{H})^{2}\right\rangle
^{2}}\text{.} \label{goLT}%
\end{equation}
Finally, to offer an alternative geometric interpretation of the unitless
curvature coefficient $\bar{\kappa}_{\mathrm{LT}}$, Laba and Tkachuk use a
classical analogy. Indeed, any classical trajectory in physical space between
two neighboring points can be approximated with a circular trajectory between
the two neighboring points. From simple trigonometric arguments, it can be
shown that the circle has a radius $R_{\mathrm{classical}}$ that satisfies%
\begin{equation}
\frac{1}{R_{\mathrm{classical}}}\overset{d\ll1}{\approx}\frac{2d}{(s/2)^{2}%
}\text{.} \label{dog}%
\end{equation}
Recall that the curvature of a circle is the reciprocal of its radius. Note
that $d$ in Eq. (\ref{dog}) is the distance between the middle point of an arc
and the cord joining the two neighboring points. The quantity $s$, instead,
denotes the length of the circular curve connecting the two points. In the
quantum setting, one replaces $d$ in Eq. (\ref{dog}) with $d_{\min}$ in Eq.
(\ref{curva}) and, in addition, $s$ in Eq. (\ref{dog}) with $s_{\mathrm{geo}%
}=2v\Delta t$ where $v\overset{\text{def}}{=}\left(  \gamma/\hslash\right)
\sqrt{\left\langle (\Delta\mathrm{H})^{2}\right\rangle }$ is the speed of
quantum evolution. Then, from Eqs. (\ref{curva}) and (\ref{dog}), one arrives
at the conclusion that $\bar{\kappa}_{\mathrm{LT}}$ can be regarded as a
quantum analogue of $1/R_{\mathrm{classical}}^{2}$ (i.e., the squared
curvature in a classical setting).

Having discussed the concept of curvature coefficient proposed in Ref.
\cite{laba17}, we discuss in the next subsection the notion of torsion
coefficient of a quantum evolution.

\subsection{Torsion}

In Ref. \cite{laba17}, Laba and Tkachuk propose a concept of torsion
coefficient for a Schr\"{o}dinger quantum-mechanical evolution specified by a
time-independent Hamiltonian \textrm{H}. The torsion coefficient quantifies
how much the evolving state vector $\left\vert \psi\left(  t\right)
\right\rangle =e^{-\left(  i/\hslash\right)  \mathrm{H}t}\left\vert
\psi\left(  0\right)  \right\rangle $ with $0\leq t\leq t_{f}$ deviates from
the plane of evolution $\Pi$ at a given time. This plane $\Pi$, in turn, is a
two-dimensional subspace spanned by two neighboring linearly independent unit
state vectors $\left\{  \left\vert \psi\left(  0\right)  \right\rangle \text{,
}\left\vert \psi\left(  \Delta t\right)  \right\rangle \right\}  $ with
$0\leq\Delta t\leq t_{f}$ that belong to the plane $\Pi$. Observe that
$\left\vert \psi\left(  0\right)  \right\rangle $ and $\left\vert \psi\left(
\Delta t\right)  \right\rangle $ are not orthogonal and, in general,
$\left\langle \psi\left(  0\right)  \left\vert \psi\left(  \Delta t\right)
\right.  \right\rangle =ae^{i\alpha}\neq0$ with $a$, $\alpha\in%
\mathbb{R}
$. From the set $\left\{  \left\vert \psi\left(  0\right)  \right\rangle
\text{, }\left\vert \psi\left(  \Delta t\right)  \right\rangle \right\}  $,
Laba and Tkachuk construct a set of orthonormal vectors $\left\{  \left\vert
\phi_{1}\right\rangle \text{, }\left\vert \phi_{2}\right\rangle \right\}  $
and use it to define an orthogonal projection operator \textrm{P}$_{\Pi
}\overset{\text{def}}{=}\left\vert \phi_{1}\right\rangle \left\langle \phi
_{1}\right\vert +\left\vert \phi_{2}\right\rangle \left\langle \phi
_{2}\right\vert $ onto the plane $\Pi$. Then, the magnitude of the deviation
of $\left\vert \psi\left(  t\right)  \right\rangle $ from the plane $\Pi$ is
characterized by the scalar quantity $1-p_{\Pi}$ with $p_{\Pi}%
\overset{\text{def}}{=}\left\langle \psi\left(  \Delta t+\Delta t^{\prime
}\right)  \left\vert \mathrm{P}_{\Pi}\right\vert \psi\left(  \Delta t+\Delta
t^{\prime}\right)  \right\rangle $, $\Delta t^{\prime}\geq0$, and $0\leq\Delta
t+\Delta t^{\prime}\leq t_{f}$. Given the definition of \textrm{P}$_{\Pi}$,
note that $p_{\Pi}=\left\vert \left\langle \phi_{1}\left\vert \psi\left(
\Delta t+\Delta t^{\prime}\right)  \right.  \right\rangle \right\vert
^{2}+\left\vert \left\langle \phi_{2}\left\vert \psi\left(  \Delta t+\Delta
t^{\prime}\right)  \right.  \right\rangle \right\vert ^{2}$. Therefore, when
$p_{\Pi}=1$, there is no deviation from the plane of evolution since the three
vectors $\left\vert \phi_{1}\right\rangle $, $\left\vert \phi_{2}\right\rangle
$, and $\left\vert \psi\left(  \Delta t+\Delta t^{\prime}\right)
\right\rangle $ are on the same plane. Therefore, in this case, the torsion
coefficient is expected to vanish. More specifically, Laba and Tkachuk
introduce the torsion coefficient $\tau_{\mathrm{LT}}$ in an approximate
setting where they Taylor expand $1-p_{\Pi}$ up to the fourth order in $\Delta
t$ and $\Delta t^{\prime}$. Assuming without loss of generality that $\Delta
t=\Delta t^{\prime}$, it is reported in Ref. \cite{laba17} that%
\begin{equation}
1-p_{\Pi}=\tau_{\mathrm{LT}}\frac{\Delta t^{4}}{\hslash^{4}}+O\left(  \Delta
t^{4}\right)  \text{,}%
\end{equation}
where $\tau_{\mathrm{LT}}$ is a constant coefficient that is independent from
$\Delta t$ and $\Delta t^{\prime}$ and is given by,%
\begin{equation}
\tau_{\mathrm{LT}}\overset{\text{def}}{=}\left\langle (\Delta\mathrm{H}%
)^{4}\right\rangle -\left\langle (\Delta\mathrm{H})^{2}\right\rangle
^{2}-\frac{\left\langle (\Delta\mathrm{H})^{3}\right\rangle ^{2}}{\left\langle
(\Delta\mathrm{H})^{2}\right\rangle } = \kappa_{LT} - \frac{\langle(\Delta
H)^{3} \rangle^{2}}{\langle(\Delta H)^{2} \rangle} \text{.}
\label{labatorsion}%
\end{equation}
Recall that $\Delta\mathrm{H}\overset{\text{def}}{=}\mathrm{H}-\left\langle
\mathrm{H}\right\rangle $ with $\left\langle \mathrm{H}\right\rangle $
denoting the expectation value of the time-independent Hamiltonian
$\mathrm{H}$. For completeness, we remark that it is possible to introduce a
unitless torsion coefficient $\bar{\tau}_{\mathrm{LT}}\overset{\text{def}%
}{=}\tau_{\mathrm{LT}}/\left\langle (\Delta\mathrm{H})^{2}\right\rangle ^{2}$,
that is%
\begin{equation}
\bar{\tau}_{\mathrm{LT}}\overset{\text{def}}{=}\frac{\left\langle
(\Delta\mathrm{H})^{4}\right\rangle -\left\langle (\Delta\mathrm{H}%
)^{2}\right\rangle ^{2}}{\left\langle (\Delta\mathrm{H})^{2}\right\rangle
^{2}}-\frac{\left\langle (\Delta\mathrm{H})^{3}\right\rangle ^{2}%
}{\left\langle (\Delta\mathrm{H})^{2}\right\rangle ^{3}} = \bar{\kappa}_{LT} -
\frac{\langle(\Delta H)^{3} \rangle^{2}}{\langle(\Delta H)^{2} \rangle^{3}}
\text{.}%
\end{equation}
Finally, to provide an additional geometric interpretation of the torsion
coefficient $\tau_{\mathrm{LT}}$ in Eq. (\ref{labatorsion}), Laba and Tkachuk
show that the scalar quantity $1-p_{\Pi}$ happens to be proportional to the
squared distance of $\left\vert \psi\left(  \Delta t+\Delta t^{\prime}\right)
\right\rangle $ to the plane $\Pi$. Therefore, they conclude that $1-p_{\Pi}$
has a clear geometrical interpretation and is a suitable indicator of the
presence of torsion during the quantum-mechanical evolution.

Having discussed this relevant background material, in the next section we
present some preliminary material on the classical Frenet-Serret apparatus.
This, in turn, will inspire our own newly proposed measures of curvature and
torsion of a quantum curve.

\section{The classical Frenet-Serret apparatus}

In this section, in preparation of our newly proposed theoretical construct in
Sec. IV, we recall the basic ingredients of a classical Frenet-Serret
apparatus with special focus on the concepts of bending and twisting as
captured by the curvature and the torsion of a curve in three-dimensional
Euclidean space.

\subsection{Essentials}

We limit our presentation here to unit speed regular curves in $%
\mathbb{R}
^{3}$. In this context, the so-called Frenet-Serret apparatus is the main tool
to study curves since it completely determines the geometry of the curve. This
apparatus consists of three unit vector fields along the curve and two
scalar-valued functions. The three vector fields are the tangent vector field
$\hat{T}$, the principal normal vector field $\hat{N}$, and the binormal
vector field $\hat{B}$. The two scalar-valued functions, instead, are given by
the curvature $\kappa_{\mathrm{FS}}$ and the torsion $\tau_{\mathrm{FS}}$. The
subscript \textquotedblleft\textrm{FS}\textquotedblright\textrm{ }means Frenet
and Serret. The set $\left\{  \hat{T}\text{, }\hat{N}\text{, }\hat{B}\right\}
$ is known as the Frenet-Serret frame. It is a very convenient set of three
orthonormal vectors that reflects the geometry of the curve and, thus, can be
used to fully characterize the geometric properties of a curve in $%
\mathbb{R}
^{3}$. Although the Frenet-Serret apparatus can be applied to non-unit speed
curves as well, we focus on unit speed curves here. In this latter case, it is
possible to show that three vectors specifying the Frenet-Serret frame satisfy
the so-called Frenet-Serret equations given by \cite{parker77}
\begin{equation}
\left(
\begin{array}
[c]{c}%
\hat{T}^{\prime}\\
\hat{N}^{\prime}\\
\hat{B}^{\prime}%
\end{array}
\right)  =\left(
\begin{array}
[c]{ccc}%
0 & \kappa_{\mathrm{FS}} & 0\\
-\kappa_{\mathrm{FS}} & 0 & \tau_{\mathrm{FS}}\\
0 & -\tau_{\mathrm{FS}} & 0
\end{array}
\right)  \left(
\begin{array}
[c]{c}%
\hat{T}\\
\hat{N}\\
\hat{B}%
\end{array}
\right)  \text{.} \label{FSeq}%
\end{equation}
The relations in Eq. (\ref{FSeq}) describe the dynamics of $\left\{  \hat
{T}\text{, }\hat{N}\text{, }\hat{B}\right\}  $ in terms of how they move and
twist as one walks
along the curve. In other words, the Frenet-Serret frame is a classical
example of a \textquotedblleft moving\textquotedblright\ frame. The prime in
Eq. (\ref{FSeq}) denotes differentiation with respect the parameter chosen to
parametrize the curve, i.e., the arc length $s$ along the curve defined as,%
\begin{equation}
s\overset{\text{def}}{=}\int_{0}^{s}\left\Vert \frac{d\vec{\gamma}}%
{dt}\right\Vert dt\text{,} \label{arc}%
\end{equation}
with $\vec{\gamma}:\left(  a\text{, }b\right)  \rightarrow%
\mathbb{R}
^{3}$ being a unit speed regular curve with $\left\Vert d\vec{\gamma
}/dt\right\Vert =1$. Note that $d\vec{\gamma}/dt$ is the velocity vector field
along $\vec{\gamma}$. The tangent vector field $\hat{T}\left(  t\right)  $,
instead, is the unit vector in the direction of the velocity vector and is
defined as $\hat{T}\left(  t\right)  \overset{\text{def}}{=}\left(
d\vec{\gamma}/dt\right)  /\left\Vert d\vec{\gamma}/dt\right\Vert $. For
completeness, we observe that curves can be parametrized with parameters other
than the arc length $s$ and, in addition, the satisfaction of the regularity
condition of the curve requires $d\vec{\gamma}/dt\neq0$. For a more general
discussion on the geometry of curves in $%
\mathbb{R}
^{3}$ extended to irregular and/or non-unit speed curves, we refer to Ref.
\cite{parker77}.

Finally, we shall define the vectors $\hat{N}$ and $\hat{B}$ in the following
subsections where the focus is on the concepts of curvature and torsion of a
curve in $%
\mathbb{R}
^{3}$.

\subsection{Curvature}

It is generally stated that the curvature of a straight line is equal to zero
and that of a circle is constant since it assumes the same value at each point
along the curve. These statements suggest that the notion of curvature should
provide a measure of the bending of a curve. Assuming a unit-speed regular
curve, a reasonable measure of the bending of the curve can be specified by
the rate of change $\hat{T}^{\prime}\left(  s\right)  $ of the tangent vector
field $\hat{T}\left(  s\right)  $ with respect to the arc length $s$. Indeed,
the curvature $\kappa_{\mathrm{FS}}\left(  s\right)  $ of a unit-speed regular
curve $\vec{\gamma}\left(  s\right)  $ is defined as%
\begin{equation}
\kappa_{\mathrm{FS}}\left(  s\right)  \overset{\text{def}}{=}\left\Vert
\hat{T}^{\prime}\left(  s\right)  \right\Vert \text{.} \label{FScurve}%
\end{equation}
Note that, in general, $\kappa_{\mathrm{FS}}$ is non-constant and depends on
$s$. Furthermore, observe that $\kappa_{\mathrm{FS}}$ equals $0$ and $1/R$ for
a straight line and a circle of radius $R$, respectively. To better understand
the significance of $\kappa_{\mathrm{FS}}$ in Eq. (\ref{FScurve}) and in
preparation of the introduction of the concept of torsion, it is convenient to
define at this point the vectors $\hat{N}$ and $\hat{B}$. The principal normal
vector field $\hat{N}$ to a unit-speed curve $\vec{\gamma}\left(  s\right)  $
is given by $\hat{N}\left(  s\right)  \overset{\text{def}}{=}\hat{T}^{\prime
}\left(  s\right)  /\kappa_{\mathrm{FS}}$ with $\hat{T}\left(  s\right)
\perp\hat{N}\left(  s\right)  $ since $\left\Vert \hat{T}\left(  s\right)
\right\Vert ^{2}=1$. Then, the binormal vector $\hat{B}\left(  s\right)  $ is
defined as the cross product between $\hat{T}\left(  s\right)  $ and $\hat
{N}\left(  s\right)  $, $\hat{B}\left(  s\right)  \overset{\text{def}}{=}%
\hat{T}\left(  s\right)  \times\hat{N}\left(  s\right)  $. We are now ready to
introduce the torsion $\tau_{\mathrm{FS}}$ in the next subsection.

\subsection{Torsion}

The torsion $\tau_{\mathrm{FS}}$ of a unit-speed curve $\vec{\gamma}\left(
s\right)  $ is formally defined as,%
\begin{equation}
\tau_{\mathrm{FS}}\left(  s\right)  \overset{\text{def}}{=}-\hat{B}^{\prime
}\left(  s\right)  \cdot\hat{N}\left(  s\right)  \text{.} \label{FStorsion}%
\end{equation}
From a geometric standpoint, $\tau_{\mathrm{FS}}$ measures how far the curve
$\vec{\gamma}\left(  s\right)  $ is from lying in the osculating plane spanned
by the orthonormal vectors $\hat{T}\left(  s\right)  $ and $\hat{N}\left(
s\right)  $. In particular, if a curve lies in a plane at all times,
$\tau_{\mathrm{FS}}=0$ and the plane is the osculating plane. When
$\tau_{\mathrm{FS}}\neq0$, $\tau_{\mathrm{FS}}$ quantifies the twisting of the
curve out of the osculating plane. Furthermore, this twisting does not need to
be constant as one walks on the curve. Unlike the curvature $\kappa
_{\mathrm{FS}}$, the torsion $\tau_{\mathrm{FS}}$ can be both positive and
negative. Its sign has a clear geometric meaning. It is positive (or,
negative) when the curve twists toward (or, toward the opposite) the side
$\hat{B}\left(  s\right)  $ points to as $s$ increases.

In the next section, taking inspiration from what we presented in Sec. III, we
propose our definitions of curvature and torsion coefficients of a quantum
curve. Remarkably, we recover the curvature and torsion coefficients as
presented in Sec. II as originally proposed by Laba and Tkachuk in Ref.
\cite{laba17}.

\section{A quantum version of the Frenet-Serret apparatus}

In this section, we propose a quantum version of the classical Frenet-Serret
apparatus. In particular, we propose novel measures of bending and twisting of
a quantum curve traced out by a parallel-transported pure quantum state that
evolves under the action of a time-independent (Hermitian) Hamiltonian operator.

\subsection{Parallel-transported unit state vectors: $\left\{  \left\vert
\Psi\right\rangle \right\}  $}

We begin by introducing the parallel-transported unit state vectors $\left\{
\left\vert \Psi\right\rangle \right\}  $ with $\left\langle \Psi\left\vert
\Psi\right.  \right\rangle =1$ and $\left\langle \Psi\left\vert \dot{\Psi
}\right.  \right\rangle =0$. Consider the normalized state vector $\left\vert
\psi\left(  t\right)  \right\rangle $ with $\left\langle \psi\left(  t\right)
\left\vert \psi\left(  t\right)  \right.  \right\rangle =1$ that satisfies the
Schr\"{o}dinger evolution equation,
\begin{equation}
i\hslash\partial_{t}\left\vert \psi\left(  t\right)  \right\rangle
=\mathrm{H}\left(  t\right)  \left\vert \psi\left(  t\right)  \right\rangle
\text{.} \label{t0}%
\end{equation}
In what follows, the Hamiltonian \textrm{H}$\left(  t\right)  $\textrm{\ }in
Eq. (\ref{t0}) is assumed to be constant in time, i.e., \textrm{H}$\left(
t\right)  =\mathrm{H}$ for any $t\geq0$. Observe that $\left\vert \dot{\psi
}\left(  t\right)  \right\rangle \overset{\text{def}}{=}\partial_{t}\left\vert
\psi\left(  t\right)  \right\rangle =-(i/\hslash)\mathrm{H}\left\vert
\psi\left(  t\right)  \right\rangle $ is not orthogonal to $\left\vert
\psi\left(  t\right)  \right\rangle $ since $\left\langle \psi\left(
t\right)  \left\vert \dot{\psi}\left(  t\right)  \right.  \right\rangle
=-(i/\hslash)\left\langle \psi\left(  t\right)  \left\vert \mathrm{H}%
\right\vert \psi\left(  t\right)  \right\rangle \neq0$, in general. Note that
$\left\langle \psi\left(  t\right)  \left\vert \mathrm{H}\right\vert
\psi\left(  t\right)  \right\rangle $ is time-independent since $\left\langle
\psi\left(  t\right)  \left\vert \mathrm{H}\right\vert \psi\left(  t\right)
\right\rangle
= \langle\psi(0)| e^{i H t/\hbar}\,H\, e^{-i H t/\hbar} |\psi(0)\rangle
=\left\langle \psi\left(  0\right)  \left\vert \mathrm{H}\right\vert
\psi\left(  0\right)  \right\rangle =E=$\textrm{constant}. Let us define
$\left\vert \Psi\left(  t\right)  \right\rangle \overset{\text{def}%
}{=}e^{i\beta\left(  t\right)  }\left\vert \psi\left(  t\right)  \right\rangle
$ with%
\begin{equation}
\beta\left(  t\right)  \overset{\text{def}}{=}\frac{1}{\hslash}\int_{0}%
^{t}\left\langle \psi\left(  t^{\prime}\right)  \left\vert \mathrm{H}%
\right\vert \psi\left(  t^{\prime}\right)  \right\rangle dt^{\prime}\text{.}%
\end{equation}
In our case, $\beta\left(  t\right)  =\frac{E}{\hslash}t$ and $\left\vert
\Psi\left(  t\right)  \right\rangle =e^{i\frac{E}{\hslash}t}\left\vert
\psi\left(  t\right)  \right\rangle $, that is, $\left\vert \psi\left(
t\right)  \right\rangle =e^{-i\frac{E}{\hslash}t}\left\vert \Psi\left(
t\right)  \right\rangle $. By construction, we have $\left\langle \Psi\left(
t\right)  \left\vert \dot{\Psi}\left(  t\right)  \right.  \right\rangle =0$
and $\left\vert \Psi\left(  t\right)  \right\rangle $ satisfies the
Schr\"{o}dinger evolution equation%
\begin{equation}
i\hslash\partial_{t}\left\vert \Psi\left(  t\right)  \right\rangle
=\Delta\mathrm{H}\left\vert \Psi\left(  t\right)  \right\rangle \text{.}
\label{t5}%
\end{equation}
In Eq. (\ref{t5}), $\Delta\mathrm{H}\overset{\text{def}}{=}\mathrm{H}%
-\left\langle \mathrm{H}\right\rangle $, with $\left\langle \mathrm{H}%
\right\rangle $ and $\Delta\mathrm{H}$ being time-independent quantities when
$\mathrm{H}$ is assumed to be time-independent. Having introduced the state
vector $\left\vert \Psi\left(  t\right)  \right\rangle $, we are ready to
introduce the unit tangent vector in the next subsection.

\subsection{Unit tangent vectors: $\left\{  \left\vert T\right\rangle
\right\}  $}

In this subsection, we introduce the unit tangent vectors $\left\{  \left\vert
T\right\rangle \right\}  $ such that $\left\langle T\left\vert T\right.
\right\rangle =1$ and $\left\langle \Psi\left\vert T\right.  \right\rangle
=0$. To begin, let us consider the time derivative of the normalized state
vector $\left\vert \Psi\left(  t\right)  \right\rangle $, $\left\vert \bar
{T}\left(  t\right)  \right\rangle \overset{\text{def}}{=}\partial
_{t}\left\vert \Psi\left(  t\right)  \right\rangle $. Note that we use the
upper bar symbol, like the one in $\left\vert \bar{T}\left(  t\right)
\right\rangle $, to indicate unnormalized vectors. Using Eq. (\ref{t5}), we
have%
\begin{equation}
\left\vert \bar{T}\left(  t\right)  \right\rangle =\partial_{t}\left\vert
\Psi\left(  t\right)  \right\rangle =-\frac{i}{\hslash}\Delta\mathrm{H}%
\left\vert \Psi\left(  t\right)  \right\rangle \text{.}%
\end{equation}
Note that $\left\vert \bar{T}\left(  t\right)  \right\rangle $ is orthogonal
to $\left\vert \Psi\left(  t\right)  \right\rangle $, $\left\langle
\Psi\left(  t\right)  \left\vert \bar{T}\left(  t\right)  \right.
\right\rangle =0$, since $\left\langle \psi\left(  t\right)  \left\vert
\Delta\mathrm{H}\right\vert \psi\left(  t\right)  \right\rangle =0$. However,
$\left\vert \bar{T}\left(  t\right)  \right\rangle $ is not normalized to one
since%
\begin{equation}
\left\langle \bar{T}\left(  t\right)  \left\vert \bar{T}\left(  t\right)
\right.  \right\rangle =-\left(  \frac{i}{\hslash}\right)  ^{2}\left\langle
\Psi\left(  t\right)  \left\vert (\Delta\mathrm{H})^{2}\right\vert \Psi\left(
t\right)  \right\rangle =\frac{\left\langle (\Delta\mathrm{H})^{2}%
\right\rangle }{\hslash^{2}}\overset{\text{def}}{=}v^{2}\text{.} \label{t6}%
\end{equation}
The quantity $v$ in Eq. (\ref{t6}) is constant in time and can be used to
define the arc length $s$ given by%
\begin{equation}
s\overset{\text{def}}{=}\int_{0}^{t}vdt=vt\text{.} \label{t7}%
\end{equation}
For completeness, note that $\left[  t\right]  _{\mathrm{MKSA}}=\sec$,
$\left[  v\right]  _{\mathrm{MKSA}}=\sec^{-1}$, and $\left[  s\right]
_{\mathrm{MKSA}}$ is adimensional (i.e., unitless). $\mathrm{MKSA}$ denotes
meters, kilograms, seconds, and amperes within the International System of
Units. From Eq. (\ref{t7}), the relation between $t$-derivatives and
$s$-derivatives is given by $\partial_{s}=(1/v)\partial_{t}$. Using
$s$-derivatives, we can introduce a properly normalized tangent vector
$\left\vert T\left(  s\right)  \right\rangle $ defined as%
\begin{equation}
\left\vert T\left(  s\right)  \right\rangle \overset{\text{def}}{=}%
\partial_{s}\left\vert \Psi\left(  s\right)  \right\rangle \equiv\left\vert
\Psi^{\prime}\left(  s\right)  \right\rangle \equiv-i\Delta h\left\vert
\Psi\left(  s\right)  \right\rangle \text{,} \label{t8}%
\end{equation}
where $\left\vert \Psi\left(  s\right)  \right\rangle \overset{\text{def}%
}{=}\left\vert \Psi\left(  t(s)\right)  \right\rangle $. Observe that
$\left\vert T\left(  s\right)  \right\rangle =-i\Delta h\left\vert \Psi\left(
s\right)  \right\rangle $, where the unitless operator $\Delta h$ is defined
as $\Delta h\overset{\text{def}}{=}\Delta\mathrm{H}/\left(  \hslash v\right)
$. By construction, $\left\vert T\left(  s\right)  \right\rangle $ in Eq.
(\ref{t8}) is such that $\left\langle T\left(  s\right)  \left\vert T\left(
s\right)  \right.  \right\rangle =1$. In addition, it can be rewritten as
$\left\vert T\left(  s\right)  \right\rangle =\mathrm{P}^{(\Psi)}\left\vert
\Psi^{\prime}\left(  s\right)  \right\rangle =\left\vert \Psi^{\prime}\left(
s\right)  \right\rangle $, with $\mathrm{P}^{(\Psi)}\overset{\text{def}%
}{=}\mathrm{I}-\left\vert \Psi\right\rangle \left\langle \Psi\right\vert $
being a Hermitian projector onto states orthogonal to the unit state vector
$\left\vert \Psi\right\rangle $ such that $\mathrm{P}^{(\Psi)}\mathrm{P}%
^{(\Psi)}=\mathrm{P}^{(\Psi)}$ and $\left(  \mathrm{P}^{(\Psi)}\right)
^{\dagger}=\mathrm{P}^{(\Psi)}$. For completeness, using this definition of
the tangent vector in terms of the projector operator $\mathrm{P}^{(\Psi)}$,
we get $\left\langle T\left(  s\right)  \left\vert T\left(  s\right)  \right.
\right\rangle =\left\langle (\Delta\mathrm{H})^{2}\right\rangle /\left(
\hslash^{2}v^{2}\right)  =1$ since $v^{2}\overset{\text{def}}{=}\left\langle
(\Delta\mathrm{H})^{2}\right\rangle /\hslash^{2}$. Lastly, notice that
$\left\vert T\left(  s\right)  \right\rangle $ is orthogonal to $\left\vert
\Psi\left(  s\right)  \right\rangle $. Indeed, using Eq. (\ref{t8}), we have
$\left\langle \Psi\left(  s\right)  \left\vert T\left(  s\right)  \right.
\right\rangle =-i\left\langle \Psi\left(  s\right)  \left\vert \Delta
h\right\vert \Psi\left(  s\right)  \right\rangle =-i\left\langle \Delta
h\right\rangle =0$ since $\Delta h\overset{\text{def}}{=}\Delta\mathrm{H}%
/\left(  \hslash v\right)  =\Delta\mathrm{H}/\sqrt{\left\langle \Delta
\mathrm{H}^{2}\right\rangle }$ with $\left\langle \Delta\mathrm{H}%
\right\rangle =0$. In summary, $\left\{  \left\vert \Psi\left(  s\right)
\right\rangle \text{, }\left\vert T\left(  s\right)  \right\rangle \right\}  $
is a pair of orthonormal state vectors. Moreover, $\left\vert T\left(
s\right)  \right\rangle $ is the projection of $\left\vert \Psi^{\prime
}\left(  s\right)  \right\rangle $ normal to $\left\vert \Psi\left(  s\right)
\right\rangle $ and specifies the so-called covariant derivative of
$\left\vert \Psi\left(  s\right)  \right\rangle $, $\left\vert T\left(
s\right)  \right\rangle =\left\vert \mathrm{D}\Psi\left(  s\right)
\right\rangle \overset{\text{def}}{=}\mathrm{P}^{(\Psi)}\left\vert
\Psi^{\prime}\left(  s\right)  \right\rangle $. \ The covariant derivative
operator $\mathrm{D}$ is defined as $\mathrm{D}\overset{\text{def}%
}{=}\mathrm{P}^{\left(  \Psi\right)  }\left(  d/ds\right)  $ and represents
the projection onto a state perpendicular to $\left\vert \Psi\left(  s\right)
\right\rangle $ of the ordinary derivative $d/ds$ with $s$ being the arc
length \cite{carlocqg23,samuel88,paulPRA23}. The set of vectors $\left\{
\left\vert \Psi\left(  s\right)  \right\rangle \text{, }\left\vert T\left(
s\right)  \right\rangle \right\}  $ can be viewed as spanning the
instantaneous \textquotedblleft osculating\textquotedblright\ plane that
contains the quantum curve traced out by the state vector $\left\vert
\Psi\left(  s\right)  \right\rangle $. This plane will play a key role when
introducing the concept of torsion of a quantum evolution. Before doing so, we
introduce in the next subsection a concept of curvature for quantum evolutions.

\subsection{Curvature}

In our approach, inspired by the classical Frenet-Serret apparatus, we propose
that the curvature coefficient $\kappa_{\mathrm{AC}}^{2}$ is given by the
magnitude squared of the covariant derivative of the tangent vector
$\left\vert T\left(  s\right)  \right\rangle $ to the state vector $\left\vert
\Psi\left(  s\right)  \right\rangle $,
\begin{equation}
\kappa_{\mathrm{AC}}^{2}\overset{\text{def}}{=}\left\langle T^{\prime}\left(
s\right)  \left\vert \left(  \mathrm{P}^{\left(  \Psi\right)  }\right)
^{\dagger}\mathrm{P}^{\left(  \Psi\right)  }\right\vert T^{\prime}\left(
s\right)  \right\rangle \text{,} \label{accurve}%
\end{equation}
with $\mathrm{D}\left\vert T(s)\right\rangle \overset{\text{def}}{=}%
\mathrm{P}^{\left(  \Psi\right)  }\left\vert T^{\prime}(s)\right\rangle $ and
$\mathrm{D}\overset{\text{def}}{=}\mathrm{P}^{\left(  \Psi\right)  }\frac
{d}{ds}=\left(  \mathrm{I}-\left\vert \Psi\right\rangle \left\langle
\Psi\right\vert \right)  \frac{d}{ds}$ \cite{carlocqg23,samuel88,paulPRA23}.
The subscript \textquotedblleft\textrm{AC}\textquotedblright\textrm{ }means
Alsing and Cafaro. Since $\left\vert T^{\prime}\left(  s\right)  \right\rangle
=\left\vert \Psi^{\prime\prime}\left(  s\right)  \right\rangle $ and
$\mathrm{P}^{\left(  \Psi\right)  }\left\vert T\left(  s\right)  \right\rangle
=\left\vert T\left(  s\right)  \right\rangle $, $\kappa_{\mathrm{AC}}^{2}$
in\ Eq. (\ref{accurve}) can be regarded as specified by the second covariant
derivative $\mathrm{D}^{2}\left\vert \Psi\left(  s\right)  \right\rangle $ of
the state vector $\left\vert \Psi\left(  s\right)  \right\rangle $ (i.e., a
form of acceleration vector) and can be recast as%
\begin{equation}
\kappa_{\mathrm{AC}}^{2}\overset{\text{def}}{=}\left\Vert \mathrm{D}\left\vert
T(s)\right\rangle \right\Vert ^{2}=\left\Vert \mathrm{D}^{2}\left\vert
\Psi\left(  s\right)  \right\rangle \right\Vert ^{2}\text{.} \label{ac12}%
\end{equation}
Note that while $\kappa_{\mathrm{AC}}^{2}$ in Eq. (\ref{ac12}) is the squared
magnitude of the second covariant derivative of the state vector $\left\vert
\Psi\left(  s\right)  \right\rangle $ that traces out the quantum
Schr\"{o}dinger trajectory, $\kappa_{\mathrm{FS}}$ in Eq. (\ref{FScurve}) is
the magnitude of the second derivative of the vector position $\vec{r}\left(
s\right)  $ that traces out the unit-speed curve $\vec{\gamma}\left(
s\right)  $ in the FS apparatus. Thus, our proposed curvature coefficient
$\kappa_{\mathrm{AC}}^{2}$ can be regarded as a quantum analogue of
$\kappa_{\mathrm{FS}}^{2}$. To find an explicit expression of $\kappa
_{\mathrm{AC}}^{2}$ in Eq. ((\ref{accurve}), we use Eq. (\ref{t8}) to obtain
$\mathrm{P}^{(\Psi)}\left\vert T^{\prime}\left(  s\right)  \right\rangle
=-\left[  \left(  \Delta h\right)  ^{2}-\left\langle \left(  \Delta h\right)
^{2}\right\rangle \right]  \left\vert \Psi\left(  s\right)  \right\rangle $.
Indeed, dropping the \textquotedblleft$s$\textquotedblright\ in $\left\vert
\Psi\left(  s\right)  \right\rangle $ and $\left\vert T\left(  s\right)
\right\rangle $, we have%
\begin{align}
\mathrm{P}^{(\Psi)}\left\vert T^{\prime}\right\rangle  &  =\mathrm{P}^{(\Psi
)}\left\vert \partial_{s}T\right\rangle \nonumber\\
&  =\mathrm{P}^{(\Psi)}\partial_{s}\left(  -i\Delta h\left\vert \Psi
\right\rangle \right) \nonumber\\
&  =-i\mathrm{P}^{(\Psi)}\left(  \Delta h\left\vert \partial_{s}%
\Psi\right\rangle \right) \nonumber\\
&  =-i\mathrm{P}^{(\Psi)}\left(  \Delta h\left\vert T\right\rangle \right)
\nonumber\\
&  =-i\mathrm{P}^{(\Psi)}\left(  -i\left(  \Delta h\right)  ^{2}\left\vert
\Psi\right\rangle \right) \nonumber\\
&  =-\left[  \mathrm{I}-\left\vert \Psi\right\rangle \left\langle
\Psi\right\vert \right]  \left[  \left(  \Delta h\right)  ^{2}\left\vert
\Psi\right\rangle \right] \nonumber\\
&  =-\left[  \left(  \Delta h\right)  ^{2}\left\vert \Psi\right\rangle
-\left\langle \Psi\left\vert \left(  \Delta h\right)  ^{2}\right\vert
\Psi\right\rangle \left\vert \Psi\right\rangle \right] \nonumber\\
&  =-\left[  \left(  \Delta h\right)  ^{2}-\left\langle \left(  \Delta
h\right)  ^{2}\right\rangle \right]  \left\vert \Psi\right\rangle \text{.}
\label{t9a}%
\end{align}
From Eq. (\ref{t9a}), we finally get an expression of the curvature
coefficient $\kappa_{\mathrm{AC}}^{2}$ in terms of expectation values of
powers of the adimensional operator $\Delta h$,
\begin{equation}
\kappa_{\mathrm{AC}}^{2}=\left\langle \left(  \Delta h\right)  ^{4}%
\right\rangle -\left\langle \left(  \Delta h\right)  ^{2}\right\rangle ^{2} =
\left\langle \left(  \Delta h\right)  ^{4}\right\rangle - 1, \label{accurve1}%
\end{equation}
where we have used $\langle(\Delta h)^{2}\rangle= \langle\left(  \Delta
H/\sqrt{\langle(\Delta H)^{2}\rangle} \right)  ^{2} \rangle= 1$.
Interestingly, we note that $\kappa_{\mathrm{AC}}^{2}$ in Eq. (\ref{accurve1})
coincides with the unitless curvature coefficient introduced by Laba and
Tkachuk given by $\bar{\kappa}_{\mathrm{LT}}\overset{\text{def}}{=}%
\kappa_{\mathrm{LT}}/\left\langle (\Delta\mathrm{H})^{2}\right\rangle ^{2}$ in
Eq. (\ref{goLT}) with $\kappa_{\mathrm{LT}}$ given in Eq. (\ref{labacurva})
once we recall that $\Delta h=\Delta\mathrm{H}/\sqrt{\left\langle
(\Delta\mathrm{H})^{2}\right\rangle }$. We are now ready to introduce our
proposal for a concept of torsion for a quantum evolution.

\subsection{Torsion}

We begin by recalling that in the classical Frenet-Serret apparatus, the
definition of the torsion coefficient requires the introduction of the
binormal vector field $\hat{B}\left(  s\right)  $ that is orthogonal to the
instantaneous osculating plane spanned by the other two orthonormal vectors
$\hat{T}\left(  s\right)  $ and $\hat{N}\left(  s\right)  $. In our quantum
framework, the instantaneous motion of the curve belongs to the plane spanned
by the orthonormal state vectors $\left\vert \Psi\left(  s\right)
\right\rangle $ and $\left\vert T\left(  s\right)  \right\rangle $. This plane
can be viewed as the quantum analogue of the osculating plane that appears in
the Frenet-Serret apparatus. Furthermore, it corresponds to the plane spanned
by the state vectors $\left\vert \psi\left(  t\right)  \right\rangle $ and
$\left\vert \psi\left(  t+\Delta t\right)  \right\rangle $ in the quantum
Laba-Tkachuk framework once one recognizes that the linear approximation of
$\left\vert \psi\left(  t+\Delta t\right)  \right\rangle $ is given by
$\left\vert \psi\left(  t\right)  \right\rangle +\left\vert \dot{\psi}\left(
t\right)  \right\rangle \Delta t+O\left(  \Delta t^{2}\right)  $ with
$\left\vert \dot{\psi}\left(  t\right)  \right\rangle =\left\vert \partial
_{t}\psi\left(  t\right)  \right\rangle $ with $\partial_{t}%
\overset{\text{def}}{=}\partial/\partial t$. However, we need to identify a
suitable quantum analogue of $\hat{B}\left(  s\right)  $. We start by
recalling that while $\left\{  \left\vert \Psi\left(  s\right)  \right\rangle
\text{, }\left\vert T\left(  s\right)  \right\rangle \right\}  $ forms an
orthonormal set of state vectors and $\mathrm{P}^{(\Psi)}\left\vert T^{\prime
}\left(  s\right)  \right\rangle $ is orthogonal to $\left\vert \Psi\left(
s\right)  \right\rangle $, we have two issues. First, $\mathrm{P}^{(\Psi
)}\left\vert T^{\prime}\left(  s\right)  \right\rangle $ and $\left\vert
T\left(  s\right)  \right\rangle $ are not orthogonal since $\left\langle
T(s)\left\vert \mathrm{P}^{(\Psi)}\right\vert T^{\prime}(s)\right\rangle
=-i\left\langle \left(  \Delta h\right)  ^{3}\right\rangle \neq0$.
Interestingly, as we shall see better later, $\left\langle \left(  \Delta
h\right)  ^{3}\right\rangle $ corresponds to the so-called skewness
coefficient in statistical mathematics \cite{ernest44,sharma15,pearson16}.
Second, $\mathrm{P}^{(\Psi)}\left\vert T^{\prime}\left(  s\right)
\right\rangle $ is not properly normalized to one. This latter matter is a
minor concern. To address the first issue, we propose to consider the (not
normalized) state vector $\left\vert \bar{N}\left(  s\right)  \right\rangle $
defined as,%
\begin{equation}
\left\vert \bar{N}\left(  s\right)  \right\rangle \overset{\text{def}%
}{=}\mathrm{P}^{(T)}\mathrm{P}^{(\Psi)}\left\vert T^{\prime}(s)\right\rangle
\text{,} \label{normale}%
\end{equation}
that is, $\left\vert \bar{N}\left(  s\right)  \right\rangle
\overset{\text{def}}{=}\mathrm{P}^{(T)}\mathrm{D}\left\vert T(s)\right\rangle
$. By construction, $\left\vert \bar{N}\left(  s\right)  \right\rangle $ in
Eq. (\ref{normale}) is orthogonal to both $\left\vert \Psi\left(  s\right)
\right\rangle $ and $\left\vert T\left(  s\right)  \right\rangle $. Clearly,
one can construct a properly normalized vector $\left\vert N\left(  s\right)
\right\rangle $ from $\left\vert \bar{N}\left(  s\right)  \right\rangle $
given by $\left\vert N\left(  s\right)  \right\rangle \overset{\text{def}%
}{=}\left\vert \bar{N}\left(  s\right)  \right\rangle /\left\Vert \bar
{N}\left(  s\right)  \right\Vert $ so that $\left\{  \left\vert \Psi\left(
s\right)  \right\rangle \text{, }\left\vert T\left(  s\right)  \right\rangle
\text{, }\left\vert N\left(  s\right)  \right\rangle \right\}  $ forms a
useful set of orthonormal state vectors. Interestingly, $\left\{  \left\vert
\Psi\left(  s\right)  \right\rangle \text{, }\left\vert T\left(  s\right)
\right\rangle \text{, }\left\vert N\left(  s\right)  \right\rangle \right\}  $
can be regarded as the outcome of an ordinary Gram-Schmidt orthonormalization
procedure applied to the input set of (linearly independent) vectors given by
$\left\{  \left\vert \Psi\left(  s\right)  \right\rangle \text{, }\left\vert
\Psi^{\prime}\left(  s\right)  \right\rangle \text{, }\left\vert \Psi
^{\prime\prime}\left(  s\right)  \right\rangle \right\}  $ or, equivalently,
$\left\{  \left\vert \Psi\left(  s\right)  \right\rangle \text{, }%
\mathrm{D}\left\vert \Psi\left(  s\right)  \right\rangle \text{, }%
\mathrm{D}^{2}\left\vert \Psi\left(  s\right)  \right\rangle \right\}  $. We
show in Fig. $1$ a graphical depiction of the Frenet-Serret frame $\left\{
\hat{T}\text{, }\hat{N}\text{, }\hat{B}\right\}  $ for the vector spaces along
a curve in three-dimensional Euclidean space $\mathbb{R}^{3}$ along with a
pictorial representation of our proposed quantum frame $\left\{  \left\vert
\Psi\left(  s\right)  \right\rangle \text{, }\left\vert T\left(  s\right)
\right\rangle \text{, }\left\vert N\left(  s\right)  \right\rangle \right\}  $
for the three-dimensional subspaces along a curve on a generalized Bloch
sphere. \begin{figure}[t]
\centering
\includegraphics[width=0.5\textwidth] {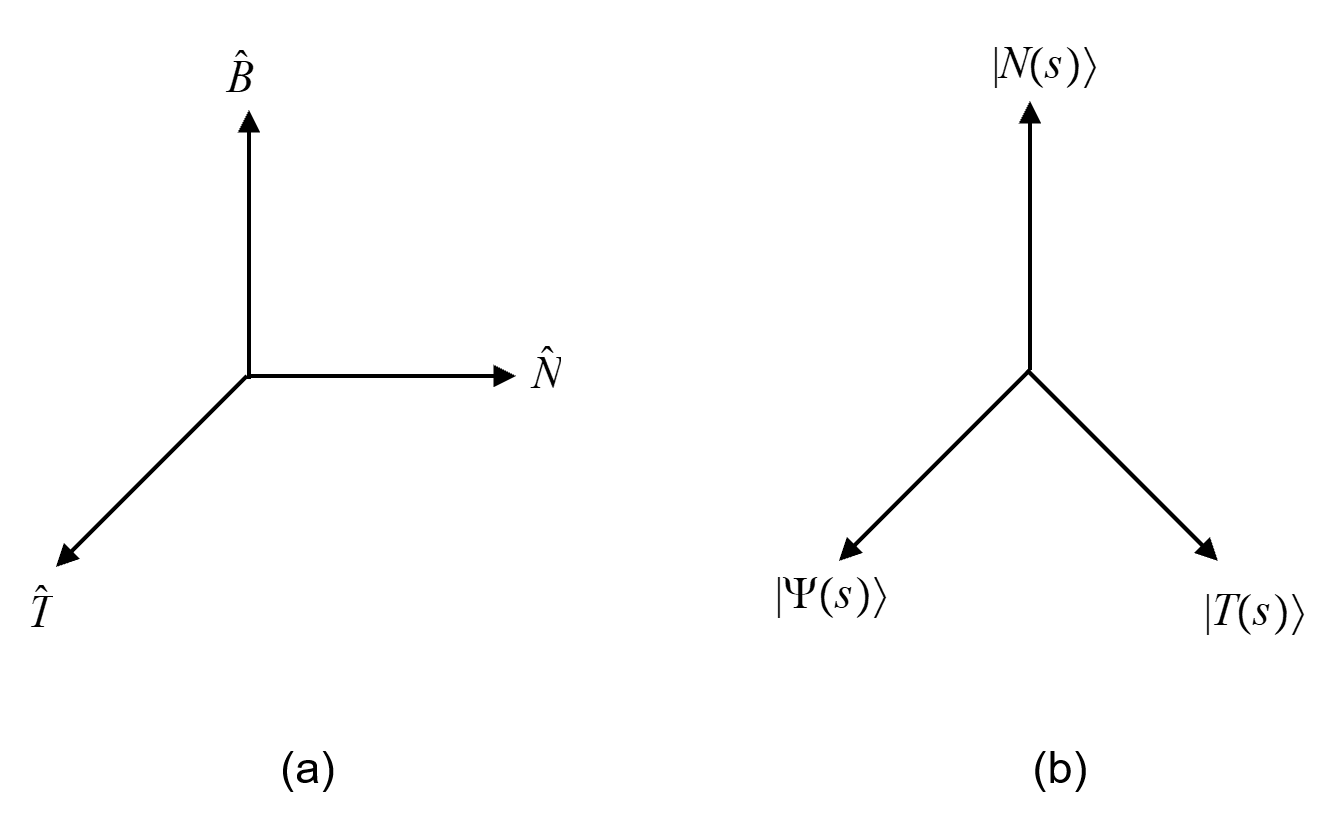}\caption{(a) Graphical depiction
of the Frenet-Serret frame $\left\{  \hat{T}\text{, }\hat{N}\text{, }\hat
{B}\right\}  $ for the vector spaces along a curve in three-dimensional
Euclidean space $\mathbb{R}^{3}$. (b) Pictorial representation of our proposed
quantum frame $\left\{  \left\vert \Psi\left(  s\right)  \right\rangle \text{,
}\left\vert T\left(  s\right)  \right\rangle \text{, }\left\vert N\left(
s\right)  \right\rangle \right\}  $ for the three-dimensional subspaces along
a curve on a generalized Bloch sphere $\mathbb{C}P^{N-1}=\mathcal{S}%
^{2N-1}/\mathcal{S}^{1}$ where $\mathcal{S}^{k}$ denotes a $k$-sphere and $N$
is the dimensionality of the complex Hilbert space. For $N=2$, the quantum
curve is on the Bloch sphere $\mathbb{C}P^{1}=\mathcal{S}^{3}/\mathcal{S}%
^{1}=\mathcal{S}^{2}$ and $\left\vert N\left(  s\right)  \right\rangle $ is
identically zero. A set of three orthonormal vectors is sufficient to define
the first two generalized curvature functions, i.e. the curvature and the
torsion coefficients.}%
\label{fig1}%
\end{figure}At this point, we can finally propose of notion of torsion
coefficient $\tau_{\mathrm{AC}}^{2}$ given by,%
\begin{equation}
\tau_{\mathrm{AC}}^{2}\overset{\text{def}}{=}|||\bar{N}(s)\rangle
||^{2}=\left\langle T^{\prime}\left(  s\right)  \left\vert \left(
\mathrm{P}^{\left(  T\right)  }\mathrm{P}^{\left(  \Psi\right)  }\right)
^{\dagger}\mathrm{P}^{\left(  T\right)  }\mathrm{P}^{\left(  \Psi\right)
}\right\vert T^{\prime}\left(  s\right)  \right\rangle =\left\langle
T^{\prime}\left(  s\right)  \left\vert \mathrm{P}^{\left(  \Psi\right)
}\mathrm{P}^{\left(  T\right)  }\mathrm{P}^{\left(  \Psi\right)  }\right\vert
T^{\prime}\left(  s\right)  \right\rangle \text{.} \label{torsionac1}%
\end{equation}
In terms of the state vector $\left\vert \bar{N}\left(  s\right)
\right\rangle \overset{\text{def}}{=}\mathrm{P}^{\left(  T\right)  }%
\mathrm{P}^{\left(  \Psi\right)  }\left\vert T^{\prime}\left(  s\right)
\right\rangle $, $\tau_{\mathrm{AC}}^{2}$ can be recast as
\begin{equation}
\tau_{\mathrm{AC}}^{2}\overset{\text{def}}{=}\left\langle \bar{N}\left(
s\right)  \left\vert \bar{N}\left(  s\right)  \right.  \right\rangle
=\left\Vert \mathrm{P}^{(T)}\mathrm{D}\left\vert T(s)\right\rangle \right\Vert
^{2}\text{.} \label{torsionac2}%
\end{equation}
Observe that since $\left\vert \bar{N}\left(  s\right)  \right\rangle $ in Eq.
(\ref{normale}) can be recast as $\mathrm{P}^{\left(  T\right)  }%
\mathrm{D}^{2}\left\vert \Psi\left(  s\right)  \right\rangle $ with
$\mathrm{D}^{2}\left\vert \Psi\left(  s\right)  \right\rangle =\mathrm{P}%
^{\left(  \Psi\right)  }\left\vert T^{\prime}\left(  s\right)  \right\rangle
$, $\left\vert \bar{N}\left(  s\right)  \right\rangle $ can be viewed as the
projection of the second covariant derivative $\mathrm{D}^{2}\left\vert
\Psi\left(  s\right)  \right\rangle $ of the state vector $\left\vert
\Psi\left(  s\right)  \right\rangle $ onto a state orthogonal to $\left\vert
T\left(  s\right)  \right\rangle $. Therefore,
by construction using $P^{T}\,P^{\Psi}$ we have that $\left\vert \bar
{N}\left(  s\right)  \right\rangle $ is a vector that is orthogonal to the
\textquotedblleft osculating\textquotedblright\ plane spanned by $\left\vert
\Psi\left(  s\right)  \right\rangle $ and $\left\vert T\left(  s\right)
\right\rangle $. Therefore, by construction, $\tau_{\mathrm{AC}}^{2}$ measures
how far the quantum curve traced out by $\left\vert \Psi\left(  s\right)
\right\rangle $ is lifting off from the instantaneous \textquotedblleft
osculating\textquotedblright\ plane spanned by the orthonormal set of vectors
$\left\vert \Psi\left(  s\right)  \right\rangle $ and $\left\vert T\left(
s\right)  \right\rangle $. Thus, $\tau_{\mathrm{AC}}^{2}$ is a measure of the
twisting of the Schr\"{o}dinger type quantum-mechanical evolution under
investigation and is, by construction, a quantum analogue of $\tau
_{\mathrm{LT}}^{2}$ with $\tau_{\mathrm{LT}}$ in\ Eq. (\ref{FStorsion})
introduced in the classical FS apparatus. To find an explicit expression of
$\tau_{\mathrm{AC}}^{2}$ in\ Eq. (\ref{torsionac1}), we first substitute Eq.
(\ref{normale}) into Eq. (\ref{torsionac1}). Then, recalling that
$\left\langle \Psi\left(  s\right)  \left\vert \Delta h\right\vert \Psi\left(
s\right)  \right\rangle =0$, we get%
\begin{equation}
\tau_{\mathrm{AC}}^{2}=\left\langle \Psi\left(  s\right)  \left\vert \left[
(\Delta h)^{2}-\left\langle (\Delta h)^{2}\right\rangle -\Delta h\left\langle
(\Delta h)^{3}\right\rangle \right]  ^{2}\right\vert \Psi\left(  s\right)
\right\rangle \geq0, \label{io1}%
\end{equation}
where the last inequality follows since the argument of the expectation value
is the square of a Hermitian operator. Then, expanding Eq. (\ref{io1}) and
noting that $\left\langle \Delta h\right\rangle =0$ and $\left\langle \left(
\Delta h\right)  ^{2}\right\rangle =1$, $\tau_{\mathrm{AC}}^{2}$ in Eq.
(\ref{io1}) reduces to%
\begin{equation}
\tau_{\mathrm{AC}}^{2}=\left\langle (\Delta h)^{4}\right\rangle -\left\langle
(\Delta h)^{3}\right\rangle ^{2}-1=\kappa_{AC}^{2}-\langle(\Delta
h)^{3}\rangle^{2}\text{.} \label{io2}%
\end{equation}
Interestingly, we note that $\tau_{\mathrm{AC}}^{2}$ in Eq. (\ref{io2})
coincides with the unitless torsion coefficient introduced by Laba and Tkachuk
given by $\bar{\tau}_{\mathrm{LT}}\overset{\text{def}}{=}\tau_{\mathrm{LT}%
}/\left\langle (\Delta\mathrm{H})^{2}\right\rangle ^{2}$ with $\tau
_{\mathrm{LT}}$ in Eq. (\ref{labatorsion}) once we observe that $\Delta
h=\Delta\mathrm{H}/\sqrt{\left\langle (\Delta\mathrm{H})^{2}\right\rangle }$.
Finally, combining Eqs. (\ref{accurve1}) and (\ref{io2}), we also recover the
same constraint relation as in the Laba-Tkachuk framework, i.e.,
$\kappa_{\mathrm{AC}}^{2}=\tau_{\mathrm{AC}}^{2}+\left\langle (\Delta
h)^{3}\right\rangle ^{2}$.

In the next section, we present several points of discussion that emerge from
our newly proposed concepts of curvature and torsion of a quantum evolution.

\section{Discussion}

We begin this section with a discussion on the statistical interpretation of
the curvature and torsion coefficients in Eqs. (\ref{accurve}) and
(\ref{torsionac1}), respectively. We then show the utility of rewriting the
expressions of these two coefficients in terms of the Bloch vector for
two-level systems. We proceed by elaborating on several challenges that can
arise in higher-dimensional Hilbert spaces with quantum evolutions governed by
nonstationary Hamiltonians. We finally conclude with a comparison between our
proposed quantum apparatus and the classical Frenet-Serret one.

\subsection{Statistical interpretation of curvature and torsion}

In mathematical statistics, the skewness and the kurtosis are two quantities
used to characterize the shape of a probability distribution. The skewness
involves the third moment of the distribution and is a measure of the
\textquotedblleft asymmetry\textquotedblright\ of the probability
distribution. It is defined as a unitless coefficient $\alpha_{3}$ given by
\cite{ernest44,sharma15},%
\begin{equation}
\alpha_{3}\overset{\text{def}}{=}\sqrt{\frac{m_{3}^{2}}{m_{2}^{3}}}%
=\frac{m_{3}}{m_{2}^{3/2}}\text{,} \label{skewness}%
\end{equation}
where $m_{r}$ is the $r$th central moment,%
\begin{equation}
m_{r}\overset{\text{def}}{=}\frac{1}{n}\sum_{i=1}^{n}\left(  x_{i}-\bar
{x}\right)  ^{r}\text{,}%
\end{equation}
and $\bar{x}$ is the arithmetic mean of $n$ real numbers $x_{i}$ with $1\leq
i\leq n$. Therefore, $m_{3}$ and $m_{2}$ in Eq. (\ref{skewness}) are the third
moment and the variance of the data set. The kurtosis, instead, involves the
fourth moment of the distribution and is a measure of the \textquotedblleft
tailedness\textquotedblright\ of the probability distribution (i.e., a
quantifier of how often outliers occur). It is defined as the unitless
coefficient $\alpha_{4}$ defined as \cite{ernest44,sharma15},%
\begin{equation}
\alpha_{4}\overset{\text{def}}{=}\frac{m_{4}}{m_{2}^{2}}\text{.}
\label{kurtosis}%
\end{equation}
A well-known inequality in mathematical statistics is the so-called Pearson
inequality \cite{pearson16},%
\begin{equation}
\alpha_{4}\geq\alpha_{3}^{2}+1\text{.} \label{pearson}%
\end{equation}
For later use, we employ Eqs. (\ref{skewness}) and (\ref{kurtosis}) to recast
the Pearson inequality $\alpha_{4}-1-\alpha_{3}^{2}\geq0$ as%
\begin{equation}
\frac{m_{4}-m_{2}^{2}}{m_{2}^{2}}-\frac{m_{3}^{2}}{m_{2}^{3}}\geq0\text{.}
\label{pearsonme}%
\end{equation}
Interestingly, our proposed curvature and torsion coefficients $\kappa
_{\mathrm{AC}}^{2}$ and $\tau_{\mathrm{AC}}^{2}$, respectively, have a neat
mathematical statistics interpretation. Indeed, we point the attention to the
following correspondences,%
\begin{equation}
\kappa_{\mathrm{AC}}^{2}=\frac{\left\langle \left(  \Delta\mathrm{H}\right)
^{4}\right\rangle -\left\langle \left(  \Delta\mathrm{H}\right)
^{2}\right\rangle ^{2}}{\left\langle \left(  \Delta\mathrm{H}\right)
^{2}\right\rangle ^{2}}\longleftrightarrow\alpha_{4}-1=\frac{m_{4}-m_{2}^{2}%
}{m_{2}^{2}}\text{,} \label{rose1}%
\end{equation}
and,%
\begin{equation}
\tau_{\mathrm{AC}}^{2}=\frac{\left\langle \left(  \Delta\mathrm{H}\right)
^{4}\right\rangle -\left\langle \left(  \Delta\mathrm{H}\right)
^{2}\right\rangle ^{2}}{\left\langle \left(  \Delta\mathrm{H}\right)
^{2}\right\rangle ^{2}}-\frac{\left\langle \left(  \Delta\mathrm{H}\right)
^{3}\right\rangle ^{2}}{\left\langle \left(  \Delta\mathrm{H}\right)
^{2}\right\rangle ^{3}}\longleftrightarrow\alpha_{4}-1-\alpha_{3}^{2}%
=\frac{m_{4}-m_{2}^{2}}{m_{2}^{2}}-\frac{m_{3}^{2}}{m_{2}^{3}}\text{.}
\label{rose2}%
\end{equation}
From Eqs. (\ref{rose1}) and (\ref{rose2}) we note that the curvature and the
torsion coefficients can be regarded in terms of statistically meaningful
quantum quantities, $\kappa_{\mathrm{AC}}^{2}\longleftrightarrow\alpha_{4}-1$
and $\tau_{\mathrm{AC}}^{2}\longleftrightarrow\alpha_{4}-1-\alpha_{3}^{2}$. In
particular, the difference $\kappa_{\mathrm{AC}}^{2}-\tau_{\mathrm{AC}}^{2}$
between the curvature and the torsion coefficient is captured by the square of
the skewness coefficient, $\alpha_{3}^{2}=\kappa_{\mathrm{AC}}^{2}%
-\tau_{\mathrm{AC}}^{2}$. This last inequality can be interpreted as follows.
Let us convey to denote a quantum state \textquotedblleft
symmetric\textquotedblright\ if $\left\langle \left(  \Delta\mathrm{H}\right)
^{3}\right\rangle $ is vanishing under the quantum evolution governed by the
stationary Hamiltonian \textrm{H}. Then, the curvature and the torsion of a
quantum curve traced out by a symmetric state are identical. The symmetry
encoded in the quantum state for a given Hamiltonian manifests itself in the
vanishing third moment $\left\langle \left(  \Delta\mathrm{H}\right)
^{3}\right\rangle $, whose presence signifies the existence of asymmetric
quantum behavior of statistical nature. Finally, the validity of the Pearson
inequality in Eq. (\ref{pearsonme}) is a straightforward consequence of the
positivity of $\tau_{\mathrm{AC}}^{2}$ in Eq. (\ref{rose2}).

\subsection{Curvature, torsion, and the Bloch vector for $2$-level systems}

Although our geometric approach is formally valid for arbitrary $d$-level
quantum systems with $d\geq2$, we focus here on two-level systems. Consider a
physical system specified by the density operator $\rho\left(  t\right)
\overset{\text{def}}{=}\left[  \mathrm{I}+\mathbf{a}\left(  t\right)
\mathbf{\cdot}\vec{\sigma}\right]  /2$ that evolves under the traceless
stationary Hamiltonian \textrm{H}$\overset{\text{def}}{=}\mathbf{m}\cdot
\vec{\sigma}$. The vectors $\mathbf{a}$ and $\mathbf{m}$ denote the Bloch
vector and the magnetic vector, respectively.
For the case of pure states considered here, $\mathbf{a}\cdot\mathbf{a}=1$ is
a unit vector on the Bloch sphere. Recall that an arbitrary qubit observable
$Q=q_{0}\mathrm{I}+\mathbf{q\cdot}\vec{\sigma}$ with $q_{0}\in%
\mathbb{R}
$ and $\mathbf{q\in%
\mathbb{R}
}^{3}$ has a corresponding expectation value given by $\left\langle
Q\right\rangle _{\rho}=q_{0}+\mathbf{a\cdot q}$. In what follows, we would
like to express the curvature and the torsion coefficients $\kappa
_{\mathrm{AC}}^{2}$ and $\tau_{\mathrm{AC}}^{2}$, respectively, in terms of
the vectors $\mathbf{a}$ and $\mathbf{m}$.

Recall that,
\begin{equation}
\kappa_{\mathrm{AC}}^{2}\overset{\text{def}}{=}\left\langle \left(  \Delta
h\right)  ^{4}\right\rangle -\left\langle \left(  \Delta h\right)
^{2}\right\rangle ^{2}=\frac{\left\langle \left(  \Delta\mathrm{H}\right)
^{4}\right\rangle -\left\langle \left(  \Delta\mathrm{H}\right)
^{2}\right\rangle ^{2}}{\left\langle \left(  \Delta\mathrm{H}\right)
^{2}\right\rangle ^{2}}\text{.} \label{invisible}%
\end{equation}
By brute force expansion and using the fact that $\mathrm{H}^{2}%
=\mathbf{m}^{2}\mathrm{I}$, we get%
\begin{align}
\left\langle \left(  \Delta\mathrm{H}\right)  ^{4}\right\rangle  &
=\left\langle \mathrm{H}^{4}\right\rangle -4\left\langle \mathrm{H}%
^{3}\right\rangle \left\langle \mathrm{H}\right\rangle +6\left\langle
\mathrm{H}^{2}\right\rangle \left\langle \mathrm{H}\right\rangle
^{2}-3\left\langle \mathrm{H}\right\rangle ^{4}\nonumber\\
&  =\left\langle \mathrm{H}^{4}\right\rangle -4\left\langle \mathrm{H}%
^{2}\right\rangle \left\langle \mathrm{H}\right\rangle ^{2}+6\left\langle
\mathrm{H}^{2}\right\rangle \left\langle \mathrm{H}\right\rangle
^{2}-3\left\langle \mathrm{H}\right\rangle ^{4}\nonumber\\
&  =\left\langle \mathrm{H}^{4}\right\rangle +2\left\langle \mathrm{H}%
^{2}\right\rangle \left\langle \mathrm{H}\right\rangle ^{2}-3\left\langle
\mathrm{H}\right\rangle ^{4}\text{,} \label{a}%
\end{align}
where we have used $H^{3} = \mathbf{m}^{2}\,H = \langle H^{2}\rangle\,H$ and
in addition,%
\begin{equation}
\left\langle \left(  \Delta\mathrm{H}\right)  ^{2}\right\rangle ^{2}%
=\left\langle \mathrm{H}^{2}\right\rangle ^{2}+\left\langle \mathrm{H}%
\right\rangle ^{4}-2\left\langle \mathrm{H}^{2}\right\rangle \left\langle
\mathrm{H}\right\rangle ^{2}\text{.} \label{b}%
\end{equation}
Combining Eqs. (\ref{a}) and (\ref{b}) and using the fact $\left\langle
\mathrm{H}^{4}\right\rangle =\left\langle \mathrm{H}^{2}\right\rangle
^{2}=\mathbf{m}^{4}$, we arrive at%
\begin{align}
\left\langle \left(  \Delta\mathrm{H}\right)  ^{4}\right\rangle -\left\langle
\left(  \Delta\mathrm{H}\right)  ^{2}\right\rangle ^{2}  &  =\left(
\left\langle \mathrm{H}^{4}\right\rangle +2\left\langle \mathrm{H}%
^{2}\right\rangle \left\langle \mathrm{H}\right\rangle ^{2}-3\left\langle
\mathrm{H}\right\rangle ^{4}\right)  -\left(  \left\langle \mathrm{H}%
^{2}\right\rangle ^{2}+\left\langle \mathrm{H}\right\rangle ^{4}-2\left\langle
\mathrm{H}^{2}\right\rangle \left\langle \mathrm{H}\right\rangle ^{2}\right)
\nonumber\\
&  =\left(  \left\langle \mathrm{H}^{4}\right\rangle -\left\langle
\mathrm{H}^{2}\right\rangle ^{2}\right)  +4\left\langle \mathrm{H}%
^{2}\right\rangle \left\langle \mathrm{H}\right\rangle ^{2}-4\left\langle
\mathrm{H}\right\rangle ^{4}\nonumber\\
&  =4\left\langle \mathrm{H}\right\rangle ^{2}\left(  \left\langle
\mathrm{H}^{2}\right\rangle -\left\langle \mathrm{H}\right\rangle ^{2}\right)
\nonumber\\
&  =4\left\langle \mathrm{H}\right\rangle ^{2}\left\langle \left(
\Delta\mathrm{H}\right)  ^{2}\right\rangle \text{,}%
\end{align}
that is,%
\begin{equation}
\left\langle \left(  \Delta\mathrm{H}\right)  ^{4}\right\rangle -\left\langle
\left(  \Delta\mathrm{H}\right)  ^{2}\right\rangle ^{2}=4\left\langle
\mathrm{H}\right\rangle ^{2}\left\langle \left(  \Delta\mathrm{H}\right)
^{2}\right\rangle \text{.} \label{invisible1}%
\end{equation}
Using Eq. (\ref{invisible1}), $\kappa_{\mathrm{AC}}^{2}$ in Eq.
(\ref{invisible}) becomes%
\begin{equation}
\kappa_{\mathrm{AC}}^{2}=4\frac{\left\langle \mathrm{H}\right\rangle ^{2}%
}{\left\langle \left(  \Delta\mathrm{H}\right)  ^{2}\right\rangle }\text{.}
\label{baby0}%
\end{equation}
Then, observing that $\left\langle \mathrm{H}\right\rangle ^{2}=\left(
\mathbf{a\cdot m}\right)  ^{2}$ and $\left\langle \left(  \Delta
\mathrm{H}\right)  ^{2}\right\rangle =\mathbf{m}^{2}-\left(  \mathbf{a\cdot
m}\right)  ^{2}$, we finally arrive at the expression for $\kappa
_{\mathrm{AC}}^{2}\left(  \mathbf{a}\text{, }\mathbf{m}\right)  $,%
\begin{equation}
\kappa_{\mathrm{AC}}^{2}\left(  \mathbf{a}\text{, }\mathbf{m}\right)
=4\frac{\left(  \mathbf{a\cdot m}\right)  ^{2}}{\mathbf{m}^{2}-\left(
\mathbf{a\cdot m}\right)  ^{2}}\text{.} \label{way1}%
\end{equation}
Remember that the torsion coefficient $\tau_{\mathrm{AC}}^{2}$ is given by,%
\begin{equation}
\tau_{\mathrm{AC}}^{2}=\left\langle \left(  \Delta h\right)  ^{4}\right\rangle
-\left\langle \left(  \Delta h\right)  ^{2}\right\rangle ^{2}-\left\langle
\left(  \Delta h\right)  ^{3}\right\rangle ^{2}=\frac{\left\langle \left(
\Delta\mathrm{H}\right)  ^{4}\right\rangle -\left\langle \left(
\Delta\mathrm{H}\right)  ^{2}\right\rangle ^{2}}{\left\langle \left(
\Delta\mathrm{H}\right)  ^{2}\right\rangle ^{2}}-\frac{\left\langle \left(
\Delta\mathrm{H}\right)  ^{3}\right\rangle ^{2}}{\left\langle \left(
\Delta\mathrm{H}\right)  ^{2}\right\rangle ^{3}}=\kappa_{\mathrm{AC}}%
^{2}-\frac{\left\langle \left(  \Delta\mathrm{H}\right)  ^{3}\right\rangle
^{2}}{\left\langle \left(  \Delta\mathrm{H}\right)  ^{2}\right\rangle ^{3}%
}\text{.} \label{torsione}%
\end{equation}
Observe that $\left\langle \left(  \Delta\mathrm{H}\right)  ^{3}\right\rangle
$ can be recast as,%
\begin{equation}
\left\langle \left(  \Delta\mathrm{H}\right)  ^{3}\right\rangle =\left\langle
\mathrm{H}^{3}\right\rangle -3\left\langle \mathrm{H}^{2}\right\rangle
\left\langle \mathrm{H}\right\rangle +2\left\langle \mathrm{H}\right\rangle
^{3}\text{.} \label{partz}%
\end{equation}
Then, noting that $\mathrm{H}^{2}=\mathbf{m}^{2}\mathrm{I}$ and $\mathrm{H}%
^{3}=\mathbf{m}^{2}\left(  \mathbf{m\cdot}\vec{\sigma}\right)  $,$\left\langle
\left(  \Delta\mathrm{H}\right)  ^{3}\right\rangle $ in Eq. (\ref{partz})
reduces to%
\begin{equation}
\left\langle \left(  \Delta\mathrm{H}\right)  ^{3}\right\rangle
=-2\left\langle \mathrm{H}\right\rangle \left\langle \left(  \Delta
\mathrm{H}\right)  ^{2}\right\rangle \text{.} \label{baby}%
\end{equation}
For later use, we emphasize that in terms of the vectors $\mathbf{a}$ and
$\mathbf{m}$, $\left\langle \left(  \Delta\mathrm{H}\right)  ^{3}\right\rangle
$ in Eq. (\ref{baby}) can be rewritten as $-2\left(  \mathbf{a\cdot m}\right)
\left[  \mathbf{m}^{2}-\left(  \mathbf{a\cdot m}\right)  ^{2}\right]  $. We
reiterate that Eq. (\ref{baby}) has been obtained when the single-qubit
quantum state evolves under a traceless stationary Hamiltonian of the form
\textrm{H}$=\mathbf{m}\cdot\vec{\sigma}$. Finally, employing Eqs.
(\ref{baby0}), (\ref{torsione}), and (\ref{baby}), we obtain%
\begin{equation}
\tau_{\mathrm{AC}}^{2}\left(  \mathbf{a}\text{, }\mathbf{m}\right)  =0\text{.}
\label{zerot}%
\end{equation}
Interestingly, the vanishing of the torsion coefficient in the case of motion
on a Bloch sphere can be explained in terms of the projector formalism as
well. Recall that $\tau_{\mathrm{AC}}^{2}$ is formally defined as the norm
squared of the vector $\left\vert \bar{N}\left(  s\right)  \right\rangle
\overset{\text{def}}{=}\mathrm{P}^{(T)}\mathrm{P}^{(\Psi)}\left\vert
T^{\prime}(s)\right\rangle $. Note that $\mathrm{I}\mathbb{=}\mathrm{I}%
_{\mathcal{H}_{2}^{1}}\overset{\text{def}}{=}\left\vert \Psi\right\rangle
\left\vert \Psi\right\vert +\left\vert T\right\rangle \left\vert T\right\vert
$ with $\dim_{%
\mathbb{C}
}\mathcal{H}_{2}^{1}=2$. Clearly, $\mathcal{H}_{2}^{1}$ denotes here the
Hilbert space of single-qubit
($k=1$, superscript) quantum states
of dimension $d=2$ (subscript). Furthermore, given that $\mathrm{P}^{(\Psi
)}\overset{\text{def}}{=}\mathrm{I}-\left\vert \Psi\right\rangle \left\langle
\Psi\right\vert $, $\mathrm{P}^{(T)}\overset{\text{def}}{=}\mathrm{I}%
-\left\vert T\right\rangle \left\langle T\right\vert $, and $\mathrm{P}%
^{(\Psi)}\mathrm{P}^{(T)}=\mathrm{P}^{(T)}\mathrm{P}^{(\Psi)}=\mathcal{O}$
(with $\mathcal{O}$ being the null operator) since $\left\vert \Psi
\right\rangle $ and $\left\vert T\right\rangle $ are orthonormal, we have%
\begin{align}
\mathrm{P}^{(T)}\mathrm{P}^{(\Psi)}  &  =\left(  \mathrm{I}-\left\vert
T\right\rangle \left\langle T\right\vert \right)  \left(  \mathrm{I}%
-\left\vert \Psi\right\rangle \left\langle \Psi\right\vert \right) \nonumber\\
&  =\mathrm{I}-\left\vert \Psi\right\rangle \left\langle \Psi\right\vert
-\left\vert T\right\rangle \left\langle T\right\vert \nonumber\\
&  =\mathrm{I}-\mathrm{I}\nonumber\\
&  =\mathcal{O}\text{.}%
\end{align}
Observe that for higher-dimensional systems, $\mathrm{P}^{(\Psi)}%
\mathrm{P}^{(T)}$ is not necessarily the null operator $\mathcal{O}$ since
since $\left\{  \left\vert \Psi\right\rangle \text{, }\left\vert
T\right\rangle \right\}  $ is not a complete set (in other words, the
resolution of the identity cannot be obtained by simply using $\left\vert
\Psi\right\rangle $ and $\left\vert T\right\rangle $, in general
higher-dimensional scenarios).

As a final remark, we observe that geodesic motion occurs when $\kappa
_{\mathrm{AC}}^{2}\left(  \mathbf{a}\text{, }\mathbf{m}\right)  =0$, i.e.,
when $\mathbf{a\perp m}$ from Eq. (\ref{way1}). In this case, the kurtosis
assumes its minimum value of one \ (since $\alpha_{4}=1\leftrightarrow
\left\langle \left(  \Delta\mathrm{h}\right)  ^{4}\right\rangle =1$) and the
skewness vanishes (since $\alpha_{3}=0\leftrightarrow\left\langle \left(
\Delta\mathrm{h}\right)  ^{3}\right\rangle =0$). Therefore, we conclude that
geodesic quantum motion specified by a stationary Hamiltonian on the Bloch
sphere occurs, from a statistical standpoint, with minimal sharpness together
with maximal symmetry.

\subsection{Extension to higher-dimensional state spaces}

In our current investigation, the expressions for $\kappa_{\mathrm{AC}}^{2}$
and $\tau_{\mathrm{AC}}^{2}$ are formally valid for arbitrary $d$-level
quantum systems evolving under stationary Hamiltonians. In our illustrative
examples, however, we shall limit the use of the notion of Bloch vectors to
Bloch spheres for $2$-level systems. Interestingly, the concept of Bloch
vector can be formally extended to $4$-level two-qubit systems \cite{jakob01}
and to arbitrary $d$-dimensional quantum systems with $d>2$ (i.e., qudits)
\cite{kimura03,krammer08,kurzy11}. For single-qubit systems, the usefulness of
the Bloch vector formalism is twofold. First, the unitary time evolution of a
single-qubit quantum state can be captured as an orbit on the Bloch spheres.
The orbit gives a clear visualization of the quantum mechanical time
evolution.\ Second, the Bloch vector has real components that can be expressed
as expectation values of experimentally measurable observables specified by
Hermitian operators (i.e., the Pauli operators in the single-qubit scenario).
When moving from single-qubit quantum systems to higher dimensional systems,
the physical interpretation of the Bloch vector preserves its usefulness.
Unfortunately, its geometric visualization is not as clear as in the $2$-level
systems. Indeed, bizarre properties of quantum theory emerge in
higher-dimensional systems, including the simplest but non-trivial case
represented by $3$-level systems (i.e., qutrits) \cite{kurzy11}. These bizarre
quantum features make it difficult to reveal the geometry of multidimensional
quantum systems \cite{xie20,siewert21}. A significant departure between
single-qubit systems and higher-dimensional quantum systems is the following.
For qubits, any point on the Bloch sphere or inside the Bloch ball corresponds
to a physical state (i.e., a pure and a mixed quantum state, respectively).
For $d$-dimensional qudit systems, instead, not every point on the
\textquotedblleft Bloch sphere\textquotedblright\ in dimensions $d^{2}-1$
corresponds to a physical state. Although it is possible to construct a unique
Bloch vector for any physical state, not every Bloch vector corresponds to a
quantum state. In particular, there is in higher-dimensions the emergence of
Bloch vectors that correspond to unphysical states specified by density
matrices with negative eigenvalues (though still of trace one, \cite{kurzy11}%
). Thus, one must also enforce the non-trivial condition of the positivity of
the corresponding density matrix for these states \cite{karol17}. For an
interesting discussion on Bloch vector representations of single-qubit
systems, single-qutrit systems, and two-qubit systems in terms of Pauli,
Gell-Mann, and Dirac matrices, respectively, we refer to Ref. \cite{gamel16}.
We shall attempt to extend our geometric intuition beyond the simplest quantum
systems together with considering the physical significance of the concepts of
curvature and torsion coefficients in relation to generalized
\textquotedblleft Bloch spheres\textquotedblright\ in forthcoming scientific endeavors.

\subsection{Extension to nonstationary quantum evolutions}

We have limited here our investigation to stationary Hamiltonian evolutions
and time-independent expressions for the coefficients $\kappa_{\mathrm{AC}%
}^{2}$ and $\tau_{\mathrm{AC}}^{2}$. In the stationary scenario, the exact
calculation of the dynamical trajectory traced out by the source state does
not present any significant problem. Although our geometric formalism yielding
time-dependent expressions for $\kappa_{\mathrm{AC}}^{2}$ and $\tau
_{\mathrm{AC}}^{2}$ can be formally extended to the time-dependent setting
\cite{alsing2}, we expect that finding analytical expressions for the
dynamical trajectories traced out by source states evolving under arbitrary
time-dependent Hamiltonians will be generally rather challenging. Indeed, it
is known that it is very difficult to find exact analytical solutions to the
time-dependent Schr\"{o}dinger equation even for two-level quantum systems.
The first examples of analytically solvable two-state time-dependent problems
were presented by Landau-Zener in Refs. \cite{landau32,zener32} and Rabi in
Refs. \cite{rabi37,rabi54}. The effort of finding analytical solutions has
been very intense throughout the years. For a partial list of relevant works
on two-state time-dependent problems that have appeared in the last ten years,
we suggest \cite{barnes12,barnes13,messina14,grimaudo18,grimaudo23} and
references therein. Finally, for a relatively recent work on finding exact
analytical solutions for specific classes of non-stationary Hamiltonians for
$d$-level quantum systems (i.e., qudits), we suggest Ref. \cite{elena20}. We
shall further investigate these issues in Ref. \cite{alsing2}.

\subsection{Comparison with the Frenet-Serret apparatus}

As previously mentioned, our proposed geometric construction of curvature and
torsion coefficients for quantum evolutions takes its original inspiration
from considering curves in three-dimensional Euclidean space framed in terms
of the Frenet-Serret frame \cite{parker77,neill06}. However, it is important
to emphasize that the classical Frenet-Serret apparatus formalism can be
extended to study curves in the higher-dimensional Euclidean space $%
\mathbb{R}
^{n}$ \cite{alvarez19} and, more interestingly, there is a freedom in
constructing frames and corresponding apparatuses when studying the local
geometry of curves in $%
\mathbb{R}
^{n}$ \cite{bishop75}. The Frenet-Serret \textquotedblleft moving
frame\textquotedblright\ is just one geometrically nice orthonormal basis for
the vector spaces along a curve, where the basis vectors move and twist as one
moves along the curve. Clearly, there is more than one way to frame a curve in
$%
\mathbb{R}
^{3}$. For instance, the canonical orthonormal basis of $%
\mathbb{R}
^{3}$, $\left\{  \hat{e}_{1}\text{, }\hat{e}_{2}\text{, }\hat{e}_{3}\right\}
$ with $\hat{e}_{1}\overset{\text{def}}{=}(1$, $0$, $0)$, $\hat{e}%
_{2}\overset{\text{def}}{=}(0$, $1$, $0)$, and $\hat{e}_{3}\overset{\text{def}%
}{=}(0$, $0$, $1)$ is a legitimate basis for the vector spaces along a curve.
However, $\left\{  \hat{e}_{1}\text{, }\hat{e}_{2}\text{, }\hat{e}%
_{3}\right\}  $ reflects the geometry of $%
\mathbb{R}
^{3}$ rather than the geometry of the curve. The Frenet-Serret frame, instead,
is an intrinsically geometric basis that reflects the geometry of the curve.
It has the peculiarity that the frame is adapted to the curve, that is, its
members are either tangent to or perpendicular to the curve. We recall that
moving frames are orthonormal basis fields that can be used to express the
derivatives of the frame with respect to the curve parameter in terms of the
frame itself. Due to orthonormality, the coefficient matrix (also known as,
the Cartan matrix) that relates the derivatives of the frame to the frame
itself is always skew-symmetric. Therefore, this matrix has three nonzero
entries, in general. The Frenet-Serret frame has only two nonzero entries
which are specified by two scalar-valued functions (i.e., the curvature
$\kappa_{\mathrm{FS}}$ and the torsion $\tau_{\mathrm{FS}}$ coefficients). As
previously mentioned, it is possible to show that there exist other adapted
frames which have only two nonzero entries in their Cartan matrices as well
\cite{bishop75}. Despite the fact that these alternative frames (constructed,
for instance, within the so-called normal development of a curve
\cite{bishop75}) yield geometric invariants that do not posses an equally nice
geometric interpretation as the one available for the Frenet invariants, they
only require $2$-times continuously differentiable curves with a nonzero first
derivative $\gamma^{\prime}$ of the curve $\gamma$ in $%
\mathbb{R}
^{3}$. A Frenet-Serret apparatus of a curve $\gamma$ in $%
\mathbb{R}
^{3}$, instead, requires $3$-times differentiability along with nondegeneracy
(i.e., the first and second derivatives of the curve, $\gamma^{\prime}$ and
$\gamma^{\prime\prime}$, respectively, must be linearly independent).

Unlike what happens in the quantum setting being considered here, the curve is
given a priori in the classical Frenet-Serret apparatus. Indeed, the curve is
not generated by any Hamiltonian and is not necessarily a solution of any
specific differential equation of motion. From the curve, one can construct
the set of orthonormal vectors (i.e., the Frenet-Serret frame) from the
derivatives of the curve. However, if the curve is a non-unit speed curve
which is parametrized by ordinary time rather than the arc length, the
curvature and the torsion coefficients can be expressed in terms of more
complicated functions of derivatives of the curve up to third order
\cite{parker77}. In general, the classical Frenet-Serret apparatus is built
for curves in $n$-dimensional (real) Euclidean spaces $%
\mathbb{R}
^{n}$ equipped with a real inner product. Our proposed quantum apparatus,
instead, is formally constructed for curves on generalized \textquotedblleft
Bloch spheres\textquotedblright\ $%
\mathbb{C}
P^{N-1}=S^{2N-1}/S^{1}$. The space $%
\mathbb{C}
P^{N-1}$ can be viewed as the quotient of the unit $\left(  2N-1\right)
$-sphere in the $N$-dimensional complex space $%
\mathbb{C}
^{N}$ under the action of $S^{1}=U\left(  1\right)  $. For instance, in the
case of the two-dimensional complex Hilbert space for single-qubit quantum
states, we consider curves on the usual Bloch sphere. Despite these
differences, in analogy to the classical Frenet-Serret apparatus, we propose a
quantum apparatus for curves to which is given (in principle, at least) a unit
speed parametrization by means of the arc length parameter. Moreover,
similarly to the Frenet-Serret frame, our proposed quantum frame is determined
by applying the Gram-Schmidt (GS) orthonormalization process. However, while
for space curves in $%
\mathbb{R}
^{3}$, the GS\ procedure is applied to $\left\{  \gamma^{\prime}\left(
s\right)  \text{, }\gamma^{\prime\prime}\left(  s\right)  \text{, }%
\gamma^{\prime\prime\prime}\left(  s\right)  \right\}  $, in our case we apply
it to the set $\left\{  \left\vert \Psi\left(  s\right)  \right\rangle \text{,
}\left\vert \Psi^{\prime}\left(  s\right)  \right\rangle \text{, }\left\vert
\Psi^{\prime\prime}\left(  s\right)  \right\rangle \right\}  $. Therefore, in
analogy to the above-mentioned alternative frames for curves in $%
\mathbb{R}
^{n}$, we only require $2$-times continuously differentiable quantum curves
with a nonvanishing first derivative $\left\vert \Psi^{\prime}\left(
s\right)  \right\rangle $ of the state vector $\left\vert \Psi\left(
s\right)  \right\rangle $. For clarity, recall that the quantum curve,
$s\rightarrow\left\vert \Psi\left(  s\right)  \right\rangle $, is the one
traced by the state $\left\vert \Psi\left(  s\right)  \right\rangle $. More
specifically, in the Frenet-Serret case of a space curve in $%
\mathbb{R}
^{3}$, we recall that the coefficient matrix $\mathrm{M}_{\partial_{s}\left(
\mathrm{Frame}\right)  \rightarrow\mathrm{Frame}}^{\left(  \mathrm{FS}\right)
}$ that expresses the derivatives of the frame with respect to the curve
parameter $s$ in terms of the frame itself is given by \cite{parker77}
\begin{equation}
\mathrm{M}_{\partial_{s}\left(  \mathrm{Frame}\right)  \rightarrow
\mathrm{Frame}}^{\left(  \mathrm{FS}\right)  }=\left(
\begin{array}
[c]{ccc}%
0 & \kappa_{\mathrm{FS}} & 0\\
-\kappa_{\mathrm{FS}} & 0 & \tau_{\mathrm{FS}}\\
0 & -\tau_{\mathrm{FS}} & 0
\end{array}
\right)  \text{.} \label{FSframe}%
\end{equation}
In our quantum scenario, instead, focusing on the three-dimensional subspace
of $%
\mathbb{C}
^{N}$ spanned by the quantum frame $\left\{  \left\vert \Psi\left(  s\right)
\right\rangle \text{, }\left\vert T\left(  s\right)  \right\rangle \text{,
}\left\vert N\left(  s\right)  \right\rangle \right\}  $, the analog of Eq.
(\ref{FSframe}) is generally given by%
\begin{equation}
\mathrm{M}_{\partial_{s}\left(  \mathrm{Frame}\right)  \rightarrow
\mathrm{Frame}}^{\left(  \mathrm{AC}\right)  }=\left(
\begin{array}
[c]{ccc}%
0 & 1 & 0\\
-1 & i\operatorname{Im}(e^{i\phi_{\left\langle T\left\vert T^{\prime}\right.
\right\rangle }}\sqrt{\kappa_{\mathrm{AC}}^{2}-\tau_{\mathrm{AC}}^{2}}) &
e^{i\phi_{\left\langle N\left\vert T^{\prime}\right.  \right\rangle }}%
\tau_{\mathrm{AC}}\\
0 & -e^{-i\phi_{\left\langle N\left\vert T^{\prime}\right.  \right\rangle }%
}\tau_{\mathrm{AC}} & i\operatorname{Im}\left(  \left\langle N\left\vert
N^{\prime}\right.  \right\rangle \right)
\end{array}
\right)  \text{.} \label{ACframe}%
\end{equation}
Note that in Eq. (\ref{ACframe}) we used the usual exponential representation
of a complex number, $z=\left\vert z\right\vert e^{i\phi_{z}}$ with $\phi_{z}$
being the (real) argument of $z$. In addition, $\tau_{\mathrm{AC}}=\left\vert
\left\langle N\left\vert T^{\prime}\right.  \right\rangle \right\vert $,
$\kappa_{\mathrm{AC}}^{2}=\left\vert \left\langle T\left\vert T^{\prime
}\right.  \right\rangle \right\vert ^{2}+\left\vert \left\langle N\left\vert
T^{\prime}\right.  \right\rangle \right\vert ^{2}$, and $\left\vert
\left\langle T\left\vert T^{\prime}\right.  \right\rangle \right\vert
=\sqrt{\kappa_{\mathrm{AC}}^{2}-\tau_{\mathrm{AC}}^{2}}$. Finally, observe
that $\operatorname{Re}\left[  \mathrm{M}_{\partial_{s}\left(  \mathrm{Frame}%
\right)  \rightarrow\mathrm{Frame}}^{\left(  \mathrm{FS}\right)  }\right]  $
is skew-symmetric and, more generally, $\mathrm{M}_{\partial_{s}\left(
\mathrm{Frame}\right)  \rightarrow\mathrm{Frame}}^{\left(  \mathrm{AC}\right)
}$ is skew-Hermitian. Essentially, skew-Hermitian matrices are the complex
versions of the real skew-symmetric matrices. Moreover, while $\mathrm{M}%
_{\partial_{s}\left(  \mathrm{Frame}\right)  \rightarrow\mathrm{Frame}%
}^{\left(  \mathrm{FS}\right)  }$ in Eq. (\ref{FSframe}) has only two nonzero
(real) entries, $\mathrm{M}_{\partial_{s}\left(  \mathrm{Frame}\right)
\rightarrow\mathrm{Frame}}^{\left(  \mathrm{AC}\right)  }$ in Eq.
(\ref{ACframe}) has generally three nonzero (complex) entries. Finally,
although using a different frame as evident from Eqs. (\ref{FSframe}) and
(\ref{ACframe}), we also propose good measures to quantify the failures of
linearity and planarity in terms of suitable curvature $\kappa_{\mathrm{AC}}$
and torsion $\tau_{\mathrm{AC}}$ coefficients, respectively. For illustrative
purposes, we provide an explicit example of the construction of a quantum
frame for a quantum curve traced by a two-qubit quantum state on a generalized
\textquotedblleft Bloch sphere\textquotedblright\ in $%
\mathbb{C}
^{4}$ in Appendix A. For clarity, we present in Fig. $2$ a sketch of two
orthonormal position vectors locating two points on a $2$ -sphere viewed as
the boundary of a $3$-ball in $R^{3}$ and, in addition, we draw two
orthonormal unit Bloch vectors locating two orthogonal quantum states.
Finally, before presenting our conclusive remarks in Sec. VII, we show some
illustrative examples in the next section.\begin{figure}[t]
\centering
\includegraphics[width=0.5\textwidth] {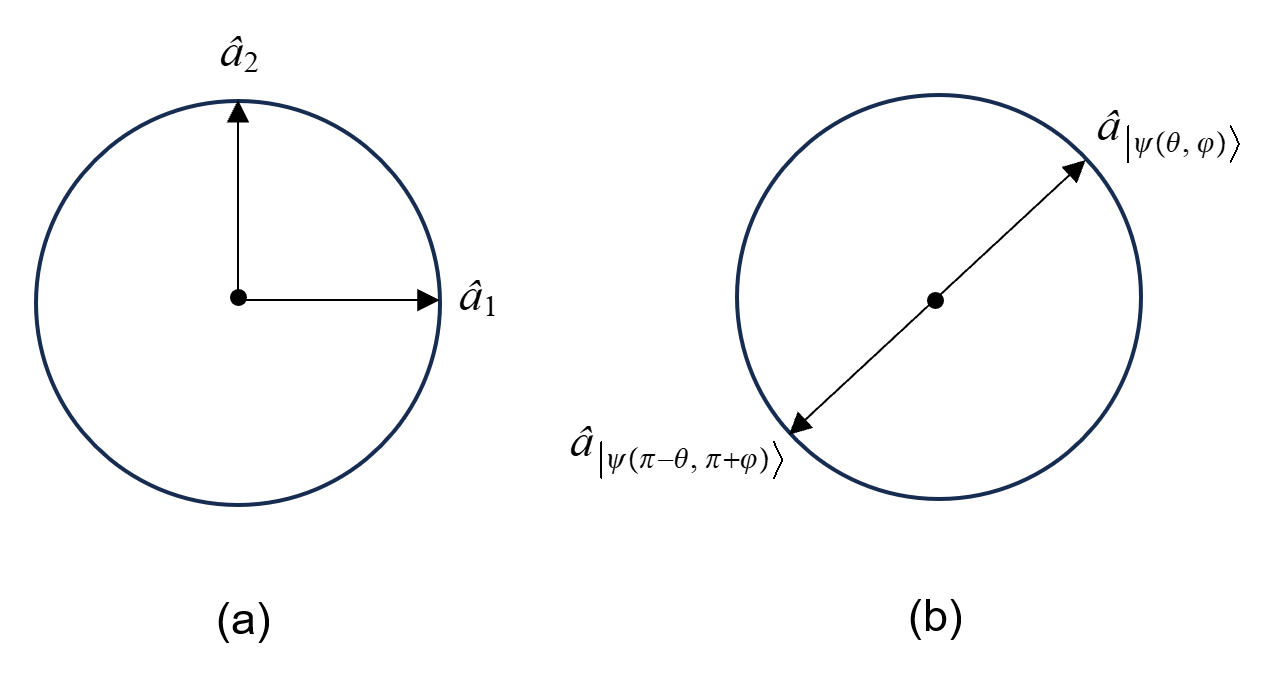}\caption{(a) Sketch of two
orthonormal position vectors locating two points on a $2$-sphere viewed as the
boundary of a $3$-ball in $\mathbb{R}^{3}$. (b) Drawing of two orthonormal
Bloch vectors locating two orthogonal quantum states. Antipodal states on the
Bloch sphere $\left\vert \psi\left(  \theta,\varphi\right)  \right\rangle
\protect\overset{\text{def}}{=}\cos\left(  \theta/2\right)  \left\vert
0\right\rangle +e^{i\varphi}\sin(\theta/2)\left\vert 1\right\rangle $ and
$\left\vert \psi\left(  \pi-\theta,\pi+\varphi\right)  \right\rangle $ with
Bloch vectors $\hat{a}_{\left\vert \psi\left(  \theta,\varphi\right)
\right\rangle }=\left(  \sin\left(  \theta\right)  \cos\left(  \varphi\right)
\text{, }\sin\left(  \theta\right)  \sin\left(  \varphi\right)  \text{, }%
\cos\left(  \theta\right)  \right)  $ and $\hat{a}_{\left\vert \psi\left(
\pi-\theta,\pi+\varphi\right)  \right\rangle }=-$ $\hat{a}_{\left\vert
\psi\left(  \theta,\varphi\right)  \right\rangle }$, respectively, are
orthogonal. The Bloch sphere is not to be regarded as the usual sphere in
three-dimensional coordinate space $\mathbb{R}^{3}$. In particular, the latter
can be viewed as the \textquotedblleft double cover\textquotedblright\ of the
former.}%
\label{fig2}%
\end{figure}

\section{Illustrative examples}

We exhibit in this section simple illustrative examples of the behavior of
curvature and torsion coefficients in Eqs. (\ref{accurve}) and
(\ref{torsionac1}), respectively, for quantum evolutions specified by
single-qubit and two-qubit time-independent Hamiltonians.

Recall that in classical physics, the force that appears in Newton's second
law depends only on the gradient of the potential and a constant additive term
is physically irrelevant. Potential differences, instead, have physical
significance. Analogously, in the quantum setting, there is no absolute energy
and the physics of \textrm{H}$_{1}\overset{\text{def}}{=}m_{0}\mathrm{I}%
+\mathbf{m}\cdot\vec{\sigma}$ and \textrm{H}$_{2}\overset{\text{def}%
}{=}\mathbf{m}\cdot\vec{\sigma}$
for a single qubit is the same. In particular, although $\left\langle
\mathrm{H}_{1}\right\rangle =m_{0}\neq0$ and $\left\langle \mathrm{H}%
_{2}\right\rangle =0$, we have that $\Delta$\textrm{H}$_{1}\overset{\text{def}%
}{=}$\textrm{H}$_{1}-\left\langle \mathrm{H}_{1}\right\rangle $ equals
$\Delta$\textrm{H}$_{2}\overset{\text{def}}{=}$\textrm{H}$_{2}-\left\langle
\mathrm{H}_{2}\right\rangle $ with $\Delta$\textrm{H}$_{2}=\mathbf{m}\cdot
\vec{\sigma}$. Therefore, the curvature coefficients that emerge from the
Hamiltonians \textrm{H}$_{1}$ and \textrm{H}$_{2}$ are identical.
Interestingly, the expectation value of a traceless Hamiltonian such as
\textrm{H}$_{2}$ can be different from zero. When this happens, we have a
nonvanishing curvature of the quantum evolution. For a traceless Hamiltonian
to generate geodesic motion, its expectation value must be zero. Indeed, the
most general single-qubit geodesic evolution defined by a traceless
time-independent Hamiltonian is given by $\mathrm{H}=E_{1}\left\vert
E_{1}\right\rangle \left\langle E_{1}\right\vert +E_{2}\left\vert
E_{2}\right\rangle \left\langle E_{2}\right\vert $, with $\mathrm{H}\left\vert
E_{i}\right\rangle =E_{i}\left\vert E_{i}\right\rangle $ for $1\leq i\leq2$
and $E_{2}=-E_{1}=E$, acting on a source state $\left\vert A\right\rangle
=(1/\sqrt{2})\left[  \left\vert E_{1}\right\rangle +e^{i\phi}\left\vert
E_{2}\right\rangle \right]  $ with $\phi\in%
\mathbb{R}
$ \cite{ali09}. In this case, the evolved state is given by%
\begin{equation}
\left\vert \psi\left(  t\right)  \right\rangle =\frac{e^{-i(E_{1}/\hslash)t}%
}{\sqrt{2}}\left[  \left\vert E_{1}\right\rangle +e^{i\phi}e^{-i\left[
\left(  E_{2}-E_{1}\right)  /\hslash\right]  t}\left\vert E_{2}\right\rangle
\right]  \text{.} \label{gibbs}%
\end{equation}
A straightforward calculation shows that $\left\langle \mathrm{H}\right\rangle
=0$ and, thus, $\kappa_{\mathrm{AC}}^{2}$ vanishes for the quantum curve
traced out by the state $\left\vert \psi\left(  t\right)  \right\rangle $
in\ Eq. (\ref{gibbs}). However, it is possible to have a non-traceless
Hamiltonian that exhibits geodesic motion even if its expectation value is
different from zero. Finally, the torsion coefficient $\tau_{\mathrm{AC}}^{2}$
vanishes for any traceless time-independent single-qubit Hamiltonian of the
form \textrm{H}$=\mathbf{m}\cdot\vec{\sigma}$, regardless of the expectation
value of the Hamiltonian.

\subsubsection{Zero curvature and zero torsion: single qubit}

Consider the spin precession that specifies the evolution of a spin-$1/2$
particle (an electron, for instance) with magnetic moment $e\hslash/(2m_{e}c)$
(with $e\leq0$, for an electron) subjected to a stationary magnetic field
$\mathbf{B}$. Clearly, $c$ is the speed of light and $m_{e}$ denotes the mass
of the electron. Then, the Hamiltonian of the system becomes \textrm{H}%
$\overset{\text{def}}{=}\mathbf{m\cdot}\vec{\sigma}$ with $\mathbf{m}%
\overset{\text{def}}{=}\left\vert e\right\vert \mathbf{B}/(2m_{e}c)$. In this
first example, let us assume $\mathbf{m}=\left(  0\text{, }0\text{, }m\right)
$ with $m\overset{\text{def}}{=}\left\vert e\right\vert B_{z}/(2m_{e}c)$. The
quantum trajectory is assumed to be traced out by the state $\left\vert
\psi\left(  t\right)  \right\rangle =U\left(  t\right)  \left\vert \psi\left(
0\right)  \right\rangle $ with $U\left(  t\right)  \overset{\text{def}}{=}%
\exp\left[  -\left(  i/\hslash\right)  \mathrm{H}t\right]  $ being the unitary
evolution operator and $\left\vert \psi\left(  0\right)  \right\rangle $ the
chosen initial state vector. We assume that the unit states $\left\vert
\psi\left(  t\right)  \right\rangle $ with $t_{A}\leq t\leq t_{B}$ are given
by%
\begin{equation}
\left\vert \psi\left(  t\right)  \right\rangle =\frac{e^{-i(m/\hslash)t}%
}{\sqrt{2}}\left[  \left\vert 0\right\rangle +e^{i(2m/\hslash)t}\left\vert
1\right\rangle \right]  \text{.} \label{geostate}%
\end{equation}
These states in Eq. (\ref{geostate}) can be regarded as states on the Bloch
sphere that can be parametrized in spherical coordinates as $\left\vert
\psi\left(  \theta\text{, }\varphi\right)  \right\rangle =\cos(\theta
/2)\left\vert 0\right\rangle +e^{i\varphi}\sin(\theta/2)\left\vert
1\right\rangle $ with a corresponding Bloch vector $\mathbf{a}=(\sin\left(
\theta\right)  \cos\left(  \varphi\right)  $, $\sin\left(  \theta\right)
\sin\left(  \varphi\right)  $, $\cos\left(  \theta\right)  )$. The initial and
terminal states are given by $\left\vert A\right\rangle \overset{\text{def}%
}{=}\left\vert \psi\left(  \theta_{A}\text{, }\varphi_{A}\right)
\right\rangle $ and $\left\vert B\right\rangle \overset{\text{def}%
}{=}\left\vert \psi\left(  \theta_{B}\text{, }\varphi_{B}\right)
\right\rangle $, respectively. From Eq. (\ref{geostate}), it is clear that we
choose $\left(  \theta_{A}\text{, }\varphi_{A}\right)  \overset{\text{def}%
}{=}\left(  \pi/2\text{, }0\right)  $, $\left(  \theta_{B}\text{, }\varphi
_{B}\right)  \overset{\text{def}}{=}\left(  \pi/2\text{, }\pi/2\right)  $, and
$0$ $\leq t\leq\hslash\pi/4m$. Therefore, the initial and final states
$\left\vert A\right\rangle =\left[  \left\vert 0\right\rangle +\left\vert
1\right\rangle \right]  /\sqrt{2}$ and $\left\vert B\right\rangle =\left[
\left\vert 0\right\rangle +i\left\vert 1\right\rangle \right]  /\sqrt{2}$ are
specified by Bloch vectors $\mathbf{a}_{\mathrm{initial}}=\left(  1\text{,
}0\text{, }0\right)  $ and $\mathbf{a}_{\mathrm{final}}=\left(  0\text{,
}1\text{, }0\right)  $, respectively. A simple calculation yields
$\kappa_{\mathrm{AC}}^{2}=4(\mathbf{a}\cdot\mathbf{m})^{2}/[\mathbf{m}%
^{2}-(\mathbf{a}\cdot\mathbf{m})^{2}]=0$ since the traceless Hamiltonian
\textrm{H}$=m\sigma_{z}$ has a vanishing expectation value, $\left\langle
\mathrm{H}\right\rangle =\mathbf{a}\cdot\mathbf{m}=0$ with $\mathbf{a}%
=\mathbf{a}_{\mathrm{initial}}$. This, in turn, is a consequence of the fact
that the Bloch vector $\mathbf{a}_{\mathrm{initial}}$ of the source state is
orthogonal to the applied magnetic field $\mathbf{B}\overset{\text{def}%
}{=}(2m_{e}c/\left\vert e\right\vert )\mathbf{m}$.

\subsubsection{Nonzero curvature and zero torsion: single qubit}

In analogy to the vanishing curvature case, we consider the quantum evolution
under the traceless Hamiltonian \textrm{H}$\overset{\text{def}}{=}%
\mathbf{m}\cdot\vec{\sigma}$ where $\mathbf{m}=\left(  0\text{, }0\text{,
}m\right)  $ with $m\overset{\text{def}}{=}\left\vert e\right\vert
B_{z}/(2m_{e}c)$. Again, the quantum trajectory is assumed to be traced out by
the state $\left\vert \psi\left(  t\right)  \right\rangle =U\left(  t\right)
\left\vert \psi\left(  0\right)  \right\rangle $ with $U\left(  t\right)
\overset{\text{def}}{=}\exp\left[  -\left(  i/\hslash\right)  \mathrm{H}%
t\right]  $ being the unitary evolution operator and $\left\vert \psi\left(
0\right)  \right\rangle $ the chosen initial state vector. Unlike the previous
case, we assume now that the unit states $\left\vert \psi\left(  t\right)
\right\rangle $ with $t_{A}\leq t\leq t_{B}$ are given by%
\begin{equation}
\left\vert \psi\left(  t\right)  \right\rangle =e^{-i(m/\hslash)t}\left[
\frac{\sqrt{2+\sqrt{2}}}{2}\left\vert 0\right\rangle +e^{i(2m/\hslash)t}%
\frac{\sqrt{2-\sqrt{2}}}{2}\left\vert 1\right\rangle \right]  \text{.}
\label{nogeo}%
\end{equation}
These states in Eq. (\ref{nogeo}) can be regarded as state on the Bloch sphere
that can be parametrized as $\left\vert \psi\left(  \theta\text{, }%
\varphi\right)  \right\rangle =\cos(\theta/2)\left\vert 0\right\rangle
+e^{i\varphi}\sin(\theta/2)\left\vert 1\right\rangle $ with corresponding
Bloch vector $\mathbf{a}=\left(  \sin\left(  \theta\right)  \cos\left(
\varphi\right)  \text{, }\sin\left(  \theta\right)  \sin\left(  \varphi
\right)  \text{, }\cos\left(  \theta\right)  \right)  $. The initial (source)
and final (target) states are $\left\vert A\right\rangle \overset{\text{def}%
}{=}\left\vert \psi\left(  \theta_{A}\text{, }\varphi_{A}\right)
\right\rangle $ and $\left\vert B\right\rangle \overset{\text{def}%
}{=}\left\vert \psi\left(  \theta_{B}\text{, }\varphi_{B}\right)
\right\rangle $, respectively. From Eq. (\ref{nogeo}), it is clear that we
choose $\left(  \theta_{A}\text{, }\varphi_{A}\right)  \overset{\text{def}%
}{=}\left(  \pi/4\text{, }0\right)  $, $\left(  \theta_{B}\text{, }\varphi
_{B}\right)  \overset{\text{def}}{=}\left(  \pi/4\text{, }\pi/2\right)  $, and
$0\leq t\leq\hslash\pi/4m$. Therefore, the initial and final states
$\left\vert A\right\rangle =(\sqrt{2+\sqrt{2}}/2)\left\vert 0\right\rangle
+(\sqrt{2-\sqrt{2}}/2)\left\vert 1\right\rangle $ and $\left\vert
B\right\rangle =(\sqrt{2+\sqrt{2}}/2)\left\vert 0\right\rangle +i(\sqrt
{2-\sqrt{2}}/2)\left\vert 1\right\rangle $ are specified by Bloch vectors
$\mathbf{a}_{\mathrm{initial}}=\left(  1/\sqrt{2}\text{, }0\text{, }1/\sqrt
{2}\right)  $ and $\mathbf{a}_{\mathrm{final}}=\left(  0\text{, }1/\sqrt
{2}\text{, }1/\sqrt{2}\right)  $, respectively. A simple calculation yields
$\kappa_{\mathrm{AC}}^{2}=4(\mathbf{a}\cdot\mathbf{m})^{2}/[\mathbf{m}%
^{2}-(\mathbf{a}\cdot\mathbf{m})^{2}]=4\neq0$ since the traceless Hamiltonian
\textrm{H}$=m\sigma_{z}$ has a nonvanishing expectation value, $\left\langle
\mathrm{H}\right\rangle =\mathbf{a}\cdot\mathbf{m}=m/\sqrt{2}\neq0$ with
$\mathbf{a}=\mathbf{a}_{\mathrm{initial}}$. This, in turn, is a consequence of
the fact that the Bloch vector $\mathbf{a}_{\mathrm{initial}}$ of the source
state is not orthogonal to the applied magnetic field $\mathbf{B}%
\overset{\text{def}}{=}(2m_{e}c/\left\vert e\right\vert )\mathbf{m}%
$.\begin{figure}[t]
\centering
\includegraphics[width=0.5\textwidth] {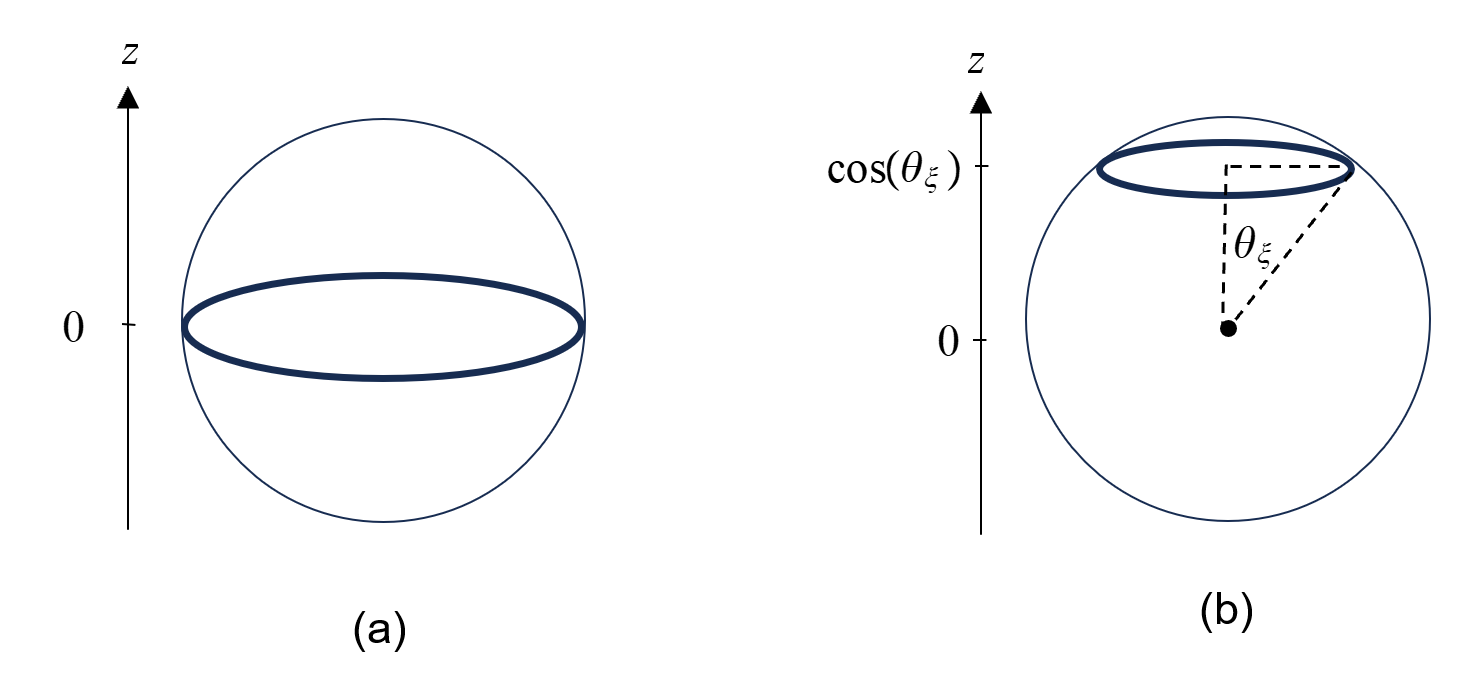}\caption{(a) The tick solid line
is a graphical depiction of the circular path on the Bloch sphere traced out
by evolving the initial state $\left\vert \psi\left(  0\right)  \right\rangle
\protect\overset{\text{def}}{=}(1/\sqrt{2})\left\vert 0\right\rangle
+(1/\sqrt{2})\left\vert 1\right\rangle $ under the Hamiltonian \textrm{H}%
$\protect\overset{\text{def}}{=}\hslash\omega_{0}\sigma_{z}$. In this case,
$\left\vert \Psi\left(  s\right)  \right\rangle =(e^{-is}\left\vert
0\right\rangle +e^{is}\left\vert 1\right\rangle )/\sqrt{2}$, $\left\vert
T\left(  s\right)  \right\rangle =\left(  -ie^{-is}\left\vert 0\right\rangle
+ie^{is}\left\vert 1\right\rangle \right)  /\sqrt{2}$, $\hat{a}_{\left\vert
\Psi\left(  s\right)  \right\rangle }=\left(  \cos\left(  2s\right)  \text{,
}\sin\left(  2s\right)  \text{, }0\right)  $, and $\hat{a}_{\left\vert
T\left(  s\right)  \right\rangle }=-\hat{a}_{\left\vert \Psi\left(  s\right)
\right\rangle }$, with $s\protect\overset{\text{def}}{=}\omega_{0}t$. In this
case, $\kappa_{\mathrm{AC}}^{2}=0$. (b) The tick solid line represents the
circular path on the Bloch sphere traced out by evolving the initial state
$\left\vert \psi\left(  0\right)  \right\rangle \protect\overset{\text{def}%
}{=}\cos\left(  \theta_{\xi}/2\right)  \left\vert 0\right\rangle +\sin
(\theta_{\xi}/2)\left\vert 1\right\rangle $ under the Hamiltonian
\textrm{H}$\protect\overset{\text{def}}{=}\hslash\omega_{0}\sigma_{z}$. Note
that case (a) assumes $\theta_{\xi}=\pi/2$. In case (b), $\left\vert
\Psi\left(  s\right)  \right\rangle =\cos\left(  \theta_{\xi}/2\right)
e^{-i\tan\left(  \theta_{\xi}/2\right)  s}\left\vert 0\right\rangle
+\sin\left(  \theta_{\xi}/2\right)  e^{i\cot\left(  \theta_{\xi}/2\right)
s}\left\vert 1\right\rangle $, $\left\vert T\left(  s\right)  \right\rangle
=-i\sin\left(  \theta_{\xi}/2\right)  e^{-i\tan\left(  \theta_{\xi}/2\right)
s}\left\vert 0\right\rangle +i\cos\left(  \theta_{\xi}/2\right)
e^{i\cot\left(  \theta_{\xi}/2\right)  s}\left\vert 1\right\rangle $, $\hat
{a}_{\left\vert \Psi\left(  s\right)  \right\rangle }=\sin\left(  \theta_{\xi
}\right)  \cos\left\{  \left[  \tan\left(  \theta_{\xi}/2\right)  +\cot\left(
\theta_{\xi}/2\right)  \right]  s\right\}  \hat{x}+\sin\left(  \theta_{\xi
}\right)  \sin\left\{  \left[  \tan\left(  \theta_{\xi}/2\right)  +\cot\left(
\theta_{\xi}/2\right)  \right]  s\right\}  \hat{y}+\cos\left(  \theta_{\xi
}\right)  \hat{z}$, and $\hat{a}_{\left\vert T\left(  s\right)  \right\rangle
}=-\hat{a}_{\left\vert \Psi\left(  s\right)  \right\rangle }$, with
$s\protect\overset{\text{def}}{=}\omega_{0}t$. Unlike case (a), in this case
the curvature coefficient is nonzero since $\kappa_{\mathrm{AC}}^{2}=4\left[
\cos^{2}(\theta_{\xi})/\sin^{2}\left(  \theta_{\xi}\right)  \right]  \neq0$
when $\theta_{\xi}\neq\pi/2$.}%
\label{fig3}%
\end{figure}

\subsubsection{Link between curvature and geodesic efficiency: single qubit}

We emphasize that the temporal evolutions in Eqs. (\ref{geostate}) and
(\ref{nogeo}) can be regarded as special cases of the evolution given by%

\begin{equation}
\left\vert \psi\left(  t\right)  \right\rangle =e^{-i\left(  h/\hslash\right)
t}\left[  \xi\left\vert 0\right\rangle +e^{i(2h/\hslash)t}\sqrt{1-\xi^{2}%
}\left\vert 1\right\rangle \right]  \text{,} \label{general}%
\end{equation}
with $\xi=\cos\left(  \theta_{\xi}/2\right)  \in\left[  0\text{,}1\right]  $
and $\theta_{\xi}\in\left[  0\text{, }\pi\right]  $ such that $\mathbf{a}%
=\left(  2\xi\sqrt{1-\xi^{2}}\text{, }0\text{, }2\xi^{2}-1\right)  $,
$\cos\left(  \theta_{\xi}\right)  =2\xi^{2}-1$, and $\sin\left(  \theta_{\xi
}\right)  =2\xi\sqrt{1-\xi^{2}}$. The vanishing and nonvanishing curvature
cases previously discussed correspond to $\xi=1/\sqrt{2}$ and $\xi
=\sqrt{2+\sqrt{2}}/2$, respectively. In this more general case, the curvature
coefficient $\kappa_{\mathrm{AC}}^{2}$ is given by%
\begin{equation}
\kappa_{\mathrm{AC}}^{2}\left(  \xi\right)  =\frac{\left(  1-2\xi^{2}\right)
^{2}}{\xi^{2}\left(  1-\xi^{2}\right)  }=4\frac{\cos^{2}\left(  \theta_{\xi
}\right)  }{\sin^{2}\left(  \theta_{\xi}\right)  }\text{.} \label{until}%
\end{equation}
We show in Fig. $3$ a graphical depiction of the circular path on the Bloch
sphere traced out by evolving the initial states $\left\vert \psi\left(
0\right)  \right\rangle \overset{\text{def}}{=}(1/\sqrt{2})\left\vert
0\right\rangle +(1/\sqrt{2})\left\vert 1\right\rangle $ and $\left\vert
\psi\left(  0\right)  \right\rangle \overset{\text{def}}{=}\cos\left(
\theta_{\xi}/2\right)  \left\vert 0\right\rangle +\sin(\theta_{\xi
}/2)\left\vert 1\right\rangle $ with $\theta_{\xi}\neq\pi/2$ under the
Hamiltonian \textrm{H}$\overset{\text{def}}{=}\hslash\omega_{0}\sigma_{z}$.
Motivated by the expression of $\kappa_{\mathrm{AC}}^{2}\left(  \xi\right)  $
in Eq. (\ref{until}), we present a discussion on the link between the concept
of geodesic curvature in classical differential geometry and our proposed
curvature coefficient for quantum evolutions in Appendix B. Clearly,
$\kappa_{\mathrm{AC}}^{2}\left(  \xi\right)  $ in Eq. (\ref{until}) assumes
its minimum value of zero at $\xi=1/\sqrt{2}\approx0.71$ (i.e., polar angle
$\theta_{\xi}=\pi/2\approx1.57$). We observe that the local behavior of
$\kappa_{\mathrm{AC}}^{2}\left(  \xi\right)  $ as a measure of the deviation
from geodesic motion is consistent with that exhibited by the so-called
geodesic efficiency as introduced in Refs. \cite{anandan90,carlopra20},%
\begin{equation}
\eta\overset{\text{def}}{=}1-\frac{\Delta s}{s}=\frac{2\cos^{-1}\left[
\left\vert \left\langle A|B\right\rangle \right\vert \right]  }{2\int_{t_{A}%
}^{t_{B}}\frac{\Delta E\left(  t^{\prime}\right)  }{\hslash}dt^{\prime}%
}\text{.} \label{efficiency}%
\end{equation}
In Eq. (\ref{efficiency}), $0\leq\eta\leq1$, $\Delta s\overset{\text{def}%
}{=}s-s_{0}$, $s_{0}$ denotes the distance along the shortest geodesic path
that connects the distinct initial $\left\vert A\right\rangle
\overset{\text{def}}{=}$ $\left\vert \psi\left(  t_{A}\right)  \right\rangle $
and final $\left\vert B\right\rangle \overset{\text{def}}{=}\left\vert
\psi\left(  t_{B}\right)  \right\rangle $ states on the projective Hilbert
space, $\Delta E\left(  t\right)  \overset{\text{def}}{=}\sqrt{\left\langle
\left(  \Delta\mathrm{H}\left(  t\right)  \right)  ^{2}\right\rangle }$ and
finally, $s$ is the distance along the dynamical trajectory traced out by the
state vector $\left\vert \psi\left(  t\right)  \right\rangle $ with $t_{A}\leq
t\leq t_{B}$. Indeed, from the evolution specified in Eq. (\ref{general}), the
efficiency $\eta$ in Eq. (\ref{efficiency}) reduces to%
\begin{equation}
\eta\left(  t\text{; }\xi\right)  =\frac{\arccos\left(  \sqrt{\cos^{2}\left(
\frac{m}{\hslash}t\right)  +\left(  2\xi^{2}-1\right)  ^{2}\sin^{2}\left(
\frac{m}{\hslash}t\right)  }\right)  }{2\xi\sqrt{1-\xi^{2}}\left(  \frac
{m}{\hslash}t\right)  }=\frac{\arccos\left(  \sqrt{\cos^{2}\left(  \frac
{m}{\hslash}t\right)  +\cos^{2}\left(  \theta_{\xi}\right)  \sin^{2}\left(
\frac{m}{\hslash}t\right)  }\right)  }{\sin\left(  \theta_{\xi}\right)
\left(  \frac{m}{\hslash}t\right)  }\text{,} \label{eff1}%
\end{equation}
assuming $t_{A}=0$ and $t_{B}=t$. In particular, taking $t_{B}=\hslash\pi/4m$
as in the previous two examples, the efficiency $\eta\left(  t\text{; }%
\xi\right)  $ in Eq. (\ref{eff1}) reduces to%
\begin{equation}
\eta\left(  \xi\right)  =\frac{2}{\pi}\frac{\arccos\left(  \sqrt{2\xi^{4}%
-2\xi^{2}+1}\right)  }{\xi\sqrt{1-\xi^{2}}}=\frac{4}{\pi}\frac{\arccos\left(
\sqrt{\frac{1+\cos^{2}\left(  \theta_{\xi}\right)  }{2}}\right)  }{\sin\left(
\theta_{\xi}\right)  }\text{.} \label{eff1b}%
\end{equation}
As expected, $\eta\left(  \xi\right)  \in\left[  0\text{, }1\right]  $ in Eq.
(\ref{eff1b}) assumes its maximum value of one at $\xi=1/\sqrt{2}$
$\approx0.71$ (i.e., polar angle $\theta_{\xi}=\pi/2\approx1.57$) where
$\kappa_{\mathrm{AC}}^{2}\left(  \xi\right)  $ in Eq. (\ref{until}) achieves
its minimum of zero. In conclusion, we reiterate that the local behavior (for
detecting geodesicity) of $\kappa_{\mathrm{AC}}^{2}\left(  \xi\right)  $ in
Eq. (\ref{until}) is consistent with that exhibited by $\eta(\xi)$ in Eq.
(\ref{eff1b}). In Fig. $4$, we present a visual comparison between the
geodesic efficiency and the curvature coefficient for the example discussed
here. Moreover, we also show a visualization of a quantum-mechanical
realization of the Pearson inequality (i.e., $\alpha_{4}\geq\alpha_{3}^{2}+1$)
in mathematical statistics between the kurtosis ($\alpha_{4}$) and the
skewness ($\alpha_{3}$) of a probability distribution function.
\begin{figure}[t]
\centering
\includegraphics[width=1\textwidth] {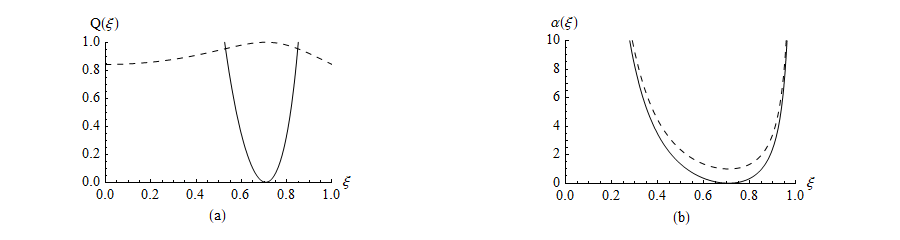}\caption{ (a) Plot of the geodesic
efficiency$\ \ \eta\left(  t;\xi\right)  \protect\overset{\text{def}%
}{=}\arccos\left(  \sqrt{\cos^{2}\left(  \omega_{0}t\right)  +\left(  2\xi
^{2}-1\right)  ^{2}\sin^{2}\left(  \omega_{0}t\right)  }\right)  /\left[
2\xi\sqrt{1-\xi^{2}}\left(  \omega_{0}t\right)  \right]  $ ($Q\left(
\xi\right)  \protect\overset{\text{def}}{=}\eta\left(  1;\xi\right)  $, black
dashed line) and the normalized curvature coefficient $\kappa_{\mathrm{AC}%
}^{2}\left(  \xi\right)  \protect\overset{\text{def}}{=}\left(  1-2\xi
^{2}\right)  ^{2}/\left[  \xi^{2}\left(  1-\xi^{2}\right)  \right]  =4\left[
\cos^{2}(\theta_{\xi})/\sin^{2}\left(  \theta_{\xi}\right)  \right]  $
($Q\left(  \xi\right)  \protect\overset{\text{def}}{=}\kappa_{\mathrm{AC}}%
^{2}\left(  \xi\right)  $, black solid line) as a function of the parameter
$\xi\in\left[  0\text{, }1\right]  $ with $\xi\protect\overset{\text{def}%
}{=}\cos(\theta_{\xi}/2)$ where $\theta_{\xi}$ is the polar angle such that
$0\leq\theta_{\xi}\leq\pi$. \ The quantum evolution is assumed to be described
by the time-independent Hamiltonian \textrm{H}$\protect\overset{\text{def}%
}{=}\hslash\omega_{0}\sigma_{z}$ that evolves the initial state $\left\vert
\psi\left(  \xi\text{, }\varphi\right)  \right\rangle
\protect\overset{\text{def}}{=}\xi\left\vert 0\right\rangle +e^{i\varphi}%
\sqrt{1-\xi^{2}}\left\vert 1\right\rangle $ with $\varphi=0$ (i.e.,
$xz$-plane). When plotting the efficiency, we set $\omega_{0}$ and $t$ equal
to one. (b) We exhibit a quantum-mechanical realization of the so-called
Pearson inequality (i.e., $\alpha_{4}\geq\alpha_{3}^{2}+1$) in mathematical
statistics between the kurtosis ($\alpha_{4}$) and the skewness ($\alpha_{3}$)
of a probability distribution function. We plot the kurtosis $\alpha
_{4}\protect\overset{\text{def}}{=}\left\langle \left(  \Delta\mathrm{H}%
\right)  ^{4}\right\rangle /\left\langle \left(  \Delta\mathrm{H}\right)
^{2}\right\rangle ^{2}=\left(  1-3\xi^{2}+3\xi^{4}\right)  /\left[  \xi
^{2}\left(  1-\xi^{2}\right)  \right]  $ ($\alpha\left(  \xi\right)
\protect\overset{\text{def}}{=}\alpha_{4}\left(  \xi\right)  $, black dashed
line) and the skewness (squared) $\alpha_{3}^{2}\protect\overset{\text{def}%
}{=}\left\langle \left(  \Delta\mathrm{H}\right)  ^{3}\right\rangle
^{2}/\left\langle \left(  \Delta\mathrm{H}\right)  ^{2}\right\rangle
^{3}=\left(  1-2\xi^{2}\right)  ^{2}/\left[  \xi^{2}\left(  1-\xi^{2}\right)
\right]  $ ($\alpha\left(  \xi\right)  \protect\overset{\text{def}}{=}%
\alpha_{3}^{2}\left(  \xi\right)  $, black solid line) as a function of the
parameter $\xi\in\left[  0\text{, }1\right]  $ with $\xi
\protect\overset{\text{def}}{=}\cos(\theta_{\xi}/2)$ where $\theta_{\xi}$ is
the polar angle such that $0\leq\theta_{\xi}\leq\pi$. \ The quantum evolution
in (b) is the same as in (a). The geodesic evolution occurs for $\xi
=1/\sqrt{2}$. From (a), we note that $\kappa_{\mathrm{AC}}^{2}\left(
1/\sqrt{2}\right)  =0$ and $\eta\left(  1;1/\sqrt{2}\right)  =1$. Finally, we
observe from (b) that $\alpha_{4}\left(  1/\sqrt{2}\right)  =1$, $\alpha
_{3}^{2}\left(  1/\sqrt{2}\right)  =0$, and the Pearson inequality saturates
becoming the identity $\alpha_{4}=\alpha_{3}^{2}+1$.}%
\label{fig4}%
\end{figure}

\subsubsection{Nonzero curvature and nonzero torsion: two qubits}

As previously mentioned, the torsion coefficient $\tau_{\mathrm{AC}}^{2}$
vanishes for any traceless time-independent single-qubit Hamiltonian of the
form \textrm{H}$=\mathbf{m}\cdot\vec{\sigma}$, regardless of the expectation
value of the Hamiltonian. In what follows, we present two examples of nonzero
torsion curves traced out by two-qubit quantum states evolving under two-qubit
time-independent Hamiltonians. Recall that a general bipartite Hamiltonian
\textrm{H }can be written as \cite{bennett02},%
\begin{equation}
\mathrm{H}\overset{\text{def}}{=}\mathrm{H}^{\left(  1\right)  }%
\otimes\mathrm{I}^{\left(  2\right)  }+\mathrm{I}^{\left(  1\right)  }%
\otimes\mathrm{H}^{\left(  2\right)  }+\sum_{i,j=1}^{3}M_{ij}\Sigma
_{i}^{\left(  1\right)  }\otimes\Sigma_{j}^{\left(  2\right)  }\text{,}
\label{g2qubit}%
\end{equation}
where $\left\{  \Sigma_{i}\right\}  _{1\leq i\leq3}$ is a basis for traceless
Hermitian operators and $\left\{  M_{ij}\right\}  _{1\leq i,j\leq3}$ are
coupling coefficients that correspond to the pairwise interaction terms
$\left\{  \Sigma_{i}^{\left(  1\right)  }\otimes\Sigma_{j}^{\left(  2\right)
}\right\}  _{1\leq i,j\leq3}$. From Eq. (\ref{g2qubit}), the most general
purely nonlocal two-qubit Hamiltonian $\mathrm{H}_{\mathrm{nonlocal}}$ (i.e.,
$\mathrm{H}_{\mathrm{nonlocal}}\neq\mathrm{H}^{\left(  1\right)  }%
\otimes\mathrm{I}^{\left(  2\right)  }+\mathrm{I}^{\left(  1\right)  }%
\otimes\mathrm{H}^{\left(  2\right)  }$) can be written as%
\begin{equation}
\mathrm{H}_{\mathrm{nonlocal}}\overset{\text{def}}{=}\sum_{i,j=1}^{3}%
m_{ij}\sigma_{i}^{\left(  1\right)  }\otimes\sigma_{j}^{\left(  2\right)
}\text{,} \label{nonlocal}%
\end{equation}
where $\left\{  \sigma_{i}\right\}  _{1\leq i\leq3}$ are the Pauli operators
with $\sigma_{1}\overset{\text{def}}{=}\sigma_{x}$, $\sigma_{2}%
\overset{\text{def}}{=}\sigma_{y}$, $\sigma_{3}\overset{\text{def}}{=}%
\sigma_{z}$. The terms $\mathrm{H}^{\left(  1\right)  }\otimes\mathrm{I}%
^{\left(  2\right)  }$ and $\mathrm{I}^{\left(  1\right)  }\otimes
\mathrm{H}^{\left(  2\right)  }$ in Eq. (\ref{g2qubit}) represent local
interaction Hamiltonians since they act non trivially only on one of the two
qubits in the two-qubit quantum state. Clearly, $\mathrm{I}^{\left(  1\right)
}$ and $\mathrm{I}^{\left(  2\right)  }$ denote the identity operators on the
first and second subsystems, respectively.

In our first example, we consider the curvature and the torsion of a quantum
curve traced out by the
initial separable state $\left\vert 00\right\rangle $ under an Hamiltonian
$\mathrm{H}$ expressed in terms of a superposition of nonlocal two-qubit
Hamiltonians,%
\begin{equation}
\mathrm{H}\overset{\text{def}}{\mathbb{=}}m_{1}\left(  \sigma_{x}^{\left(
1\right)  }\otimes\sigma_{x}^{\left(  2\right)  }\right)  +m_{2}\left(
\sigma_{z}^{\left(  1\right)  }\otimes\sigma_{z}^{\left(  2\right)  }\right)
+m_{3}\left(  \sigma_{x}^{\left(  1\right)  }\otimes\sigma_{z}^{\left(
2\right)  }\right)  +m_{4}\left(  \sigma_{z}^{\left(  1\right)  }\otimes
\sigma_{x}^{\left(  2\right)  }\right)  \text{.} \label{eye1}%
\end{equation}
A straightforward calculation yields distinct curvature and torsion
coefficients $\kappa_{\mathrm{AC}}^{2}=\kappa_{\mathrm{AC}}^{2}\left(
m_{1}\text{, }m_{2}\text{, }m_{3}\text{, }m_{4}\right)  $ and $\tau
_{\mathrm{AC}}^{2}=\tau_{\mathrm{AC}}^{2}\left(  m_{1}\text{, }m_{2}\text{,
}m_{3}\text{, }m_{4}\right)  $, respectively, given by%
\begin{equation}
\kappa_{\mathrm{AC}}^{2}=4\frac{\allowbreak\left(  m_{2}^{2}m_{3}^{2}%
+m_{2}^{2}m_{4}^{2}+m_{3}^{2}m_{4}^{2}\right)  }{\left(  m_{1}^{2}+m_{3}%
^{2}+m_{4}^{2}\right)  ^{2}}\text{, and }\tau_{\mathrm{AC}}^{2}=4\frac{\left(
m_{3}^{2}+m_{4}^{2}\right)  \left(  m_{1}m_{2}-m_{3}m_{4}\right)  ^{2}%
}{\left(  m_{1}^{2}+m_{3}^{2}+m_{4}^{2}\right)  ^{3}}\text{,} \label{nzc}%
\end{equation}
with $0\leq\tau_{\mathrm{AC}}^{2}\leq\kappa_{\mathrm{AC}}^{2}$ for any choice
of the real coupling parameters $m_{1}$, $m_{2}$, $m_{3}$, and $m_{4}$.
Interestingly, we remark that the quantum curve traced out by the maximally
entangled Bell state $\left\vert \Phi_{\mathrm{Bell}}^{+}\right\rangle
\overset{\text{def}}{=}\left(  \left\vert 00\right\rangle +\left\vert
11\right\rangle \right)  /\sqrt{2}$ under the Hamiltonian in Eq. (\ref{eye1})
is characterized by a nonvanishing curvature coefficient $\kappa_{\mathrm{AC}%
}^{2}=4\left[  \left(  m_{1}+m_{2}\right)  /\left(  m_{3}-m_{4}\right)
\right]  ^{2}$ and a vanishing torsion coefficient $\tau_{\mathrm{AC}}^{2}=0$.
Indeed, it can be checked that under the Hamiltonian in Eq. (\ref{eye1}), no
curve traced out by any of the remaining three maximally entangled Bell states
$\left\vert \Phi_{\mathrm{Bell}}^{-}\right\rangle \overset{\text{def}%
}{=}\left(  \left\vert 00\right\rangle -\left\vert 11\right\rangle \right)
/\sqrt{2}$, $\left\vert \Psi_{\mathrm{Bell}}^{+}\right\rangle
\overset{\text{def}}{=}\left(  \left\vert 01\right\rangle +\left\vert
10\right\rangle \right)  /\sqrt{2}$, and $\left\vert \Psi_{\mathrm{Bell}}%
^{-}\right\rangle \overset{\text{def}}{=}\left(  \left\vert 01\right\rangle
-\left\vert 10\right\rangle \right)  /\sqrt{2}$ exhibits nonzero torsion. For
this reason, one may wonder which might be a suitable two-qubit Hamiltonian to
observe some twisting of a quantum curve traced out by evolving a maximally
entangled Bell state. We address this question in the next illustrative example.

In our second example, we consider a quantum evolution governed by a two-qubit
Hamiltonian $\mathrm{H}$ expressed as a superposition of local two-qubit
Hamiltonians,%
\begin{equation}
\mathrm{H}\overset{\text{def}}{=}m_{1}\left(  \mathrm{I}^{\left(  1\right)
}\otimes\sigma_{x}^{\left(  2\right)  }\right)  +m_{2}\left(  \sigma
_{x}^{\left(  1\right)  }\otimes\mathrm{I}^{\left(  2\right)  }\right)
+m_{3}\left(  \mathrm{I}^{\left(  1\right)  }\otimes\sigma_{z}^{\left(
2\right)  }\right)  +m_{4}\left(  \sigma_{z}^{\left(  1\right)  }%
\otimes\mathrm{I}^{\left(  2\right)  }\right)  \text{.} \label{eye2}%
\end{equation}
We consider the curvature and the torsion of a quantum curve traced out by the
maximally entangled Bell state $\left\vert \Phi_{\mathrm{Bell}}^{+}%
\right\rangle \overset{\text{def}}{=}\left(  \left\vert 00\right\rangle
+\left\vert 11\right\rangle \right)  /\sqrt{2}$. A simple calculation leads to
identical curvature and torsion coefficients $\kappa_{\mathrm{AC}}^{2}%
=\kappa_{\mathrm{AC}}^{2}\left(  m_{1}\text{, }m_{2}\text{, }m_{3}\text{,
}m_{4}\right)  $ and $\tau_{\mathrm{AC}}^{2}=\tau_{\mathrm{AC}}^{2}\left(
m_{1}\text{, }m_{2}\text{, }m_{3}\text{, }m_{4}\right)  $, respectively, given
by%
\begin{equation}
\kappa_{\mathrm{AC}}^{2}=4\frac{\allowbreak\left(  m_{1}m_{4}-m_{2}%
m_{3}\right)  ^{2}}{\left[  \left(  m_{1}+m_{2}\right)  ^{2}+\left(
m_{3}+m_{4}\right)  ^{2}\right]  ^{2}\allowbreak}\text{, and }\tau
_{\mathrm{AC}}^{2}=4\frac{\allowbreak\left(  m_{1}m_{4}-m_{2}m_{3}\right)
^{2}}{\left[  \left(  m_{1}+m_{2}\right)  ^{2}+\left(  m_{3}+m_{4}\right)
^{2}\right]  ^{2}\allowbreak}\text{.} \label{street1}%
\end{equation}
The presence of identical expressions for $\kappa_{\mathrm{AC}}^{2}$ and
$\tau_{\mathrm{AC}}^{2}$ in Eq. (\ref{street1}) is a consequence of the fact
that the skewness coefficient vanishes in this case (i.e., $\alpha
_{3}=0\leftrightarrow\left\langle \left(  \Delta\mathrm{h}\right)
^{3}\right\rangle =0$). For completeness, we remark that the curvature
$\kappa_{\mathrm{AC}}^{2}=\kappa_{\mathrm{AC}}^{2}\left(  m_{1}\text{, }%
m_{2}\text{, }m_{3}\text{, }m_{4}\right)  $ and the torsion $\tau
_{\mathrm{AC}}^{2}=\tau_{\mathrm{AC}}^{2}\left(  m_{1}\text{, }m_{2}\text{,
}m_{3}\text{, }m_{4}\right)  $ coefficients of the quantum curve obtained by
traced out by the separable state $\left\vert 00\right\rangle $ under the
Hamiltonian in Eq. (\ref{eye2}) are distinct. They are given by%
\begin{equation}
\kappa_{\mathrm{AC}}^{2}=4\frac{\left(  m_{1}^{2}m_{2}^{2}+m_{1}^{2}m_{3}%
^{2}+m_{2}^{2}m_{4}^{2}\right)  }{\left(  m_{1}^{2}+m_{2}^{2}\right)  ^{2}%
}\text{, and }\tau_{\mathrm{AC}}^{2}=\allowbreak4\frac{m_{1}^{2}m_{2}%
^{2}\left[  m_{1}^{2}+m_{2}^{2}+\left(  m_{3}-m_{4}\right)  ^{2}\right]
}{\left(  m_{1}^{2}+m_{2}^{2}\right)  ^{3}}\text{,} \label{street2}%
\end{equation}
respectively, with $\kappa_{\mathrm{AC}}^{2}-\tau_{\mathrm{AC}}^{2}=4\left(
m_{3}m_{1}^{2}+m_{4}m_{2}^{2}\right)  ^{2}/\left(  m_{1}^{2}+m_{2}^{2}\right)
^{3}$ and $0\leq\tau_{\mathrm{AC}}^{2}\leq\kappa_{\mathrm{AC}}^{2}$. As
evident from Eqs. (\ref{street1}) and (\ref{street2}), the way the curvature
and torsion coefficients vary with respect to the tunable (Hamiltonian)
parameters depends on both the chosen source state and the selected driving
Hamiltonian. This observation paves the way to several intriguing exploratory
questions. For instance, in addition to being interested in (shortest length)
geodesic quantum evolutions as pointed out in the previous subsubsection, one
may be interested in driving a source state along a pre-selected (nongeodesic)
path \cite{uzdin12,campaioli19}. In this case, the coefficients $\kappa
_{\mathrm{AC}}^{2}$ and $\mathrm{\tau}_{\mathrm{AC}}^{2}$ can be of help in
providing useful insights into the behavior of the quantum evolution from a
geometric perspective and, in principle, can assist in singling out the proper
driving Hamiltonian for the chosen source state that is optimal for the task
at hand. Roughly speaking, by tuning the Hamiltonian parameters, one can bend
and twist the trajectory traced out by the source state so to avoid obstacles
along the path and reach the target state in an optimal manner. Moreover, one
could use $\kappa_{\mathrm{AC}}^{2}$ and $\mathrm{\tau}_{\mathrm{AC}}^{2}$ to
geometrically characterize the effects of different degrees of entanglement of
quantum states under different quantum Hamiltonian evolutions. In particular,
one could find out how difficult and/or easy is to bend and/or twist quantum
curves traced out by quantum states with different degrees of entanglement. In
Appendix C, we present an example that considers the curvature and torsion
coefficients for a quantum curve traced by evolving a three-qubit quantum
state, including the $\left\vert \mathrm{GHZ}\right\rangle $-state and the
$\left\vert \mathrm{W}\right\rangle $-state, under a three-qubit stationary
Hamiltonian belonging to the family of the quantum Heisenberg models. For the
time being, keeping in mind Feynman's attitude on the use of geometric methods
in the study of quantum evolutions \cite{dick57}, we leave a quantitative
discussion of these intriguing exploratory lines of investigations to future
scientific efforts.

\section{Final Remarks}

In this paper, we proposed a quantum version of the Frenet-Serret apparatus
for a quantum trajectory on the
generalized Bloch sphere traced out by a parallel-transported pure quantum
state $\left\vert \Psi\left(  s\right)  \right\rangle $
of arbitrary dimension parametrized in terms of the arc length $s$ that
evolves unitarily under a time-independent Hamiltonian specifying the
Schr\"{o}dinger equation (Eq. (\ref{t5})). Given this parallel-transported
unit state vector, we introduced in a proper sequential fashion a suitable
pair of two projector operators (i.e., \textrm{P}$^{\left(  \Psi\right)  }$
and \textrm{P}$^{\left(  T\right)  }$\textrm{P}$^{\left(  \Psi\right)  }$) to
construct an intrinsically geometric set of three orthonormal vectors
specifying a moving frame associated with the quantum curve $\gamma\left(
s\right)  :s\mapsto\left\vert \Psi\left(  s\right)  \right\rangle $ with
$s_{i}\leq s\leq s_{f}$. First, we introduce the unit tangent vector as the
covariant derivative of the parallel-transported unit state vector (Eq.
(\ref{t8})), $\left\vert T(s)\right\rangle \overset{\text{def}}{=}$%
\textrm{P}$^{\left(  \Psi\right)  }\left\vert \Psi^{\prime}\left(  s\right)
\right\rangle =\mathrm{D}\left\vert \Psi\left(  s\right)  \right\rangle $. The
covariant derivative operator $\mathrm{D}\overset{\text{def}}{=}$%
\textrm{P}$^{\left(  \Psi\right)  }(d/ds)$ is defined as the composition of
\textrm{P}$^{\left(  \Psi\right)  }$ and $d/ds$. The quantity \textrm{P}%
$^{\left(  \Psi\right)  }$ is a projection operator onto a state orthogonal to
the parallel-transported state vector $\left\vert \Psi\left(  s\right)
\right\rangle $, while $d/ds$ denotes the ordinary differential operator with
respect to the arc length $s$. Then, the parallel-transported unit state
vector $\left\vert \Psi\left(  s\right)  \right\rangle $ and the unit tangent
vector $\left\vert T(s)\right\rangle $ span a two-dimensional space that can
be regarded as the quantum analogue of the instantaneous osculating plane in
the Frenet-Serret apparatus. Finally, the curvature $\kappa_{\mathrm{AC}}%
^{2}\left(  s\right)  $ (Eqs. (\ref{accurve}) and (\ref{ac12})) of a quantum
curve is defined as the squared magnitude $\left\Vert \mathrm{D}\left\vert
T(s)\right\rangle \right\Vert ^{2}$ of the covariant derivative of the unit
tangent vector $\left\vert T(s)\right\rangle $ and measures the bending of a
quantum curve. Second, to introduce the concept of torsion, we construct a
non-normalized vector $\left\vert \tilde{N}\left(  s\right)  \right\rangle $
(Eq. (\ref{normale})) that is orthogonal to the instantaneous quantum version
of the osculating plane. This third vector is defined as $\mathrm{P}%
^{(T)}\mathrm{P}^{(\Psi)}\left\vert T^{\prime}(s)\right\rangle =\mathrm{P}%
^{(T)}\mathrm{D}\left\vert T(s)\right\rangle $, the projection of the
derivative with respect to the arc length of the unit tangent vector onto a
state that is orthogonal to the plane spanned by the parallel-transported unit
state $\left\vert \Psi\left(  s\right)  \right\rangle $ and the unit tangent
vector $\left\vert T(s)\right\rangle $. Finally, the torsion $\tau
_{\mathrm{AC}}^{2}$ (Eqs. (\ref{torsionac1}) and (\ref{torsionac2})) of a
quantum curve is defined as the squared magnitude $\left\Vert \mathrm{P}%
^{(T)}\mathrm{D}\left\vert T(s)\right\rangle \right\Vert ^{2}$ of the
projection onto a state orthogonal to the unit tangent vector of the covariant
derivative of the unit tangent vector and measures the twisting of a quantum
curve. In conclusion, the parallel-transported unit state vector $\left\vert
\Psi\left(  s\right)  \right\rangle $, the unit tangent vector $\left\vert
T(s)\right\rangle $, and the normalized version of the projection onto a state
orthogonal to the unit tangent vector of the covariant derivative of the unit
tangent vector, $\left\vert N\left(  s\right)  \right\rangle
\overset{\text{def}}{=}\left\vert \tilde{N}\left(  s\right)  \right\rangle
/\left\Vert \left\vert \tilde{N}\left(  s\right)  \right\rangle \right\Vert $,
form a quantum version of the Frenet-Serret frame $\left\{  \hat{T}\text{,
}\hat{N}\text{, }\hat{B}\right\}  $. When adding our proposed concepts of
curvature $\kappa_{\mathrm{AC}}^{2}$ and torsion $\tau_{\mathrm{AC}}^{2}$ to
the three orthonormal vectors $\left\{  \left\vert \Psi\left(  s\right)
\right\rangle \text{, }\left\vert T(s)\right\rangle \text{, }\left\vert
N\left(  s\right)  \right\rangle \right\}  $, we have the set $\left\{
\left\vert \Psi\left(  s\right)  \right\rangle \text{, }\left\vert
T(s)\right\rangle \text{, }\left\vert N\left(  s\right)  \right\rangle \text{,
}\kappa_{\mathrm{AC}}^{2}\left(  s\right)  \text{, }\tau_{\mathrm{AC}}%
^{2}\left(  s\right)  \right\}  $, a quantum version of the Frenet-Serret
apparatus $\left\{  \hat{T}\text{, }\hat{N}\text{, }\hat{B}\text{, }%
\kappa_{\mathrm{FS}}\left(  s\right)  \text{, }\tau_{\mathrm{FS}}\left(
s\right)  \right\}  $. Recall that $\kappa_{\mathrm{FS}}^{2}\left(  s\right)
=\left\Vert \hat{T}^{\prime}\left(  s\right)  \right\Vert ^{2}$,
$\tau_{\mathrm{FS}}^{2}\left(  s\right)  =\left\Vert \hat{B}^{\prime}\left(
s\right)  \right\Vert ^{2}$ are formally defined in Eqs. (\ref{FScurve}) and
(\ref{FStorsion}), respectively. We emphasize that unlike the
Frenet-Serret\ apparatus, we do not have in our theoretical proposal an exact
quantum version of the Frenet-Serret equations (Eq. (\ref{FSeq})) specified by
a closed set of dynamical equations for the FS frame $\left\{  \hat{T}\text{,
}\hat{N}\text{, }\hat{B}\right\}  $. However, we do have a clear
correspondence between our pair $\left(  \kappa_{\mathrm{AC}}^{2}%
\overset{\text{def}}{=}\left\Vert \mathrm{D}\left\vert T(s)\right\rangle
\right\Vert ^{2}=\left\Vert \mathrm{D}^{2}\left\vert \Psi\left(  s\right)
\right\rangle \right\Vert ^{2}\text{, }\tau_{\mathrm{AC}}^{2}%
\overset{\text{def}}{=}\left\Vert \mathrm{P}^{(T)}\mathrm{D}\left\vert
T(s)\right\rangle \right\Vert ^{2}=\left\Vert \mathrm{P}^{(T)}\mathrm{D}%
^{2}\left\vert \Psi\left(  s\right)  \right\rangle \right\Vert ^{2}\right)  $
and the Frenet-Serret pair $\left(  \kappa_{\mathrm{FS}}^{2}=\left\Vert
\hat{T}^{\prime}\left(  s\right)  \right\Vert ^{2}\text{, }\tau_{\mathrm{FS}%
}^{2}=\left\Vert \hat{B}^{\prime}\left(  s\right)  \right\Vert ^{2}\right)  $.
Remarkably, when focusing on time-independent Hamiltonian quantum evolutions,
our curvature coefficient $\kappa_{\mathrm{AC}}^{2}$ and our torsion
coefficient $\tau_{\mathrm{AC}}^{2}$ correspond to the dimensionless curvature
and torsion coefficients $\bar{\kappa}_{\mathrm{LT}}\overset{\text{def}%
}{=}\kappa_{\mathrm{LT}}/\left\langle (\Delta\mathrm{H})^{2}\right\rangle
^{2}$ (with $\kappa_{\mathrm{LT}}$ in\ Eq. (\ref{labacurva})) and $\bar{\tau
}_{\mathrm{LT}}\overset{\text{def}}{=}\tau_{\mathrm{LT}}/\left\langle
(\Delta\mathrm{H})^{2}\right\rangle ^{2}$ (with $\tau_{\mathrm{LT}}$
in\ Eq.~(\ref{labatorsion})), respectively, proposed by Laba and Tkachuk in
Ref. \cite{laba17}.

\bigskip

Keeping in mind the limitations of our proposed approach as discussed in part
of Section V, what is the relevance of our investigation?

\begin{enumerate}
\item[{[i]}] It provides a useful set of tools (i.e., quantum frames (Appendix
A) along with curvature and torsion coefficients) to characterize quantum
evolutions in terms of geometrically intuitive concepts such as bending
(related to curvature, Eqs. (\ref{accurve}) and (\ref{ac12})) and twisting
(related to torsion, Eqs. (\ref{torsionac1}) and (\ref{torsionac2})).

\item[{[ii]}] It leads (Eqs. (\ref{rose1}) and (\ref{rose2})) to relevant
physical insights into the underlying statistical structure of quantum theory
via concepts like kurtosis, skewness, and Pearson inequality. In particular,
geodesic quantum motion specified by a stationary Hamiltonian on the Bloch
sphere occurs, from a statistical standpoint, with minimal sharpness (i.e.,
minimal kurtosis) together with maximal symmetry (i.e., zero skewness). For
illustrative details, we refer to Fig. $4$.

\item[{[iii]}] It provides an alternative way of quantifying geodesic motion
in projective Hilbert space in the framework of geometric quantum mechanics.
For instance, curvature (Eq. (\ref{ac12})) as well as geodesic efficiency (Eq.
(\ref{efficiency})) can be equally used to detect geodesic motion (Eqs.
(\ref{until}) and (\ref{eff1b})). For a discussion on the link between the
concept of geodesic curvature in classical differential geometry and our
proposed curvature coefficient for quantum evolutions, we refer to Appendix B.

\item[{[iv]}] It can provide a way to help to control and manipulate
higher-dimensional quantum spin systems by analyzing how the bending and the
twisting caused via the application of a given driving Hamiltonian specified
by experimentally tunable driving parameters. This can be accomplished by
examining the changes in the curvature and torsion depending on the specific
degree of entanglement of the quantum state that traces out the curve in
projective Hilbert space (Eqs. (\ref{nzc}), (\ref{street1}), and
(\ref{street2})).

\item[{[v]}] It can represent a useful platform to study the effects of
entanglement on the behavior of curvature and torsion coefficients of curves
traced out by quantum states with different degrees of entanglement. For
example, from our exploratory investigations (for instance, Eqs. (\ref{nzc}),
(\ref{street1}), (\ref{street2}), and Appendix C), it seems to be generally
false that it is harder to bend and/or twist curves traced out by highly
entangled quantum states as one may be inclined to believe. Indeed, it seems
to be the case that the degree of bending and twisting of a quantum curve
depends on the specific matching of the pair (source state being driven,
driving Hamiltonian).

\item[{[vi]}] Unlike the existing approaches \cite{laba17,laba22}, our
theoretical construct allows to extend the notions of curvature and torsion
coefficients to nonstationary quantum Hamiltonian evolutions in a relatively
straightforward manner \cite{alsing2}.
\end{enumerate}

\bigskip

In conclusion, regardless of its current limitations, we hope our work will
stimulate other researchers and open the way toward further explicit
investigations on the interplay between geometry and quantum mechanics. For
the time being, we leave a more in-depth quantitative discussion on these
potential geometric extensions of our analytical findings, including
generalizations to mixed state geometry and time-dependent quantum evolutions
of higher-dimensional physical systems, to future scientific endeavors.

\begin{acknowledgments}
P.M.A. acknowledges support from the Air Force Office of Scientific Research
(AFOSR). C.C. is grateful to the United States Air Force Research Laboratory
(AFRL) Summer Faculty Fellowship Program for providing support for this work.
Any opinions, findings and conclusions or recommendations expressed in this
material are those of the author(s) and do not necessarily reflect the views
of the Air Force Research Laboratory (AFRL).
\end{acknowledgments}

\bigskip\pagebreak

\appendix

\section{Framing a quantum curve}

In this Appendix, following our suggestion in Section V, we present an
illustrative example on how to frame a quantum curve. Specifically, we wish to
construct the frame for the quantum curve traced by evolving the two-qubit
quantum state $\left\vert \psi\left(  0\right)  \right\rangle =\left\vert
00\right\rangle $ under the stationary Hamiltonian \textrm{H}%
$\overset{\text{def}}{=}\sigma_{x}^{\left(  1\right)  }\otimes\sigma
_{z}^{\left(  2\right)  }+\sigma_{z}^{\left(  1\right)  }\otimes\sigma
_{x}^{\left(  2\right)  }$. Using the statistical approach based upon the
calculation of expectation values in Eqs. (\ref{accurve1}) and (\ref{io2}), it
is straightforward to verify that $\kappa_{\mathrm{AC}}^{2}=\tau_{\mathrm{AC}%
}^{2}=1$ in this case. Since the Hilbert space $\mathcal{H}_{2}^{2}$ of
two-qubit quantum states is four-dimensional, we want to get a frame specified
by a set of orthonormal vectors given by $\left\{  \left\vert \Psi\left(
s\right)  \right\rangle \text{, }\left\vert T\left(  s\right)  \right\rangle
\text{, }\left\vert N\left(  s\right)  \right\rangle \text{, }\left\vert
V\left(  s\right)  \right\rangle \right\}  $. We begin by observing that the
unitary time evolution operator $\mathcal{U}\left(  t\right)
\overset{\text{def}}{=}e^{-i\mathrm{H}t}$ that corresponds to \textrm{H}%
$\overset{\text{def}}{=}\sigma_{x}^{\left(  1\right)  }\otimes\sigma
_{z}^{\left(  2\right)  }+\sigma_{z}^{\left(  1\right)  }\otimes\sigma
_{x}^{\left(  2\right)  }$ is given by,
\begin{equation}
\mathcal{U}\left(  t\right)  =\left(
\begin{array}
[c]{cccc}%
\frac{1}{4e^{2it}}\left(  e^{2it}+1\right)  ^{2} & -\frac{1}{4e^{2it}}\left(
e^{4it}-1\right)  & -\frac{1}{4e^{2it}}\left(  e^{4it}-1\right)  & -\frac
{1}{4e^{2it}}\left(  e^{2it}-1\right)  ^{2}\\
-\frac{1}{4e^{2it}}\left(  e^{4it}-1\right)  & \frac{1}{4e^{2it}}\left(
e^{2it}+1\right)  ^{2} & \frac{1}{4e^{2it}}\left(  e^{2it}-1\right)  ^{2} &
\frac{1}{4e^{2it}}\left(  e^{4it}-1\right) \\
-\frac{1}{4e^{2it}}\left(  e^{4it}-1\right)  & \frac{1}{4e^{2it}}\left(
e^{2it}-1\right)  ^{2} & \frac{1}{4e^{2it}}\left(  e^{2it}+1\right)  ^{2} &
\frac{1}{4e^{2it}}\left(  e^{4it}-1\right) \\
-\frac{1}{4e^{2it}}\left(  e^{2it}-1\right)  ^{2} & \frac{1}{4e^{2it}}\left(
e^{4it}-1\right)  & \frac{1}{4e^{2it}}\left(  e^{4it}-1\right)  & \frac
{1}{4e^{2it}}\left(  e^{2it}+1\right)  ^{2}%
\end{array}
\right)  \text{,} \label{uni}%
\end{equation}
where we have set $\hslash$ equal to one. From Eq. (\ref{uni}), the evolved
state $\left\vert \psi\left(  t\right)  \right\rangle =\mathcal{U}\left(
t\right)  \left\vert \psi\left(  0\right)  \right\rangle $ becomes%
\begin{equation}
\left\vert \psi\left(  t\right)  \right\rangle =\cos^{2}\left(  t\right)
\left\vert 00\right\rangle -\frac{i}{2}\sin\left(  2t\right)  \left\vert
01\right\rangle -\frac{i}{2}\sin\left(  2t\right)  \left\vert 10\right\rangle
+\sin^{2}\left(  t\right)  \left\vert 11\right\rangle \text{.} \label{state}%
\end{equation}
We note that since $\left\langle \mathrm{H}\right\rangle =0$, we have
$\left\vert \Psi\left(  t\right)  \right\rangle =\left\vert \psi\left(
t\right)  \right\rangle $. Moreover, since $\left\langle (\Delta
\mathrm{H})^{2}\right\rangle =2$, we have $s=\sqrt{2}t$. Therefore,
$t=s/\sqrt{2}$ and $\left\vert \Psi\left(  s\right)  \right\rangle $ becomes%
\begin{equation}
\left\vert \Psi\left(  s\right)  \right\rangle =\cos^{2}\left(  \frac{s}%
{\sqrt{2}}\right)  \left\vert 00\right\rangle -\frac{i}{2}\sin\left(  \sqrt
{2}s\right)  \left\vert 01\right\rangle -\frac{i}{2}\sin\left(  \sqrt
{2}s\right)  \left\vert 10\right\rangle +\sin^{2}(\left(  \frac{s}{\sqrt{2}%
}\right)  \left\vert 11\right\rangle \text{.} \label{state2}%
\end{equation}
Using Eq. (\ref{state2}), the expressions for $\left\vert T\left(  s\right)
\right\rangle $ and $\left\vert T^{\prime}\left(  s\right)  \right\rangle $
are given by
\begin{equation}
\left\vert T\left(  s\right)  \right\rangle =-\frac{1}{\sqrt{2}}\sin\left(
\sqrt{2}s\right)  \left\vert 00\right\rangle -\frac{i}{\sqrt{2}}\cos\left(
\sqrt{2}s\right)  \left\vert 01\right\rangle -\frac{i}{\sqrt{2}}\cos\left(
\sqrt{2}s\right)  \left\vert 10\right\rangle +\frac{1}{\sqrt{2}}\sin\left(
\sqrt{2}s\right)  \left\vert 11\right\rangle \text{,} \label{state3}%
\end{equation}
and%
\begin{equation}
\left\vert T^{\prime}\left(  s\right)  \right\rangle =-\cos(\sqrt
{2}s)\left\vert 00\right\rangle +i\sin(\sqrt{2}s)\left\vert 01\right\rangle
+i\sin(\sqrt{2}s)\left\vert 10\right\rangle +\cos\left(  \sqrt{2}s\right)
\left\vert 11\right\rangle \text{,}%
\end{equation}
respectively. For completeness, we remark that $\left\vert \Psi\left(
t\right)  \right\rangle $ and $\left\vert T\left(  s\right)  \right\rangle $
are normalized to one and they are orthogonal. We need to find now $\left\vert
N\left(  s\right)  \right\rangle \overset{\text{def}}{=}\mathrm{P}^{\left(
T\right)  }\mathrm{P}^{\left(  \Psi\right)  }\left\vert T^{\prime}\left(
s\right)  \right\rangle /\left\Vert \mathrm{P}^{\left(  T\right)  }%
\mathrm{P}^{\left(  \Psi\right)  }\left\vert T^{\prime}\left(  s\right)
\right\rangle \right\Vert $. Therefore, we need to calculate the projectors
$\mathrm{P}^{\left(  \Psi\right)  }\overset{\text{def}}{=}\mathrm{I}%
_{\mathcal{H}_{2}^{2}}-\left\vert \Psi\right\rangle \left\langle
\Psi\right\vert $ and $\mathrm{P}^{\left(  T\right)  }\overset{\text{def}%
}{=}\mathrm{I}_{\mathcal{H}_{2}^{2}}-\left\vert T\right\rangle \left\langle
T\right\vert $. Using Eqs. (\ref{state2}) and (\ref{state3}), we get that
$\mathrm{P}^{\left(  \Psi\right)  }$ and $\mathrm{P}^{\left(  T\right)  }$ are
given by
\begin{equation}
\mathrm{P}^{\left(  \Psi\right)  }=\left(
\begin{array}
[c]{cccc}%
1-\cos^{4}\left(  \frac{s}{\sqrt{2}}\right)  & -\frac{i}{2}\sin\left(
\sqrt{2}s\right)  \cos^{2}\left(  \frac{s}{\sqrt{2}}\right)  & -\frac{i}%
{2}\sin\left(  \sqrt{2}s\right)  \cos^{2}\left(  \frac{s}{\sqrt{2}}\right)  &
-\sin^{2}\left(  \frac{s}{\sqrt{2}}\right)  \cos^{2}\left(  \frac{s}{\sqrt{2}%
}\right) \\
\frac{i}{2}\sin\left(  \sqrt{2}s\right)  \cos^{2}\left(  \frac{s}{\sqrt{2}%
}\right)  & 1-\frac{1}{4}\sin^{2}\left(  \sqrt{2}s\right)  & -\frac{1}{4}%
\sin^{2}\left(  \sqrt{2}s\right)  & \frac{i}{2}\sin\left(  \sqrt{2}s\right)
\sin^{2}\left(  \frac{s}{\sqrt{2}}\right) \\
\frac{i}{2}\sin\left(  \sqrt{2}s\right)  \cos^{2}\left(  \frac{s}{\sqrt{2}%
}\right)  & -\frac{1}{4}\sin^{2}\left(  \sqrt{2}s\right)  & 1-\frac{1}{4}%
\sin^{2}\left(  \sqrt{2}s\right)  & \frac{i}{2}\sin\left(  \sqrt{2}s\right)
\sin^{2}\left(  \frac{s}{\sqrt{2}}\right) \\
-\sin^{2}\left(  \frac{s}{\sqrt{2}}\right)  \cos^{2}\left(  \frac{s}{\sqrt{2}%
}\right)  & -\frac{i}{2}\sin\left(  \sqrt{2}s\right)  \sin^{2}(\left(
\frac{s}{\sqrt{2}}\right)  & -\frac{i}{2}\sin\left(  \sqrt{2}s\right)
\sin^{2}(\left(  \frac{s}{\sqrt{2}}\right)  & 1-\sin^{4}\left(  \frac{s}%
{\sqrt{2}}\right)
\end{array}
\right)  \text{,}%
\end{equation}
and%
\begin{equation}
\mathrm{P}^{\left(  T\right)  }=\left(
\begin{array}
[c]{cccc}%
1-\frac{1}{2}\sin^{2}\left(  \sqrt{2}s\right)  & \frac{i}{2}\sin\left(
\sqrt{2}s\right)  \cos\left(  \sqrt{2}s\right)  & \frac{i}{2}\sin\left(
\sqrt{2}s\right)  \cos\left(  \sqrt{2}s\right)  & \frac{1}{2}\sin^{2}\left(
\sqrt{2}s\right) \\
-\frac{i}{2}\sin\left(  \sqrt{2}s\right)  \cos\left(  \sqrt{2}s\right)  &
1-\frac{1}{2}\cos^{2}\left(  \sqrt{2}s\right)  & -\frac{1}{2}\cos^{2}\left(
\sqrt{2}s\right)  & \frac{i}{2}\sin\left(  \sqrt{2}s\right)  \cos\left(
\sqrt{2}s\right) \\
-\frac{i}{2}\sin\left(  \sqrt{2}s\right)  \cos\left(  \sqrt{2}s\right)  &
-\frac{1}{2}\cos^{2}\left(  \sqrt{2}s\right)  & 1-\frac{1}{2}\cos^{2}\left(
\sqrt{2}s\right)  & \frac{i}{2}\sin\left(  \sqrt{2}s\right)  \cos\left(
\sqrt{2}s\right) \\
\frac{1}{2}\sin^{2}\left(  \sqrt{2}s\right)  & -\frac{i}{2}\sin\left(
\sqrt{2}s\right)  \cos\left(  \sqrt{2}s\right)  & -\frac{i}{2}\sin\left(
\sqrt{2}s\right)  \cos\left(  \sqrt{2}s\right)  & 1-\frac{1}{2}\sin^{2}\left(
\sqrt{2}s\right)
\end{array}
\right)  \text{,}%
\end{equation}
respectively. As a consistency check, we verified that $\mathrm{P}^{\left(
\Psi\right)  }$ and $\mathrm{P}^{\left(  T\right)  }$ are proper orthogonal
projectors. Additionally, to evaluate $\kappa_{\mathrm{AC}}^{2}$ by using the
definition itself of curvature coefficient (i.e., $\kappa_{\mathrm{AC}}%
^{2}=\left\Vert \mathrm{P}^{\left(  \Psi\right)  }\left\vert T^{\prime}\left(
s\right)  \right\rangle \right\Vert ^{2}$), we note that%
\begin{equation}
\mathrm{P}^{\left(  \Psi\right)  }\left\vert T^{\prime}\left(  s\right)
\right\rangle =\left(
\begin{array}
[c]{c}%
\frac{1}{2}-\frac{1}{2}\cos\left(  \sqrt{2}s\right) \\
\frac{1}{2}i\sin\left(  \sqrt{2}s\right) \\
\frac{1}{2}i\sin\left(  \sqrt{2}s\right) \\
\frac{1}{2}\cos\left(  \sqrt{2}s\right)  +\frac{1}{2}%
\end{array}
\right)  \text{,}%
\end{equation}
and, thus, $\kappa_{\mathrm{AC}}^{2}=1$. In addition, to evaluate
$\tau_{\mathrm{AC}}^{2}$ by using the definition itself of torsion coefficient
(i.e., $\tau_{\mathrm{AC}}^{2}=\left\Vert \mathrm{P}^{\left(  T\right)
}\mathrm{P}^{\left(  \Psi\right)  }\left\vert T^{\prime}\left(  s\right)
\right\rangle \right\Vert ^{2}$), we note that%
\begin{equation}
\mathrm{P}^{\left(  T\right)  }\mathrm{P}^{\left(  \Psi\right)  }\left\vert
T^{\prime}\left(  s\right)  \right\rangle =\left(
\begin{array}
[c]{c}%
\frac{1}{2}-\frac{1}{2}\cos\left(  \sqrt{2}s\right) \\
\frac{1}{2}i\sin\left(  \sqrt{2}s\right) \\
\frac{1}{2}i\sin\left(  \sqrt{2}s\right) \\
\frac{1}{2}\cos\left(  \sqrt{2}s\right)  +\frac{1}{2}%
\end{array}
\right)  \text{,}%
\end{equation}
$\allowbreak$that is, $\tau_{\mathrm{AC}}^{2}=1$. Finally, the vector
$\left\vert N\left(  s\right)  \right\rangle $ reduces to%
\begin{equation}
\left\vert N\left(  s\right)  \right\rangle \overset{\text{def}}{=}%
\frac{\mathrm{P}^{\left(  T\right)  }\mathrm{P}^{\left(  \Psi\right)
}\left\vert T^{\prime}\left(  s\right)  \right\rangle }{\left\Vert
\mathrm{P}^{\left(  T\right)  }\mathrm{P}^{\left(  \Psi\right)  }\left\vert
T^{\prime}\left(  s\right)  \right\rangle \right\Vert }=\left(
\begin{array}
[c]{c}%
\frac{1}{2}-\frac{1}{2}\cos\left(  \sqrt{2}s\right) \\
\frac{1}{2}i\sin\left(  \sqrt{2}s\right) \\
\frac{1}{2}i\sin\left(  \sqrt{2}s\right) \\
\frac{1}{2}\cos\left(  \sqrt{2}s\right)  +\frac{1}{2}%
\end{array}
\right)  \text{,}%
\end{equation}
that is%
\begin{equation}
\left\vert N\left(  s\right)  \right\rangle =\left[  \frac{1}{2}-\frac{1}%
{2}\cos\left(  \sqrt{2}s\right)  \right]  \left\vert 00\right\rangle +\frac
{i}{2}\sin\left(  \sqrt{2}s\right)  \left\vert 01\right\rangle +\frac{i}%
{2}\sin\left(  \sqrt{2}s\right)  \left\vert 10\right\rangle +\left[  \frac
{1}{2}\cos\left(  \sqrt{2}s\right)  +\frac{1}{2}\right]  \left\vert
11\right\rangle \text{.} \label{state4}%
\end{equation}
We point out that we checked that $\left\vert N\left(  s\right)  \right\rangle
$ is normalized to one and orthogonal to both $\left\vert \Psi\left(
s\right)  \right\rangle $ and $\left\vert T\left(  s\right)  \right\rangle $.
To find an orthonormal basis of $\mathcal{H}_{2}^{2}$, we note that (by
educated guess) $\left\{  \left\vert \Psi\left(  s\right)  \right\rangle
\text{, }\left\vert T\left(  s\right)  \right\rangle \text{, }\left\vert
N\left(  s\right)  \right\rangle \text{, }\left\vert 01\right\rangle \right\}
$ is a set of linearly independent vectors. Then, applying the Gram-Schmidt
procedure, we find that an orthonormal basis of $\mathcal{H}_{2}^{2}$ is given
by $\left\{  \left\vert \Psi\left(  s\right)  \right\rangle \text{,
}\left\vert T\left(  s\right)  \right\rangle \text{, }\left\vert N\left(
s\right)  \right\rangle \text{, }\left\vert V\left(  s\right)  \right\rangle
\right\}  $ with $\left\vert V\left(  s\right)  \right\rangle $ given by%
\begin{equation}
\left\vert V\left(  s\right)  \right\rangle \overset{\text{def}}{=}%
\frac{\left\vert 01\right\rangle -\left\langle \Psi\left(  s\right)
\left\vert 01\right.  \right\rangle \left\vert \Psi\left(  s\right)
\right\rangle +\left\langle T\left(  s\right)  \left\vert 01\right.
\right\rangle \left\vert T\left(  s\right)  \right\rangle +\left\langle
N\left(  s\right)  \left\vert 01\right.  \right\rangle \left\vert N\left(
s\right)  \right\rangle }{\left\Vert \left\vert 01\right\rangle -\left\langle
\Psi\left(  s\right)  \left\vert 01\right.  \right\rangle \left\vert
\Psi\left(  s\right)  \right\rangle +\left\langle T\left(  s\right)
\left\vert 01\right.  \right\rangle \left\vert T\left(  s\right)
\right\rangle +\left\langle N\left(  s\right)  \left\vert 01\right.
\right\rangle \left\vert N\left(  s\right)  \right\rangle \right\Vert }%
=\frac{\left\vert 01\right\rangle -\left\vert 10\right\rangle }{\sqrt{2}%
}\text{.}%
\end{equation}
For completeness, we point out that we checked explicitly that $\left\{
\left\vert \Psi\left(  s\right)  \right\rangle \text{, }\left\vert T\left(
s\right)  \right\rangle \text{, }\left\vert N\left(  s\right)  \right\rangle
\text{, }\left\vert V\left(  s\right)  \right\rangle \right\}  $ is a set of
four orthonormal vectors. Finally, restricting our attention to the
three-dimensional subspace of $\mathcal{H}_{2}^{2}$ spanned by the orthonormal
quantum frame $\left\{  \left\vert \Psi\left(  s\right)  \right\rangle
,\left\vert T\left(  s\right)  \right\rangle \text{, }\left\vert N\left(
s\right)  \right\rangle \right\}  $ with $\left\vert \Psi\left(  s\right)
\right\rangle $ in Eq. (\ref{state2}), $\left\vert T\left(  s\right)
\right\rangle $ in Eq. (\ref{state3}), and $\left\vert N\left(  s\right)
\right\rangle $ in Eq. (\ref{state4}), we have that the coefficient matrix
$\mathrm{M}_{\partial_{s}\left(  \mathrm{Frame}\right)  \rightarrow
\mathrm{Frame}}^{\left(  \mathrm{FS}\right)  }$ that expresses the derivatives
of the frame with respect to the curve parameter $s$ in terms of the frame
itself is given by%
\begin{equation}
\left(
\begin{array}
[c]{c}%
\partial_{s}\left\vert \Psi\left(  s\right)  \right\rangle \\
\partial_{s}\left\vert T\left(  s\right)  \right\rangle \\
\partial_{s}\left\vert N\left(  s\right)  \right\rangle
\end{array}
\right)  =\left(
\begin{array}
[c]{ccc}%
0 & 1 & 0\\
-1 & 0 & 1\\
0 & -1 & 0
\end{array}
\right)  \left(
\begin{array}
[c]{c}%
\left\vert \Psi\left(  s\right)  \right\rangle \\
\left\vert T\left(  s\right)  \right\rangle \\
\left\vert N\left(  s\right)  \right\rangle
\end{array}
\right)  \text{.} \label{YO}%
\end{equation}
Interestingly, we note that $\mathrm{M}_{\partial_{s}\left(  \mathrm{Frame}%
\right)  \rightarrow\mathrm{Frame}}^{\left(  \mathrm{FS}\right)  }$ in Eq.
(\ref{YO}) is real and exhibits the skew-symmetric property, despite the fact
that curvature and torsion coefficients are nonzero in this case. Indeed, the
generally skew-Hermitian matrix $\mathrm{M}_{\partial_{s}\left(
\mathrm{Frame}\right)  \rightarrow\mathrm{Frame}}^{\left(  \mathrm{FS}\right)
}$ in Eq. (\ref{ACframe}) becomes the skew-symmetric one in Eq. (\ref{YO}) in
this scenario because $\kappa_{\mathrm{AC}}^{2}=\tau_{\mathrm{AC}}^{2}=1$,
$\phi_{\left\langle N\left\vert T^{\prime}\right.  \right\rangle }=0$, and
$\operatorname{Im}\left(  \left\langle N\left\vert N^{\prime}\right.
\right\rangle \right)  =0$. Finally, we point out that \textrm{Span}$\left\{
\left\vert \Psi\left(  s\right)  \right\rangle \text{, }\left\vert T\left(
s\right)  \right\rangle \text{, }\left\vert N\left(  s\right)  \right\rangle
\right\}  $ is equal to \textrm{Span}$\left\{  \partial_{s}\left\vert
\Psi\left(  s\right)  \right\rangle \text{, }\partial_{s}\left\vert T\left(
s\right)  \right\rangle \text{, }\partial_{s}\left\vert N\left(  s\right)
\right\rangle \right\}  $. This is an important remark since we are formally
considering curves on \textquotedblleft generalized Bloch
spheres\textquotedblright\ embedded in the four-dimensional complex Hilbert
space $\mathcal{H}_{2}^{2}$ and the three-dimensional subspace spanned by the
frame fields is big enough to accommodate the derivatives of the frame fields
as well.

\section{Link between geodesic curvature and $\kappa_{\mathrm{AC}}^{2}$}

In this Appendix, we point out the connection between the concept of geodesic
curvature $\kappa_{\mathrm{geo}}$ in differential geometry and our proposed
curvature coefficient $\kappa_{\mathrm{AC}}^{2}$. We begin by remarking that
care is required when studying curvature aspects of a curve \cite{krey91}. For
example, the curvature of a circle on a plane is an intrinsic property of the
circle. Instead, the curvature of a circle on a spherical surface is an
extrinsic property of the circle. Consider a circle of radius $R$. When this
circle is considered as a great circle on a sphere, it has geodesic curvature
$\kappa_{\mathrm{geo}}$ equal to zero.\ Clearly, if the circle lies on the
sphere but is not a great sphere, its geodesic curvature $\kappa
_{\mathrm{geo}}$ differs from zero. For a discussion on spherical curves on a
sphere embedded in three-dimensions, we refer to Refs.
\cite{kazaras12,erdos00}. Moreover, when the circle is viewed as a curve in a
plane, its curvature $\kappa_{\mathrm{FS}}$ is $1/R$, with $\kappa
_{\mathrm{FS}}$ as introduced within the Frenet-Serret apparatus for a curve
in three-dimensional physical space \cite{krey91}. The curvature
$\kappa_{\mathrm{FS}}$ and torsion coefficients $\tau_{\mathrm{FS}}$ for a
curve $\gamma$ in three-dimensional Euclidean space defined by the vector
relation $\vec{r}=$ $\vec{r}\left(  s\right)  $ with $s$ being the arc length
of $\gamma$ are given by
\begin{equation}
\kappa_{\mathrm{FS}}\left(  s\right)  \overset{\text{def}}{=}\frac{\left\Vert
\vec{r}^{\prime}\times\vec{r}^{\prime\prime}\right\Vert }{\left\Vert \vec
{r}^{\prime}\right\Vert ^{3}}\text{, and }\tau_{\mathrm{FS}}\left(  s\right)
\overset{\text{def}}{=}\frac{\left(  \vec{r}^{\prime}\times\vec{r}%
^{\prime\prime}\right)  \cdot\vec{r}^{\prime\prime\prime}}{\left\Vert \vec
{r}^{\prime}\times\vec{r}^{\prime\prime}\right\Vert ^{2}}\text{,} \label{FSCT}%
\end{equation}
respectively, with $\vec{r}^{\prime}\overset{\text{def}}{=}d\vec{r}/ds$.
Alternatively, consider a curve $\tilde{\gamma}$ on a surface $S$ embedded in
a three-dimensional Euclidean space specified by the vectorial relation
$\vec{r}=\vec{r}\left(  u^{1}\left(  s\right)  \text{, }u^{2}\left(  s\right)
\right)  $ with $s$ being the arc length of $\tilde{\gamma}$. Let $\hat{n}$
and $\hat{t}$ be the unit normal and unit tangent vectors to $S$ at a point
$P$ of $\tilde{\gamma}$, respectively. Let $\hat{b}\overset{\text{def}}{=}%
\hat{n}\times\hat{t}$ be the unit vector orthogonal to both $\hat{n}$ and
$\hat{t}$. Then, the curvature vector $\vec{\kappa}$ of $\tilde{\gamma}$ at
$P$ is the vector sum of the normal curvature vector $\kappa_{n}\hat{n}$ and
the geodesic curvature vector $\kappa_{\mathrm{geo}}\hat{b}$ and is defined
as,%
\begin{equation}
\vec{\kappa}\overset{\text{def}}{=}\kappa_{n}\hat{n}+\kappa_{\mathrm{geo}}%
\hat{b}\text{.} \label{curvaturevector}%
\end{equation}
The scalar quantities $\kappa_{n}\left(  s\right)  \overset{\text{def}}{=}$
$\vec{r}^{\prime\prime}\cdot\hat{n}$ and $\kappa_{\mathrm{geo}}\left(
s\right)  \overset{\text{def}}{=}\vec{r}^{\prime\prime}\cdot$ $\hat{b}$ are
the normal and the geodesic curvatures, respectively. As an illustrative
example, consider a spherical curve $\gamma$ on a sphere of radius $R>0$ and
centered at the origin. More specifically, using spherical coordinates (with
$\theta_{\xi}\in\left[  0,\pi\right]  $ being the polar angle as in the main
paper), assume to consider a circle at $z=\cos\left(  \theta_{\xi}\right)  $
and parametrized by the arc length $s$ as%
\begin{equation}
\vec{r}\left(  s\right)  =\left(  R\sin\left(  \theta_{\xi}\right)  \cos
(\frac{s}{R\sin\left(  \theta_{\xi}\right)  })\text{, }R\sin\left(
\theta_{\xi}\right)  \sin(\frac{s}{R\sin\left(  \theta_{\xi}\right)  })\text{,
}R\cos\left(  \theta_{\xi}\right)  \right)  \text{.} \label{stuckonyou}%
\end{equation}
Note that $\vec{r}\left(  s\right)  $ in Eq. (\ref{stuckonyou}) yields a
unit-speed parametrization since $\left\Vert \vec{r}^{\prime}\left(  s\right)
\right\Vert ^{2}=1$. Substituting Eq. (\ref{stuckonyou}) into Eq. (\ref{FSCT})
leads to $\tau_{\mathrm{FS}}\left(  s\right)  =0$ and $\kappa_{\mathrm{FS}%
}\left(  s\right)  =1/\left[  R\sin\left(  \theta_{\xi}\right)  \right]  $.
Moreover, use of $\vec{r}\left(  s\right)  $ into $\kappa_{\mathrm{geo}%
}\left(  s\right)  \overset{\text{def}}{=}\vec{r}^{\prime\prime}\cdot$
$\hat{b}$ with $\hat{b}=\vec{r}/\left\Vert \vec{r}\right\Vert $ gives
$\kappa_{\mathrm{geo}}=\cot(\theta_{\xi})/R$, that is, $\kappa_{\mathrm{geo}%
}^{2}\sim\cos^{2}\left(  \theta_{\xi}\right)  /\sin^{2}\left(  \theta_{\xi
}\right)  $. Thus, $\kappa_{\mathrm{geo}}^{2}\sim\kappa_{\mathrm{AC}}^{2}$ in
Eq. (\ref{until}). With this interesting conclusive remark, we end our
discussion here.

\section{Curvature, torsion, and the quantum Heisenberg model}

In this Appendix, in addition to the illustrative examples discussed in
Section VI, we report an example that considers the curvature and torsion
coefficients for a quantum curve traced by evolving a three-qubit quantum
state, including the $\left\vert GHZ\right\rangle $-state
(Greenberger-Horne-Zeilinger) and the $\left\vert W\right\rangle $-state
(Wolfgang D\"{u}r), under a three-qubit stationary Hamiltonian belonging to
the family of the quantum Heisenberg models.

The quantum Heisenberg model (QHM) is a model of magnetic spin systems used in
statistical mechanics to study thermodynamics aspects of the system
\cite{landau05}. In this model, magnetic spins are treated
quantum-mechanically and, therefore, they are expressed by quantum operators
and not by classical vectors. For a system of $N$ spin-$1/2$ particles with a
nearest-neighbours interaction and immersed in a uniform magnetic field
$\vec{B}=h\hat{z}$, the QHM is given by%
\begin{equation}
\mathrm{H}\overset{\text{def}}{=}\sum_{j=1}^{N}\left(  J_{x}\sigma
_{x}^{\left(  j\right)  }\sigma_{x}^{\left(  j+1\right)  }+J_{y}\sigma
_{y}^{\left(  j\right)  }\sigma_{y}^{\left(  j+1\right)  }+J_{z}\sigma
_{z}^{\left(  j\right)  }\sigma_{z}^{\left(  j+1\right)  }+h\sigma
_{z}^{\left(  j\right)  }\right)  \text{,} \label{QHM}%
\end{equation}
where $\sigma_{a}^{\left(  j\right)  }=\mathrm{I}^{\otimes j-1}\otimes
\sigma_{a}^{\left(  j\right)  }\otimes\mathrm{I}^{\otimes N-j}$, $a\in\left\{
x\text{, }y\text{, }z\right\}  $, $1\leq j\leq N$, $\mathrm{I}$ is the
$2\times2$ identity matrix, and $\sigma_{N+1}=\sigma_{1}$ is the usual
periodic boundary condition. The real quantities $J_{x}$, $J_{y}$, and $J_{z}$
are the coupling constants. When $J_{x}\neq J_{y}\neq J_{z}$, one has the
Heisenberg $XYZ$ model. Furthermore, when $J_{x}=J_{y}\neq J_{z}$, one has the
Heisenberg XXZ model. Finally, when $J_{x}=J_{y}=J_{z}$, one has the
Heisenberg $XXX$ model. For the $XXX$ model, $J_{x}>0$ and $J_{x}<0$
correspond to the antiferromagnetic and ferromagnetic scenarios, respectively.
In our explicit example, we focus on a special case of Eq. (\ref{QHM})
specified by the Hamiltonian%
\begin{align}
&  \mathrm{H}\overset{\text{def}}{=}J_{x}\left(  \sigma_{x}^{\left(  1\right)
}\otimes\sigma_{x}^{\left(  2\right)  }\otimes\mathrm{I}^{\left(  3\right)
}+\sigma_{x}^{\left(  1\right)  }\otimes\mathrm{I}^{\left(  2\right)  }%
\otimes\sigma_{x}^{\left(  3\right)  }+\mathrm{I}^{\left(  1\right)  }%
\otimes\sigma_{x}^{\left(  2\right)  }\otimes\sigma_{x}^{\left(  3\right)
}\right) \nonumber\\
&  +J_{y}\left(  \sigma_{y}^{\left(  1\right)  }\otimes\sigma_{y}^{\left(
2\right)  }\otimes\mathrm{I}^{\left(  3\right)  }+\sigma_{y}^{\left(
1\right)  }\otimes\mathrm{I}^{\left(  2\right)  }\otimes\sigma_{y}^{\left(
3\right)  }+\mathrm{I}^{\left(  1\right)  }\otimes\sigma_{y}^{\left(
2\right)  }\otimes\sigma_{y}^{\left(  3\right)  }\right) \nonumber\\
&  +J_{z}\left(  \sigma_{z}^{\left(  1\right)  }\otimes\sigma_{z}^{\left(
2\right)  }\otimes\mathrm{I}^{\left(  3\right)  }+\sigma_{z}^{\left(
1\right)  }\otimes\mathrm{I}^{\left(  2\right)  }\otimes\sigma_{z}^{\left(
3\right)  }+\mathrm{I}^{\left(  1\right)  }\otimes\sigma_{z}^{\left(
2\right)  }\otimes\sigma_{z}^{\left(  3\right)  }\right) \nonumber\\
&  +h\left(  \sigma_{z}^{\left(  1\right)  }\otimes\mathrm{I}^{\left(
2\right)  }\otimes\mathrm{I}^{\left(  3\right)  }+\mathrm{I}^{\left(
1\right)  }\otimes\sigma_{z}^{\left(  2\right)  }\otimes\mathrm{I}^{\left(
3\right)  }+\mathrm{I}^{\left(  1\right)  }\otimes\mathrm{I}^{\left(
2\right)  }\otimes\sigma_{z}^{\left(  3\right)  }\right)  \text{,}
\label{model}%
\end{align}
that is, in the canonical $\left(  8\times8\right)  $-matrix notation,%
\begin{equation}
\mathrm{H}=\left(
\begin{array}
[c]{cccccccc}%
3h+3J_{z} & 0 & 0 & J_{x}-J_{y} & 0 & J_{x}-J_{y} & J_{x}-J_{y} & 0\\
0 & h-J_{z} & J_{x}+J_{y} & 0 & J_{x}+J_{y} & 0 & 0 & J_{x}-J_{y}\\
0 & J_{x}+J_{y} & h-J_{z} & 0 & J_{x}+J_{y} & 0 & 0 & J_{x}-J_{y}\\
J_{x}-J_{y} & 0 & 0 & -h-J_{z} & 0 & J_{x}+J_{y} & J_{x}+J_{y} & 0\\
0 & J_{x}+J_{y} & J_{x}+J_{y} & 0 & h-J_{z} & 0 & 0 & J_{x}-J_{y}\\
J_{x}-J_{y} & 0 & 0 & J_{x}+J_{y} & 0 & -h-J_{z} & J_{x}+J_{y} & 0\\
J_{x}-J_{y} & 0 & 0 & J_{x}+J_{y} & 0 & J_{x}+J_{y} & -h-J_{z} & 0\\
0 & J_{x}-J_{y} & J_{x}-J_{y} & 0 & J_{x}-J_{y} & 0 & 0 & 3J_{z}-3h
\end{array}
\right)  \text{.} \label{hmodel}%
\end{equation}
In our first sub-case, assume that $\mathrm{H}$ in Eq. (\ref{hmodel}) is the
driving Hamiltonian and the source state to be driven is the entangled quantum
state of three qubits given by $\left\vert GHZ\right\rangle
\overset{\text{def}}{=}\left[  \left\vert 000\right\rangle +\left\vert
111\right\rangle \right]  /\sqrt{2}$. Then, after some tedious but
straightforward calculations, it happens that the curvature $\kappa
_{\mathrm{AC}}^{2}$ and torsion $\tau_{\mathrm{AC}}^{2}$ coefficients are
given by$\allowbreak$%
\begin{equation}
\kappa_{\mathrm{AC}}^{2}\left(  J_{x}\text{, }J_{y}\text{, }J_{z}\text{,
}h\right)  =\frac{4}{3}\left(  J_{x}-J_{y}\right)  ^{2}\frac{h^{2}+\left[
\left(  J_{x}+J_{y}\right)  -2J_{z}\right]  ^{2}}{\left[  3h^{2}+\left(
J_{x}-J_{y}\right)  ^{2}\right]  ^{2}}\text{,} \label{curvaturaH}%
\end{equation}
and,%
\begin{equation}
\tau_{\mathrm{AC}}^{2}\left(  J_{x}\text{, }J_{y}\text{, }J_{z}\text{,
}h\right)  =\frac{4}{3}\left(  J_{x}-J_{y}\right)  ^{2}\frac{h^{2}+\left[
\left(  J_{x}+J_{y}\right)  -2J_{z}\right]  ^{2}}{\left[  3h^{2}+\left(
J_{x}-J_{y}\right)  ^{2}\right]  ^{2}}-\allowbreak\frac{4}{3}\left(
J_{x}-J_{y}\right)  ^{4}\frac{\left[  \left(  J_{x}+J_{y}\right)
-2J_{z}\right]  ^{2}}{\left[  3h^{2}+\left(  J_{x}-J_{y}\right)  ^{2}\right]
^{3}}\text{,} \label{tensioneH}%
\end{equation}
respectively. As a sanity check, note from Eqs. (\ref{curvaturaH}) and
(\ref{tensioneH}) that $0\leq\tau_{\mathrm{AC}}^{2}\left(  J_{x}\text{, }%
J_{y}\text{, }J_{z}\text{, }h\right)  \leq\kappa_{\mathrm{AC}}^{2}\left(
J_{x}\text{, }J_{y}\text{, }J_{z}\text{, }h\right)  $ since we have the
following chain of inequalities,%
\begin{equation}
\frac{4}{3}\left(  J_{x}-J_{y}\right)  ^{2}\frac{h^{2}+\left[  \left(
J_{x}+J_{y}\right)  -2J_{z}\right]  ^{2}}{\left[  3h^{2}+\left(  J_{x}%
-J_{y}\right)  ^{2}\right]  ^{2}}\geq\frac{4}{3}\left(  J_{x}-J_{y}\right)
^{2}\frac{\left[  \left(  J_{x}+J_{y}\right)  -2J_{z}\right]  ^{2}}{\left[
3h^{2}+\left(  J_{x}-J_{y}\right)  ^{2}\right]  ^{2}}\geq\frac{4}{3}\left(
J_{x}-J_{y}\right)  ^{4}\frac{\left[  \left(  J_{x}+J_{y}\right)
-2J_{z}\right]  ^{2}}{\left[  3h^{2}+\left(  J_{x}-J_{y}\right)  ^{2}\right]
^{3}}\text{.} \label{again}%
\end{equation}
The last inequality in Eq. (\ref{again}) is true since it is equivalent to%
\begin{equation}
\frac{\left(  J_{x}-J_{y}\right)  ^{2}}{3h^{2}+\left(  J_{x}-J_{y}\right)
^{2}}\leq1\text{,}%
\end{equation}
which is obviously satisfied since $3h^{2}\geq0$. In our second sub-case,
assume that $\mathrm{H}$ in Eq. (\ref{hmodel}) is the driving Hamiltonian and
the source state to be driven is the entangled quantum state of three qubits
given by $\left\vert W\right\rangle \overset{\text{def}}{=}\left[  \left\vert
001\right\rangle +\left\vert 010\right\rangle +\left\vert 100\right\rangle
\right]  /\sqrt{3}$. Again, after some tedious but straightforward
calculations, it turns out that the curvature $\kappa_{\mathrm{AC}}^{2}$ and
torsion $\tau_{\mathrm{AC}}^{2}$ coefficients become%
\begin{equation}
\kappa_{\mathrm{AC}}^{2}\left(  J_{x}\text{, }J_{y}\text{, }J_{z}\text{,
}h\right)  =\frac{4}{3}\frac{\left(  2h+J_{x}+J_{y}-2J_{z}\right)  ^{2}%
}{\left(  J_{x}-J_{y}\right)  ^{2}}\text{, and }\tau_{\mathrm{AC}}^{2}\left(
J_{x}\text{, }J_{y}\text{, }J_{z}\text{, }h\right)  =0\text{, } \label{go1}%
\end{equation}
respectively. From the expressions of the curvature coefficients in Eq.
(\ref{curvaturaH}) and the first relation in Eq. (\ref{go1}), we see that
curves traced by the $\left\vert GHZ\right\rangle $-state and the $\left\vert
W\right\rangle $-state are differently bent by the driving Hamiltonian.
Moreover, from Eq. (\ref{tensioneH}) and the second relation in Eq.
(\ref{go1}) we observe that while the driving Hamiltonian in Eq.
(\ref{hmodel}) can twist the quantum curve traced by the $\left\vert
GHZ\right\rangle $-state, it cannot twist the curve traced by the $\left\vert
W\right\rangle $-state. These distinct curvature and torsion behaviors under
the same driving Hamiltonian are interesting. Indeed, it is known that the
$\left\vert GHZ\right\rangle $-state and the $\left\vert W\right\rangle
$-state cannot be transformed into each other by local quantum operations.
Therefore, they represent very different types of tripartite entanglement. The
curvature and torsion coefficients seem to detect this difference. However, a
more comprehensive investigation would be necessary to fully understand these behaviors.

\bigskip

\bigskip

\end{document}